\documentclass[a4paper, fleqn,usenatbib]{mnras}
\usepackage{natbib}

\usepackage{newtxtext,newtxmath}

\usepackage[T1]{fontenc}
\usepackage{ae,aecompl}
\usepackage{epsfig}

\usepackage{graphicx}	
\usepackage{amsmath}	
\usepackage{amssymb}	

\def\kms{km s$^{-1}$}

\title[Formation of S0s in extreme environments I]{Formation of S0s in extreme environments I: clues from kinematics and stellar populations.}
\author[L. Coccato et al.]
{Lodovico Coccato$^{1}$\thanks{E-mail: lcoccato@eso.org},
Yara L. Jaff\'e$^2$,
Arianna Cortesi$^{3,4}$,
Michael Merrifield$^{5}$, \and
Evelyn Johnston$^{6}$,
Bruno Rodr\'iguez del Pino$^{7}$, 
Boris Haeussler$^{8}$, 
Ana L. Chies-Santos$^{9}$, \and
Claudia L. Mendes de Oliveira$^{3}$,
Yun-Kyeong Sheen$^{10}$,
Kar\'in Men\'endez-Delmestre$^{4}$.
\bigskip
\\
$^{1}$European Southern Observatory, Karl-Schwarzchild-str., 2, 85748 Garching b. M\"unchen, Germany.\\
$^{2}$Instituto de F\'isica y Astronom\'ia, Facultad de Ciencias, Universidad de Valpara\'iso, Gran Breta$\tilde{\it n}$a 1111, 5030 Casilla, Valpara\'iso, Chile.\\
$^{3}$Instituto de Astronomia, Geof\'isica e Ci\^encias Atmosf\'ericas (IAG), Universidade de S$\tilde{\it a}$o Paulo (USP), R. do Mat$\tilde{\it a}$o 1226, 05508-090 
S$\tilde{\it a}$o Paulo, Brazil.\\
$^{4}$ Observatorio de Valongo, Universidade Federal do Rio de Janeiro, Ladeira do Pedro Ant\^onio, 43, Centro, Rio de Janeiro - RJ, 
20080-090, Brazil. \\
$^{5}$School of Physics and Astronomy, University of Nottingham, University Park, Nottingham NG7 2RD, UK.\\
$^{6}$Pontificia Universidad Catolica de Chile, Av. Vicuna Mackenna 4860, 7820436 Macul, Santiago, Chile.\\
$^{7}$Centro de Astrobiolog\'ia (CSIC-INTA), Torrej\'on de Ardoz, E-28850 Madrid, Spain.\\
$^{8}$European Southern Observatory, Alonso de Cordova 3107 Vitacura Casilla 7630355 Santiago, Chile.\\ 
$^{9}$Departamento de Astronomia, Universidade Federal do Rio Grande do Sul, Av. Bento Gon\c{c}alves 9500, 91501-970 Porto Alegre, RS, Brazil.\\
$^{10}$Korea Astronomy and Space Science Institute, 776, Daedeokdae-ro, Yuseong-gu, Daejeon 34055, Korea.
}

\date{Accepted XXX. Received YYY; in original form ZZZ}

\pubyear{2019}

\begin{document}
\label{firstpage}
\pagerange{\pageref{firstpage}--\pageref{lastpage}}
\maketitle

\begin{abstract}
Despite numerous efforts, it is still unclear whether lenticular galaxies (S0s) evolve from spirals whose star formation was suppressed, or formed trough mergers or disk instabilities. 
In this paper we present a pilot study of 21 S0 galaxies in extreme environments (field and cluster), and compare their spatially-resolved kinematics and global stellar populations. 
Our aim is to identify whether there are different mechanisms that form S0s in different environments. 
Our results show that the kinematics of S0 galaxies in field and cluster are, indeed, different. Lenticulars in the cluster are more rotationally supported, suggesting that they are formed through processes that involve the rapid consumption or removal of gas (e.g. starvation, ram pressure stripping). 
In contrast, S0s in the field are more pressure supported, suggesting that minor mergers served mostly to shape their kinematic properties. 
These results are independent of total mass, luminosity, or disk-to-bulge ratio. 
On the other hand, the mass-weighted age, metallicity, and star formation time-scale of the galaxies correlate more with mass than with environment, in agreement with known relations from previous work such as the one between mass and metallicity. 
Overall, our results re-enforce the idea that there are multiple mechanisms that produce S0s, and that both mass {\it and} environment play key roles. A larger sample is highly desirable to confirm or refute the results and the interpretation of this pilot study.

\end{abstract}




\section{Introduction}
\label{sec:introduction}
Lenticular galaxies (S0s) outnumber galaxies of other morphological types in the local Universe \citep{Bernardi+10}, and their number density has increased with time in clusters and groups since $z\sim1$ at the expenses of spirals (e.g., \citealt{Dressler80, Poggianti+09}).  
Therefore, understanding their formation is a key element in understanding galaxy evolution. 

S0s have long been thought of as quenched spirals since they share the disky morphology of spirals, but have the redder colours of older stellar populations. 
Evidence for this evolutionary scenario comes from studies such as \citet{Dressler80, Dressler+97, Cappellari+11b}, which show that the fraction of S0s increases towards higher density environments and lower redshift, while spirals show the opposite trend.  
A popular explanation is that spiral galaxies falling into clusters lose their gas via ram-pressure stripping (e.g., \citealt{Gunn+72,Steinhauser+12} or tidal interactions (e.g., \citealt{Merluzzi+16}), thus enhancing the population of lenticulars. Quenching of star formation in spiral galaxies is also a possible mechanism: among spiral galaxies hosting a classical bulge, the fraction of quenched galaxies increases with denser environments, which then lead to an increase in the fraction of S0s \citep{Mishra+18}.
However, observations of nearby isolated S0s \citep{vandenBerg+09} have suggested that there must be alternative evolutionary pathways to form S0s.  
One possibility is that isolated S0 galaxies are the end product of past minor mergers that used up the gas in the disc, concentrating it to the central parts of the galaxy to form a classical bulge, analogous to ``fossil groups'' \citep{Ponman+94, Arnold+11}.  
Similarly, mergers could induce the formation of a bar in the disc that can, in turn, build up a pseudo-bulge.  
N-body simulations \citep{Bournaud+05, Eliche-Moral+12} have indeed suggested that dry intermediate (mass ratios of 1:4$-$1:7) and minor ($<$1:7) mergers can induce global structural evolution resulting in S0 systems.  
Alternatively, S0s could simply be faded field spiral galaxies that exhausted their gas reservoirs and lost their spiral structures through disc instabilities (starvation, \citealt{Eliche-Moral+13}). Finally, low mass S0s could also have originated from primordial galaxies formed at redshift $\sim 2$ through violent disk instability and fragmentation \citep{Saha+18}.

Much observational effort has been made to distinguish between these scenarios. 
For instance studies, of nearby S0s (either isolated or in groups) have revealed that their discs are dynamically hotter than those in spirals of similar luminosity \citep{Cortesi+13}, and that their blue globular cluster population has a wide range of ages \citep{Chies-Santos+11, Lee+19}, suggesting past interaction with other galaxies, and therefore favouring minor-merger scenarios for the formation of S0s. 
In contrast, other recent studies on globular cluster kinematics \citep{Bellstedt+17} do not favour mergers as formation mechanism for lenticular galaxies, although they cannot entirely rule them out. 
Moreover, using 3D spectroscopy, \citet{Katkov+14} found that the number of field S0s with
counter-rotating gas kinematics is higher than in denser environments, implying that this gas could have been accreted from dwarf satellites. 
However, the S0 Tully-Fisher relation \citep{Tully+77} has been found to be systematically fainter at fixed rotational velocity than the spiral relation \citep{Williams+10, Cortesi+13, Jaffe+14}, suggesting that S0s are faded spirals that consumed their gas reservoirs. 
Other studies have further shown that the luminosity of bulges in S0s, relative to their associated discs, are brighter than expected from a simple cessation of star formation in the disc \citep{Christlein+04}. 
In addition, \citet{Johnston+12, Johnston+14} found evidence of younger stellar populations in the bulge regions of Fornax and Virgo S0s, triggered by residual disc gas that was channelled into the centre of the galaxy, inducing star formation and thus increasing the luminosity of the bulge relative to the disc. Similar behaviour is observed in 13 post-starburst spiral galaxies in the Abell Cluster S1077 (AC114), which had their last episode of star formation towards the central parts of the galaxies \citep{DelPino+14}, suggesting indeed a link between S0s and spirals.

Broadly speaking, the main formation mechanisms of lenticular galaxies discussed above related either to mass accretions (e.g. merger) or gas-related effects (e.g. ram pressure stripping, gas accretion, quenching of star formation). 
Undoubtedly, different environments can contribute in different ways to these mechanisms (see for example \citealt{Aguerri12} and \citealt{Vollmer13} for reviews). 
It is therefore fair to ask whether or not S0s in the field have different properties from S0s in clusters.  
The best way to characterise the global properties of galaxies is to combine morphological and kinematic information, with a spatially resolved analysis of their stellar kinematics and populations out to large radii.  
To this end, we have embarked on a project aimed at studying the properties of lenticular galaxies in extreme environments. 
Recent attempts (e.g. \citealt{Fraser+18, Rizzo+18}) found no significant dependency with the environment; in particular \citet{Fraser+18} indicates mass as the main driver for S0 formation.
However their sample did not contain galaxies in cluster, whereas in this study, we seek to optimise any signal by select disk galaxies from the extremes of environments, field and cluster.  
The main scientific drivers of our study are: i) investigate whether or not S0s living in extreme environments have different properties. Then, if differences are indeed present, ii) establish a link between these differences and the formation mechanisms, and hence iii) identify the dominant formation mechanism that acts in a given environment.

This paper presents the first results of the project, based on tailored pilot MUSE \citep[The Multi Unit Spectroscopic Explorer][]{Bacon+10} observations of a small sample of galaxies and complementary integral-field data from the literature. 
Here, we concentrate mainly on the spatially-resolved kinematic analysis (stellar $v/\sigma$ and specific angular momentum), presence of ionised gas, and on the average properties of the stellar population ({mass-weighted age, metallicity and star-formation time-scale). 
Future works will present a detailed analysis of the stellar populations of the galaxy structural components (such as bulge and disk).  
The sample of galaxies studied in this paper is described in Section \ref{sec:sample}. 
Data reduction and analysis of new galaxies are discussed in Section \ref{sec:muse_sample}. 
The measurement of global properties both for new and literature data and the results of their comparison is discussed in Section \ref{sec:Results}. 
Finally, Section \ref{sec:conclusions} presents the initial conclusions from this study.

\begin{figure}
\vbox{
\psfig{file=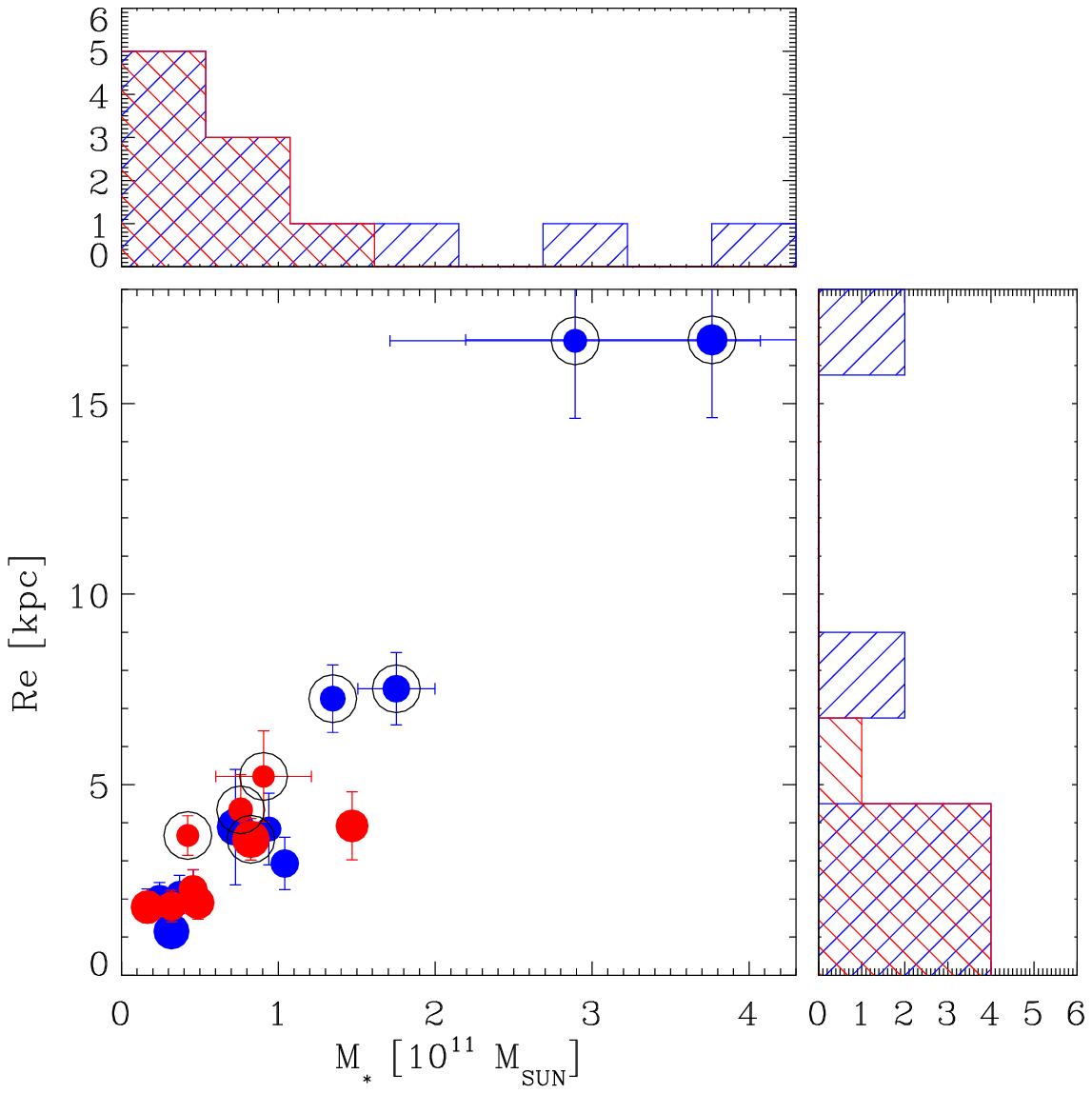,clip=,width=8.2cm}
\psfig{file=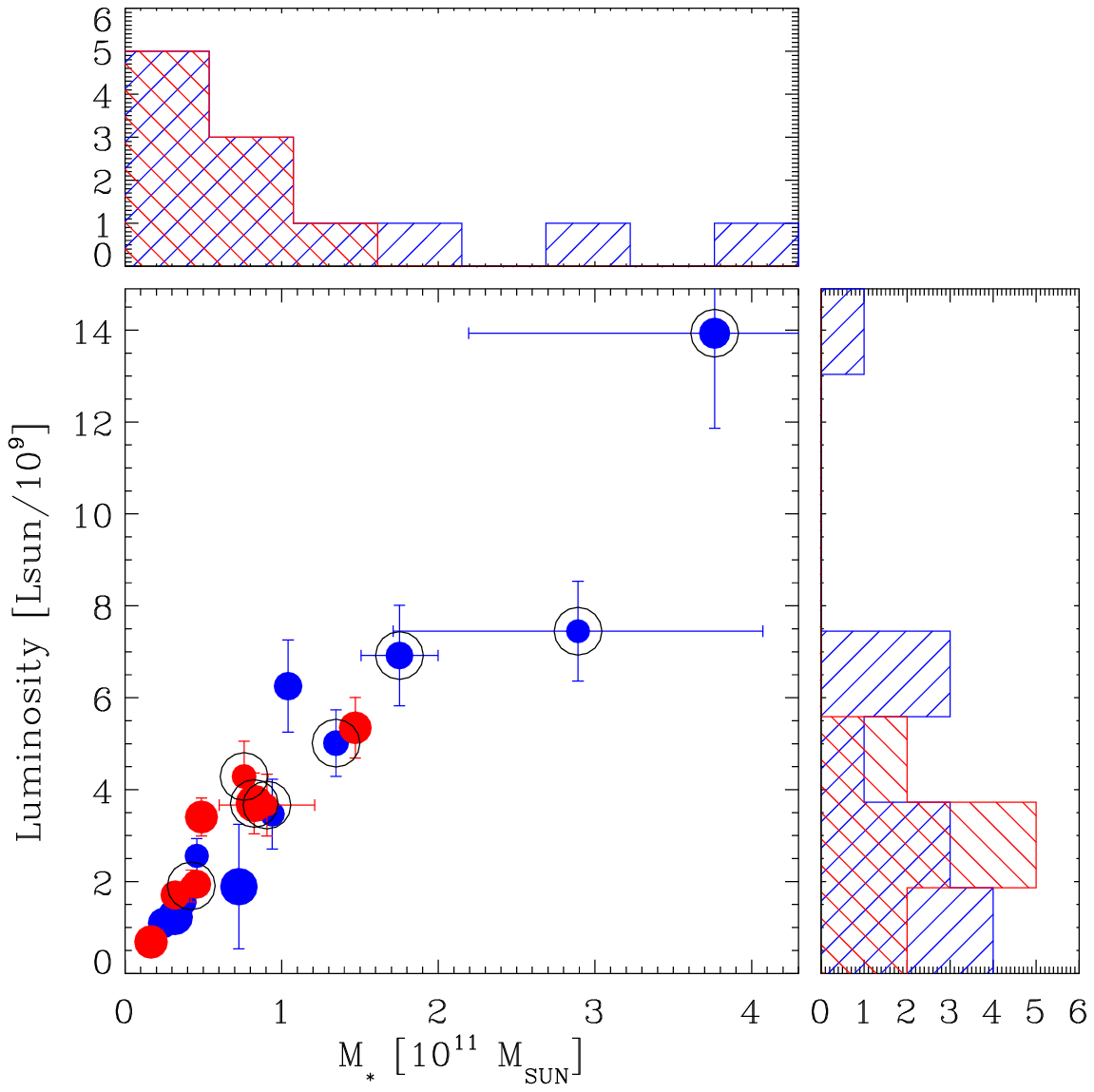,clip=,width=8.2cm,}}
\caption{Distribution of the basic properties of the sample galaxies. The plots show the total mass versus the effective radius (upper plot) and the total luminosity (lower plot), the histogram of their distribution on the sides of the axis. Blue indicates field galaxies, whereas red indicates cluster galaxies. The size of the symbols is proportional to the Disk-to-total light ratio, disky galaxies are represented by bigger symbols. Galaxies in the MUSE sample are highlighted by a black circle.}
\label{fig:sample}
\end{figure}

\section{The sample of galaxies} 
\label{sec:sample}

Galaxies in our sample were chosen from a pilot MUSE  program that includes dedicated observations of 8 S0s in extreme environments (i.e., field and cluster)\footnote{Prog ID: 096.B-0325, P.I. Jaff\'e.}, plus 13 galaxies from publicly-available data of the ATLAS3D survey \citep{Cappellari+11a} to augment the sample.   

One part of the sample comprises isolated galaxies, drawn from the 2MASS Isolated Galaxies Catalogue (2MIG, \citealt{Karachentseva+10}). 
For the MUSE campaign we selected 4 2MIG targets that were visually classified as S0s by the team, and that were observable from Paranal. 
To these were added the 8 S0 galaxies studied by the ATLAS 3D survey that are present in the 2MIG catalogue or that are in an environment with galaxy three-dimensional density\footnote{\citet{Cappellari+11b} defines $\rho_{10}$  as the mean density of galaxies inside a sphere centred on the galaxy and containing the 10 nearest neighbours.} lower than $\log \rho_{10}/[Mpc^{-3}] \leq -2.5$.

The rest of the sample galaxies belong to dense galaxy clusters. 
For the MUSE campaign we selected 4 S0s in the Centaurus cluster, from the sample of \citet{Jerien+97}.
Tho these were added the 5 galaxies in the ATLAS 3D survey that live in an environment with a galaxy three-dimensional density greater than $\log \rho_{10}/[Mpc^{-3}] \geq 1$\footnote{The adopted thresholds on $\rho_{10}$ (1 and --2.5) represent the densest and less dense environment bins of the ATLAS3D survey (e.g. Figure 8 in \citealt{Cappellari+11b})}.

All the galaxies in this study have morphological Hubble type $-3.5<T<-.5$, from LEDA and do not show clear evidence of strong bars or tidal interactions in their visual appearance.  
These morphological criteria were the same as those used in the ATLAS 3D survey to disentangle S0s from ellipticals. However, because the classification of S0s in the past was very much influenced by the limited dynamic range of photographic plates, we have also visually inspected the sample galaxies. In addition, for the MUSE sample, the presence of a disk-like rotating stellar structure is also reflected in the kinematics (see Appendix \ref{sec:app_a}). 
For the ATLAS 3D sample, their S0 nature was also confirmed by \citet{Krajnovic+13}, with the exception of 8 galaxies (marked with asterisks in the last column of Table  \ref{tab:best_fit_parameters}). 
The morphological structure of these 8 galaxies was uncertain to \citet{Krajnovic+13} and therefore these galaxies were classified as ``single Sersic component objects'' (although some of them, e.g. NGC 6548, are clearly dominated by a disk component). 
We therefore performed an independent image decomposition on archival images (VLT-VIMOS or 2MASS archives) to verify the presence of a disk component in the photometric profile
  \footnote{The reduced images used by \citet{Krajnovic+13} are not publicly available.}. 
The inclusion or exclusion of these 8 objects and their impact on our results will be discussed in Section \ref{sec:morph_bias}.

The total number of lenticular galaxies considered in this study is 21, 12 isolated and 9 in cluster (although not the densest environments in the Universe, the Virgo and Centaurus Clusters from which these objects were selected are very much cluster environments, and have the key benefit of observability and plentiful data from the literature). Table~\ref{tab:sample} and Figure \ref{fig:sample} summarise their main properties. One can immediately see from Fig. \ref{fig:sample} that the sample in this pilot study is not homogeneous: massive and luminous galaxies are under-represented in the cluster sample. The effects of selection biases will be discussed in Sect. \ref{sec:observational_biases}. 
The scatter on the relations shown in Fig. \ref{fig:sample} is mainly dominated by measurement errors (uncertainty on galaxy distance has been included in the luminosity error computation) and by the degeneracy between baryonic matter and dark matter that is present in the dynamical models.
\begin{table*}
\caption{Main properties of the sample galaxies.} 
\begin{scriptsize}
\begin{tabular}{lcccccccccccc}
\hline
GALAXY     & Alt. NAME &  Sample& Type  &Distance      &$M_K$            & $M_B$             &   W1            &  $\log\left(\frac{R_e}{\rm ["]}\right)$ & $\frac{V_{\rm rot}}{\sigma\left(R_e\right)}$&$\lambda(R_e)$ & Gas    & Exposure \\
           &           &               & &   [Mpc]      & [mag]           & [mag]             &   [mag]         &                                             &                                    &                & &    [s] \\
 (1)       &    (2)    &  (3)          &   (4) &   (5)        & (6)             &  (7)              &   (8)           &                 (9)                         &                (10)                 &   (11)         & (12)   & (13)\\
\hline                                                                                                                                                      
field      &           &               &       &              &                 &                   &                 &                                             &                                    &                &        &                \\
PGC 004187 &2MIG  131  &  M         & --2.9 &$106.5\pm 7.5$&$-24.67\pm 0.16$ &  --               &$-22.2 \pm 0.2 $ &             1.508                           &               0.93                 &     0.36       &   n    & $670\times 3$ (O)                     \\
IC  1989   &2MIG  445  &  M         & --2.9 &$157.1\pm11.1$&$-25.36\pm 0.16$ & $-21.6 \pm 0.3 $ &$-22.8 \pm 0.2 $ &             1.340                           &               0.73                 &     0.28       &   i    & $990\times 9$ (O) + $90\times 5$ (S)  \\
NGC 3546   &2MIG 1546  &  M         & --2.5 &$64.0\pm 4.5$&$-24.25\pm 0.16$ &  --               &$-21.6 \pm 0.2 $ &             1.369                           &               2.43                 &     0.60       &   y    & $670\times 4$ (O)                    \\
PGC 045474 &2MIG 1814  &  M         & --3.4 &$93.8\pm 7.2$&$-24.60\pm 0.17$ &  --               &$-22.0 \pm 0.2 $ &             1.218                           &               0.41                 &     0.19       &   y    & $550\times 4$ (O) + $60\times 2$ (S)  \\
NGC 2880   &2MIG 1275  &A$^{\rm 3D}$& --2.6 &$21.3\pm 1.9$&$-22.98\pm 0.20$ & $-19.2 \pm 0.3 $ &$-20.0 \pm 0.2 $ &             1.312                           &               2.03                 &     0.55       &   y    & -- \\
NGC 3098   &2MIG 1374  &A$^{\rm 3D}$& --1.5 &$23.0\pm 1.8$&$-22.72\pm 0.18$ & $-19.0 \pm 0.2 $ &$-19.9 \pm 0.2 $ &             1.013                           &               1.27                 &     0.35       &   y    & -- \\
NGC 6149   &UGC 10391  &A$^{\rm 3D}$& --2.0 &$37.2\pm 3.0$&$-22.60\pm 0.18$ &   --              &$-19.9 \pm 0.2 $ &             1.039                           &               1.52                 &     0.52       &   y    & -- \\
NGC 6278   &UGC 10656  &A$^{\rm 3D}$& --1.8 &$42.9\pm 3.4$&$-24.49\pm 0.17$ & $-20.0 \pm 0.2 $ &$-21.3 \pm 0.2 $ &             1.149                           &               2.01                 &     0.38       &   y    & -- \\
NGC 6548   &NGC4549    &A$^{\rm 3D}$& --1.9 &$22.4\pm 7.2$&$-23.19\pm 0.72$ & $-19.6 \pm 0.8 $ &$-19.2 \pm 0.7 $ &             1.554                           &               2.18                 &     0.36       &   y    & -- \\
NGC 6703   &UGC 11356  &A$^{\rm 3D}$& --2.8 &$25.9\pm 2.8$&$-23.85\pm 0.24$ & $-20.1 \pm 0.3 $ &$-20.9 \pm 0.2 $ &             1.485                           &               0.05                 &     0.06       &   i    & -- \\
NGC 6798   &2MIG 2649  &A$^{\rm 3D}$& --1.9 &$37.5\pm 2.7$&$-23.52\pm 0.16$ &   --              &$-20.7 \pm 0.2 $ &             1.093                           &               0.75                 &     0.31       &   y    & -- \\
PGC 056772 &NSA 073372 &A$^{\rm 3D}$& --1.0 &$39.5\pm 3.2$&$-22.06\pm 0.18$ &   --              &$-19.6 \pm 0.2 $ &             0.982                           &               0.47                 &     0.33       &   y    & -- \\
cluster    &           &               &              &                 &                   &                 &                                             &                                    &                &        &   \\
NGC 4696D  &CCC   43   &  M         & --2.1 &$48.7\pm 4.4$&$-23.91\pm 0.20$ & $-20.0 \pm 0.3 $ &$-21.4 \pm 0.2 $ &             1.344                           &               2.37                 &     0.46       &   n    & $520\times 4$ (O) + $90\times 2$ (S) \\
NGC 4706   &CCC  122   &  M         & --1.9 &$48.7\pm 4.3$&$-24.08\pm 0.19$ & $-20.1 \pm 0.3 $ &$-21.3 \pm 0.2 $ &             1.264                           &               2.71                 &     0.62       &   i    & $850\times 6$ (O) + $80\times 3$ (S)  \\
PGC 043435 &CCC  137   &  M         & --2.1 & $48.7\pm 4.3$&$-23.92\pm 0.19$ & $-20.0 \pm 0.3 $ &$-21.3 \pm 0.2 $ &             1.179                           &               1.84                 &     0.40       &   n    & $520\times 4$ (O)                     \\
PGC 043466 &CCC  158   &  M         & --2.1 &  $48.7\pm 4.3$&$-23.20\pm 0.20$ & $-19.4 \pm 0.3 $ &$-20.5 \pm 0.2 $ &             1.191                           &               1.35                 &     0.42       &   i    & $890\times 6$ (O)                   \\
NGC 4425   &VCC 0984   &A$^{\rm 3D}$ &--0.6 & $16.5\pm 1.0$&$-22.09\pm 0.13$ & $-18.5 \pm 0.2 $ &$-19.0 \pm 0.1 $ &             1.349                           &               0.98                 &     0.38       &   n    & -- \\
NGC 4429   &VCC 1003   &A$^{\rm 3D}$ &--0.8 & $16.5\pm 1.0$&$-24.32\pm 0.13$ & $-20.1 \pm 0.1 $ &$-19.0 \pm 0.1 $ &             1.690                           &               1.29                 &     0.40       &   y    & -- \\
NGC 4435   &VCC 1030   &A$^{\rm 3D}$ &--2.1 & $16.7\pm 1.0$&$-23.83\pm 0.13$ & $-19.5 \pm 0.1 $ &$-20.4 \pm 0.1 $ &             1.371                           &               1.45                 &     0.54       &   y    & -- \\
NGC 4461   &VCC 1158   &A$^{\rm 3D}$ &--0.7 & $16.5\pm 1.0$&$-23.08\pm 0.13$ & $-19.1 \pm 0.1 $ &$-20.00\pm 0.1 $ &             1.356                           &               1.65                 &     0.49       &   n    & -- \\
NGC 4503   &VCC 1412   &A$^{\rm 3D}$ &--1.7 &  $16.5\pm 1.0$&$-23.22\pm 0.13$ & $-19.0 \pm 0.2 $ &$-20.  \pm 0.1 $ &             1.449                           &               2.12                 &     0.45       &   n    & -- \\
\hline
\end{tabular}
\end{scriptsize}
\label{tab:sample}
\begin{minipage}{18cm}
  Notes: Columns 1-2: name of the galaxy. 
  Column 3: dataset the galaxy belongs to: M=MUSE dataset, A$^{\rm 3D}$= Atlas$^{\rm 3D}$.
  Column 4: Morphological type code according to LEDA (http://leda.univ-lyon1.fr/).
  Columns 5-6: distance and total $M_k$ luminosity. For galaxies in the MUSE sample, magnitudes are from the 2MASS Extended Source Image server (Cluster sample) and from \citet[field sample]{Karachentseva+10}, distances from NED (NASA/IPAC Extragalactic Database), assuming $H_0=70$ km s$^{-1}$ Mpc$^{-1}$, $\Omega_\Lambda = 0.3$, $\Omega_M=0.7$. For galaxies in the ATLAS 3D sample, magnitudes and distances are as reported in \citet{Cappellari+11a}. Error on magnitude includes the contribution from the error on the distance.
  Column 7: Total apparent "face-on" magnitude corrected for galactic and internal extinction, and for redshift (from \citealt{RC3}). Error on magnitude includes the contribution from the error on the distance. 
  Column 8: WISE \citep{Wright+10} W1 magnitudes at 3.4 $\mu$, corrected for internal and galactic extinction, and with aperture and $k$ corrections \citep{Sorce+12}. Error on magnitude includes the contribution from the error on the distance.
  Column 9: log$_{10}$ of the effective radius. For galaxies in the MUSE sample, the value is fitted with {\tt galfit} \citep{Peng+02} on the reconstructed image obtained from the datacubes its associated error is $\sim$ 10\%. For the ATLAS 3D sample the value of $R_e$ is as reported by \citet{Cappellari+13}, its associated error is $\approx$ 22\%, according to Section 4.1 of \citet{Cappellari+13}.
  Columns 10 and 11: values of  $V_{\rm rot}/\sigma$ and the $\lambda$ parameter within 1 effective radius, as computed in Sections \ref{sec:vsigma} and \ref{sec:lambda}. 
  Column 12: flag indicating the presence of ionised-gas in the spectra (y=yes, n=no, i=only in the innermost
  regions). 
  Column 13 (for MUSE sample only): observational set-up, indicating the exposure time in seconds for each exposure, the number of exposures on target (O) and on sky offset (S, if acquired).
\end{minipage}
\end{table*}

\begin{table*}
  \caption{Best fit parameters of the kinemetry and Jeans axisimmetric models.}
\begin{tabular}{lccccccccccc}
\hline
GALAXY  &$\langle PA_{\rm KIN} \rangle$&$\langle q \rangle$&$V_{\rm circ}$ &$M_{\rm star}$        & $M/L_{\rm star}$   &$Incl$  &  $\beta$     & $M_{\rm DM}$      & $r_s$       &$D/T$\\
            &                  &                    &  [km s$^{-1}$]    &[$M_{\odot}$/$10^{10}$]&                  &[deg]  &               &[$M_{\odot}$/$10^{10}$]&   [pc]  &    \\
 (1)        &        (2)       &        (3)         &        (4)        &     (5)            &     (6)          & (7)   &  (8)          &    (9)          &  (10)       &(11) \\
\hline                                                                                                                             
field       &                  &                    &                   &                    &                  &       &               &                 &             &     \\     
PGC 004187  & 121.7 $\pm$ 0.4  &  0.44 $\pm$ 0.05   &      --           &   28.92 $\pm$11.8  & 7.14$\pm$0.04    &    90 &  0.09$\pm$0.01&        --       &      --     & 0.49 \\
IC  1989    & 136.5 $\pm$ 0.4  &  0.59 $\pm$ 0.07   &      --           &   37.64 $\pm$15.70 & 5.53$\pm$0.42    &    90 &  0.13$\pm$0.03&  6.00 $\pm$ 1.28&  2009$\pm$11& 0.76 \\
NGC 3546    & 100.0 $\pm$ 0.4  &  0.73 $\pm$ 0.11   &    262$\pm$ 26    &   13.47 $\pm$0.40  & 5.34$\pm$0.03    &    90 &  0.09$\pm$0.01& 81.39 $\pm$ 0.76& 20007$\pm$40& 0.56 \\
PGC 045474  & 141.9 $\pm$ 0.9  &  0.62 $\pm$ 0.08   &    271$\pm$ 27    &   17.52 $\pm$2.46  & 5.00$\pm$0.14    &    90 &  0.16$\pm$0.03& 10.42 $\pm$ 1.28&  2000$\pm$13& 0.63 \\
NGC 2880    & 142.8 $\pm$ 0.6  &  0.71 $\pm$ 0.10   &    201$\pm$ 20    &   3.70  $\pm$0.12  & 4.28$\pm$0.03    &    51 &--0.09$\pm$0.03&        --       &      --     & 0.62$^{(**)}$ \\
NGC 3098    & 269.9 $\pm$ 1.0  &  0.50 $\pm$ 0.06   &    193$\pm$ 19    &   3.19  $\pm$0.11  & 4.89$\pm$0.03    &    90 &  0.18$\pm$0.01&        --       &      --     & 0.94 \\
NGC 6149    & 200.2 $\pm$ 0.5  &  0.71 $\pm$ 0.10   &    147$\pm$ 15    &   2.43  $\pm$0.06  & 4.11$\pm$0.03    &    66 &  0.01$\pm$0.02&        --       &      --     & 0.73 \\
NGC 6278    & 306.5 $\pm$ 0.2  &  0.56 $\pm$ 0.07   &    268$\pm$ 27    &   10.41 $\pm$0.29  & 5.43$\pm$0.03    &    66 &  0.13$\pm$0.01&        --       &      --     & 0.66 \\
NGC 6548    &  63.1 $\pm$ 0.5  &  0.47 $\pm$ 0.05   &    234$\pm$ 23    &   7.27  $\pm$0.44  & 7.25$\pm$0.06    &    19 &--0.96$\pm$0.39&        --       &      --     & 1.00$^{(**)}$ \\
NGC 6703    & 125.7 $\pm$25.6  &  0.38 $\pm$ 0.04   &    263$\pm$ 26    &   9.40  $\pm$0.07  & 5.92$\pm$0.01    &    19 &--0.08$\pm$0.02&        --       &      --     & 0.53$^{(**)}$ \\
NGC 6798    & 138.5 $\pm$ 1.1  &  0.28 $\pm$ 0.03   &    191$\pm$ 19    &   4.58  $\pm$0.12  & 4.29$\pm$0.03    &    84 &  0.09$\pm$0.01&        --       &      --     & 0.50 \\
PGC 056772  & 190.9 $\pm$26.6  &  0.40 $\pm$ 0.04   &    129$\pm$ 13    &   1.64  $\pm$0.04  & 3.91$\pm$0.03    &    64 &  0.46$\pm$0.01&        --       &      --     & 0.37$^{(**)}$ \\
cluster     &                  &                    &                   &                    &                  &       &               &                 &             &      \\
NGC 4696D   & 318.4 $\pm$ 0.4  &  0.44 $\pm$ 0.05   &      --           &   9.06  $\pm$0.38  & 5.37$\pm$0.04    &    90 &  0.07$\pm$0.01&        --       &      --     & 0.44 \\
NGC 4706    &  27.7 $\pm$ 0.2  &  0.48 $\pm$ 0.06   &    205$\pm$ 21    &   7.60  $\pm$0.25  & 4.81$\pm$0.03    &    90 &--0.03$\pm$0.01&        --       &      --     & 0.52 \\
PGC 043435  &  13.2 $\pm$ 0.3  &  0.52 $\pm$ 0.06   &      --           &   8.25  $\pm$0.47  & 6.93$\pm$0.06    &    90 &  0.05$\pm$0.01&        --       &      --     & 1.00 \\
PGC 043466  & 325.7 $\pm$ 0.4  &  0.43 $\pm$ 0.05   &      --           &   4.23  $\pm$0.60  & 5.00$\pm$0.14    &    90 &  0.00$\pm$0.02&        --       &      --     & 0.45 \\
NGC 4425    & 207.0 $\pm$ 0.3  &  0.38 $\pm$ 0.04   &    117$\pm$ 12    &   1.66  $\pm$0.01  & 4.05$\pm$0.01    &    90 &  0.30$\pm$0.00&        --       &      --     & 0.85$^{(**)}$ \\
NGC 4429    &  85.3 $\pm$ 1.3  &  0.68 $\pm$ 0.09   &    283$\pm$ 28    &   14.70 $\pm$3.05  & 6.14$\pm$0.21    &    70 &  0.00$\pm$0.01&        --       &      --     & 0.82$^{(*)}$ \\
NGC 4435    & 192.8 $\pm$ 0.6  &  0.45 $\pm$ 0.05   &    237$\pm$ 24    &   4.88  $\pm$0.27  & 3.91$\pm$0.06    &    68 &  0.00$\pm$0.02&        --       &      --     & 0.83 \\
NGC 4461    &  12.1 $\pm$ 0.5  &  0.57 $\pm$ 0.07   &    190$\pm$ 19    &   3.20  $\pm$0.08  & 3.93$\pm$0.02    &    71 &  0.12$\pm$0.01&        --       &      --     & 0.69$^{(**)}$ \\
NGC 4503    & 182.5 $\pm$ 0.8  &  0.70 $\pm$ 0.10   &    208$\pm$ 21    &   4.57  $\pm$0.13  & 5.07$\pm$0.03    &    67 &  0.24$\pm$0.01&        --       &      --     & 0.68$^{(**)}$ \\
\hline                                                                                                        
\end{tabular}                                                                                                 
\label{tab:best_fit_parameters}
\begin{minipage}{18cm}
Notes: Columns 1 name of the galaxy. Columns 2- 3: median values of the kinematic position angle and flattening computed by the kinemetry analysis. Column 4: Circular velocity computed as mean of the values in the last 75\% of the radial extend of the $V_{\rm circ}$ curve. Columns 5-10: best fit parameters of the Jeans axisimmetric model. Column 11: dist-to-total luminosity in the R-band. Values of the MUSE sample are from our photometric bulge-disk decomposition of MUSE observations. Values for the ATLAS3D sample are from \citet{Krajnovic+13}, except for $^{(*)}$ for which we applied photometric decomposition with {\tt galfit} on reduced ESO archive data (P.I. Puzia, Prog. ID: 090.B-0498(A), Dataset ID: ADP.2018-03-26T11\_15\_10.464), and for $^{(**)}$, for which we applied photometric decomposition with {\tt galfit} on 2MASS images \citep{Skrutskie+06}.
\end{minipage}
\end{table*}

In the following we describe the MUSE observations and data reduction. 

\subsection{The MUSE observations and data reduction}
\label{sec:muse_sample}

Observations were executed with the Multi Unit Spectroscopic Explorer
(MUSE) mounted on Unit Telescope 4 a the ESO La Silla Paranal
observatory (Chile). MUSE was configured in WFM-NOAO-N mode (wide
field of view, no adaptive optics, nominal wavelength coverage) that
ensured a spatial sampling of 0.2 arcsec per spaxel and a spectral
coverage of 4750 -- 9350 \AA\ with a nominal resolving power of
$R\sim 2000$.

Observations were executed in service mode between October 2015 and
February 2016 (Period 96) and organised in a series of observing
blocks. Each observing block contains several exposures on target that
were slightly dithered and rotated by 90 degrees with respect to each
other to minimise the signature of the IFU geometry in the final
reduced product. For the targets that had a size comparable to the
field of view, dedicated offset sky exposures were included in each
observing block.

Data reduction was performed with the MUSE pipeline
\citep{Weilbacher+12} version 2.2 executed under the EsoReflex
environment \citep{Freudling+13}. Depending on the projected size of
each target, the contribution of the sky background was computed
either from dedicated exposures executed close in time to the science
observations (within the same observing block) or from the edges of
the field of view, where the contribution of the galaxy was
negligible. After removing the sky background, the exposures were then aligned and co-added using bright
sources as reference. 
Whenever available, the cubes of dedicated sky exposures were also sky subtracted and co-added following the same
dither pattern as the science observations, creating ``master cubes'' of sky residuals.
Sky residuals were further reduced using the ZAP
algorithm \citep{Soto+16}; this method models the sky residuals
with a series of principal components and then it fits and subtracts
them from the science cubes. The sky-residual components were
evaluated on the master sky residual cubes or on the border of the
field of view, depending on the dimension of the target.

Adjacent spaxels in the datacube were co-added using Voronoi  tessellation as implemented by \citet{Cappellari+03}, seeking a target signal-to-noise ratio of 50 per pixel.
Spectra in each spatial bin were fitted using the {\tt ppxf} \citep{Cappellari+04} and {\tt
gandalf} \citep{Sarzi+06} spectral fitting procedures to extract stellar and ionized-gas kinematics. 
Stellar templates from the MIUSCAT spectral library \citep{Vazdekis+12} at the nominal MUSE spectral resolution of $FWHM=2.51$ \AA\ were used in the fit. 
The spectral range running from 4750--7000 \AA\ was used, with regions affected by high residuals from sky or telluric lines masked and excluded from the fit.
The two-dimensional maps of velocity, velocity dispersion and higher Gauss-Hermite moments are shown in Figures \ref{kinemaps_i}-\ref{kinemaps_cc}.

\section{Analysis}
\label{sec:Results}

In this section we define the global galaxy properties we use to study the differences between cluster and field S0s and the analysis carried out to measure them.  
In particular, we will focus our attention on $V_{\rm rot}/\sigma$ and the proxy for specific angular momentum $\lambda$. 
Then, we compute the Tully-Fisher relation between the circular velocity and the total luminosity in K-band, B-band and W1 (3.4 $\mu$m) band. Finally, we derived the mass-weighted values of age, metallicity and star formation time-scale within 1$R_e$.

\begin{figure*}
\psfig{file=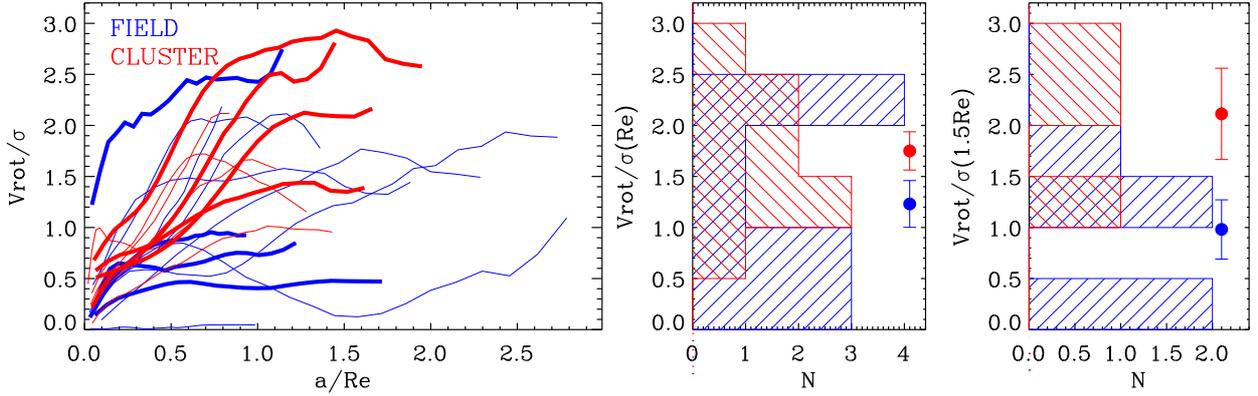,clip=,width=17cm}
\caption{Comparison between the $V_{\rm rot} / \sigma$ radial profiles
  of field (blue) and cluster (red) galaxies in our sample. Thick
  lines identify the MUSE sample, thin lines the ATLAS3D sample. The
  right-hand side of the figure shows the histograms of the value of
  $V_{\rm rot} / \sigma$ at 1 $R_e$ and $1.5 R_e$.}
\label{fig:vsigma}
\end{figure*}

\begin{figure*}
\psfig{file=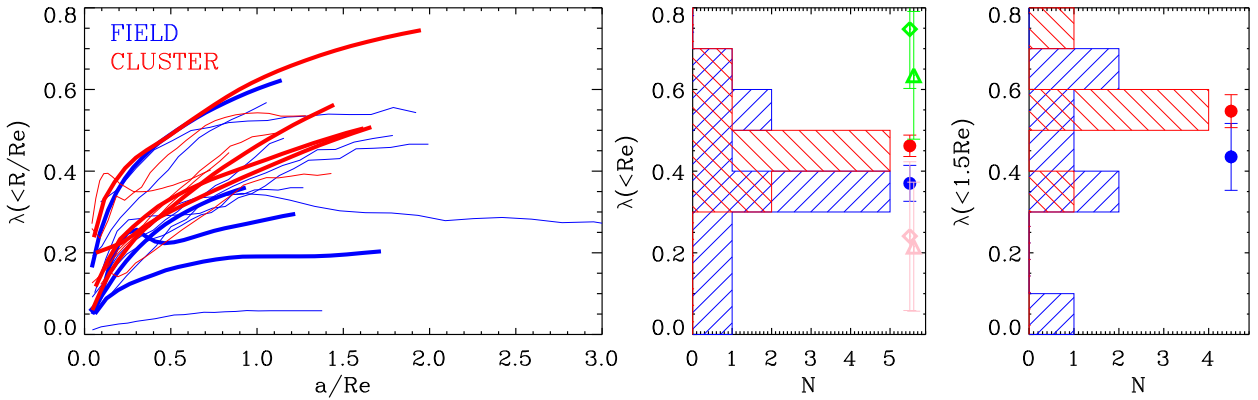,clip=,width=17cm}
\caption{Comparison between the radial profiles of the proxy for specific angular momentum ($\lambda$) in field galaxies (blue) and cluster galaxies (red) in our sample. Thick lines identify the MUSE sample, thin lines the ATLAS3D sample. The right-hand side of the figure shows the histograms of the value of $\lambda$ at 1 $R_e$ and $1.5 R_e$. The green symbol shows the mean value for spiral galaxies at $1 R_e$
(diamonds: \citealt{Graham+18}, triangles: \citealt{barroso+19}), the error bar is the standard deviation of the values. The pink symbol shows the mean value of ellipticals (diamonds: galaxies in Table B1 of \citealt{Emsellem+11} with morphological type  T$\leq-3.5$; triangles: \citealt{barroso+19}), the error bar is the standard deviation of the values.}
\label{fig:lambda}
\end{figure*}

\subsection{$V_{\rm rot}/\sigma$ radial profiles }
\label{sec:vsigma}
The $V_{\rm rot}/\sigma$ radial profile is computed as the ratio of
the stellar rotation velocity and velocity dispersion calculated along
the kinematic semi-major axis. We performed a harmonic first-term
expansion of the observed two-dimensional velocity and velocity
dispersion maps using the {\tt kinemetry} code by
\citet{Krajnovic+06}.  At each radius $a$ we fit the amplitude of
rotation $V_{\rm rot}(a)$ (corrected for inclination), the kinematic position angle $PA(a)$, and
the projected axial ratio $q(a)$. Errors are directly computed by the
Levenberg-Marquardt least-squares minimisation algorithm used by the
{\tt kinemetry} code. The systemic velocity ($V_{\rm syst}$) is
assumed constant with radius. For each galaxy we then define the
median kinematic position angle $\langle PA_{\rm KIN} \rangle$ and
axial ratio $\langle q \rangle$. Their errors are computed as standard
deviation of the computed values, divided by the square root of the
number of radial bins. Values for each galaxy are reported in Table~\ref{tab:best_fit_parameters}.
The velocity dispersion profile $\sigma(a)$ is obtained as weighted average of the measured velocity dispersion computed on concentric ellipses with constant position angle and ellipticity (from $\langle PA_{\rm KIN} \rangle$ and $\langle q \rangle$), as determined by the fit to the velocity two-dimensional field.

Figure \ref{fig:vsigma} shows the radial profiles of $V_{\rm rot} / \sigma$ as function of distance along the semi-major axis expressed in units of effective radii, and the distribution of $V_{\rm rot}/\sigma$ interpolated at fiducial radii.
The figure shows differences in the distributions of $V_{\rm rot} / \sigma$ radial profiles of the two families of galaxies; in particular, there is indication that lenticulars in the field reach lower values of $V_{\rm rot} / \sigma$.  
If we use 1$R_e$ as reference radius, the mean $V_{\rm rot} / \sigma$ for field galaxies is: $1.2 \pm 0.2$, whereas for cluster galaxies is: $1.8 \pm 0.2$.
The mean values differ by $\sim 3 \sigma$; if we consider the errorbars, the null-hypothesis that the difference of mean values is consistent with 0 is discarded at $\sim 1.5 \sigma$ level (i.e the errobars touch each other if multiplied by $\sim 1.5$). If we use 1.5$R_e$ as reference radius, the separation of $V_{\rm rot}/\sigma$ is more evident, but error bars get slightly larger because the number of galaxies for which we obtain kinematics out to that radius is smaller. While the $V_{\rm rot} /\sigma$ of cluster lenticular galaxies increases from 1 $R_e$ to 1.5 $R_e$, it decreases for field lenticulars. 
The difference in the $V_{\rm rot} / \sigma$ profiles of cluster and field S0s suggests that cluster S0s are more rotationally supported than field ones, which could in turn indicate different formation mechanisms (see discussion in Section \ref{sec:conclusions}).

\subsection{$\lambda$ radial profiles}
\label{sec:lambda}

We compare the radial profile of the $\lambda$ parameter, which is a proxy for the specific angular momentum as defined in \citet{Emsellem+07}.  
The $\lambda$ profile is computed as function of semi-major axis $a$ of concentric ellipses that have the same flattening and orientation as those used for the computation of $V/\sigma$ in Section \ref{sec:vsigma}. 
Note that $\lambda$ is a cumulative and luminosity-weighted quantity, whereas $V/\sigma$ is a "local" kinematic measurement.

Figure \ref{fig:lambda} shows the $\lambda(a/R_e)$ profiles for the sample galaxies and the distribution of values at fiducial radii.  The distributions of $\lambda$ profiles of the two families of galaxies look different, suggesting that lenticulars in the field tend to reach lower values of $\lambda$.
Indeed, if we use 1$R_e$ as reference radius, the mean $\lambda$ of field galaxies is $0.36 \pm 0.04$, whereas it is $0.46 \pm 0.03$ for cluster galaxies. The mean values differ by $\sim 3\sigma$; if we consider the errorbars, the null-hypothesis that the difference of mean values is consistent with 0 is discarded at $\sim 1.43\sigma$ level (i.e the errorbars touch each other if multiplied by $\sim 1.43$).
If we consider 1.5$R_e$ as reference radius, the separation between field and cluster is $\sim 1\sigma$ (i.e., the error bars of the two measurements touch). Both field and cluster population show an increase of $\lambda$ if computed at 1$R_e$ or 1.5$R_e$.
As in Section~\ref{sec:vsigma}, the different radial profiles  of the angular momentum parameter $\lambda$ found in this section indicate that the field S0s tend to be less dominated by rotation than their cluster counterparts.
 From inspection of the values of $\lambda(R_e)$ available in the literature for ellipticals and spirals, we note that our field S0s have specific angular momentum closer to the typical values of ellipticals.  By contrast, our cluster S0s have angular momenta closer to, although not as high as, the typical values of spirals (see Fig.~\ref{fig:lambda}). This suggests that field S0s have formation processes more similar to those of elliptical galaxies (e.g., \citealt{Bournaud+05}), and that cluster S0s have formation processes more similar to those of spirals (e.g., \citealt{Eliche-Moral+13}).

\subsection{Tully--Fisher relation}
\label{sec:tf}

The Tully--Fisher relation between the circular velocity and the luminosity of disk galaxies is known to be a very tight correlation for spirals, especially at near-infrared wavelengths \citep[e.g.,][]{Sorce+13}. It has been successfully used not only as an extra-galactic distance indicator\citep[e.g.,][]{Tully+92, Courtois+11},  but also to study morphological transformation of disk galaxies  \citep[e.g.,][]{Williams+10, Cortesi+13} and environmental effects on such transformations \citep[e.g.,][]{jaffe+11b, Jaffe+14}.

In this section we derive the Tully--Fisher relation of field and cluster S0 galaxies in our sample, and compare it to that of other S0s and spirals from the literature.  
To construct the  Tully--Fisher relation we first had to derive the circular velocity in our galaxy sample. For this we constructed axisymmetric models by fitting the second moment Jeans equations (e.g., \citealt{Binney+08}) to the data.  
To do that, we adopted the procedures set out in the Multi-Gaussian Expansion (MGE, \citealt{Cappellari02}) and Jeans Axisymmetric Model (JAM, \citealt{Cappellari+08}) packages. Details of the procedure are outlined in Section \ref{sec:jam}. 
Only those galaxies for which we reached a relatively flat regime in the circular velocity curve are included in the analysis (see end of Section \ref{sec:jam}). 
The error on $V_{\rm circ}$ accounts for errors in the JAM best fit parameter plus a 10\% uncertainty due to differences in various methods of evaluating $V_{\rm circ}$ (via Jeans or Schwarzschild models, asymmetric drift, or CO rotation curves), as advocated by \citet{leung+18}. 

Figure~\ref{fig:tf} shows the Tully--Fisher relations obtained for field (blue) and cluster (red) S0 galaxies in K, B, and 3.4 $\mu$m bands. B-band is more sensitive to star formation, while M$_k$ and W1 are better tracers for the underlying older stellar populations. Overall, there is a good agreement within error bars with the relation determined for S0s by other authors \citep{Williams+10}. 
Lenticular galaxies lie below the Tully--Fisher relation found for spiral galaxies at all bands: at similar $V_{\rm circ}$, S0s are less luminous that their spiral counterparts. In the figure, we fit the data by fixing the slope of the Tully--Fisher relation to that determined for spiral galaxies.
The scatter on $V_{\rm circ}$, which is mostly dominated by the uncertainties in the model, does not allow us to determine any significant difference in the relations obtained for field and cluster environments. Also, the measured scatter is different at different bands; however not all the same galaxies are used in different bands, therefore a direct comparison of the scatter is not straightforward. A number of outliers are observed in the B-band relation (NGC 4425 and NGC 4706 are over-luminous with respect to the rest of S0s of similar $V_{\rm circ}$) and the 3.4 $\mu$m-band relation (NGC 6548 and NGC 4429 are under-luminous with respect to the rest of the S0 of similar $V_{\rm circ}$).  Uncertainties in photometry and distance alone do not explain these outliers; the most plausible explanation for the discrepancies is an additional source of error in their catalogue magnitudes. These outliers are excluded from the fit. 

\begin{figure}
\psfig{file=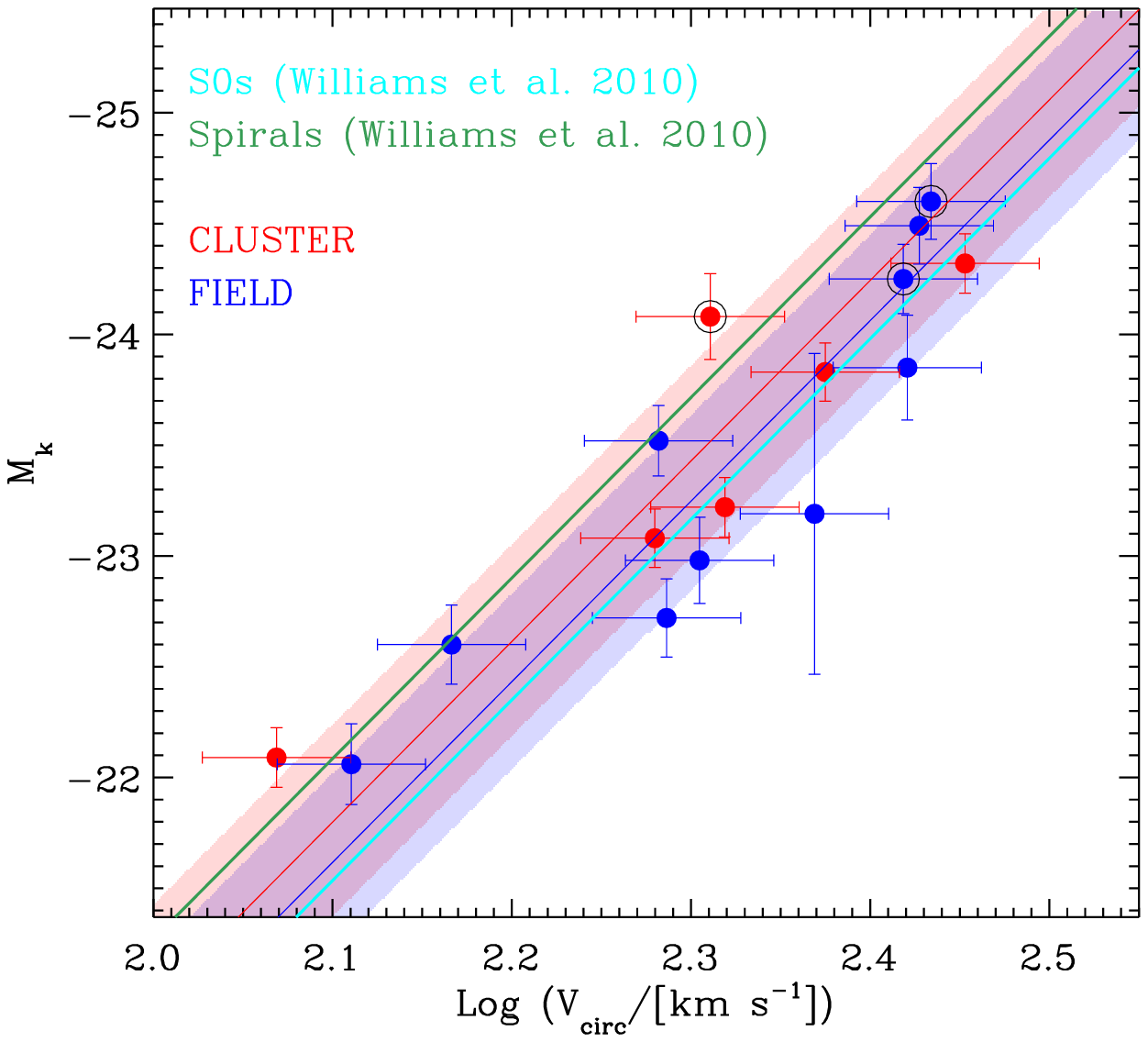,clip=,width=8.cm,bb=28 398 393 687}
\psfig{file=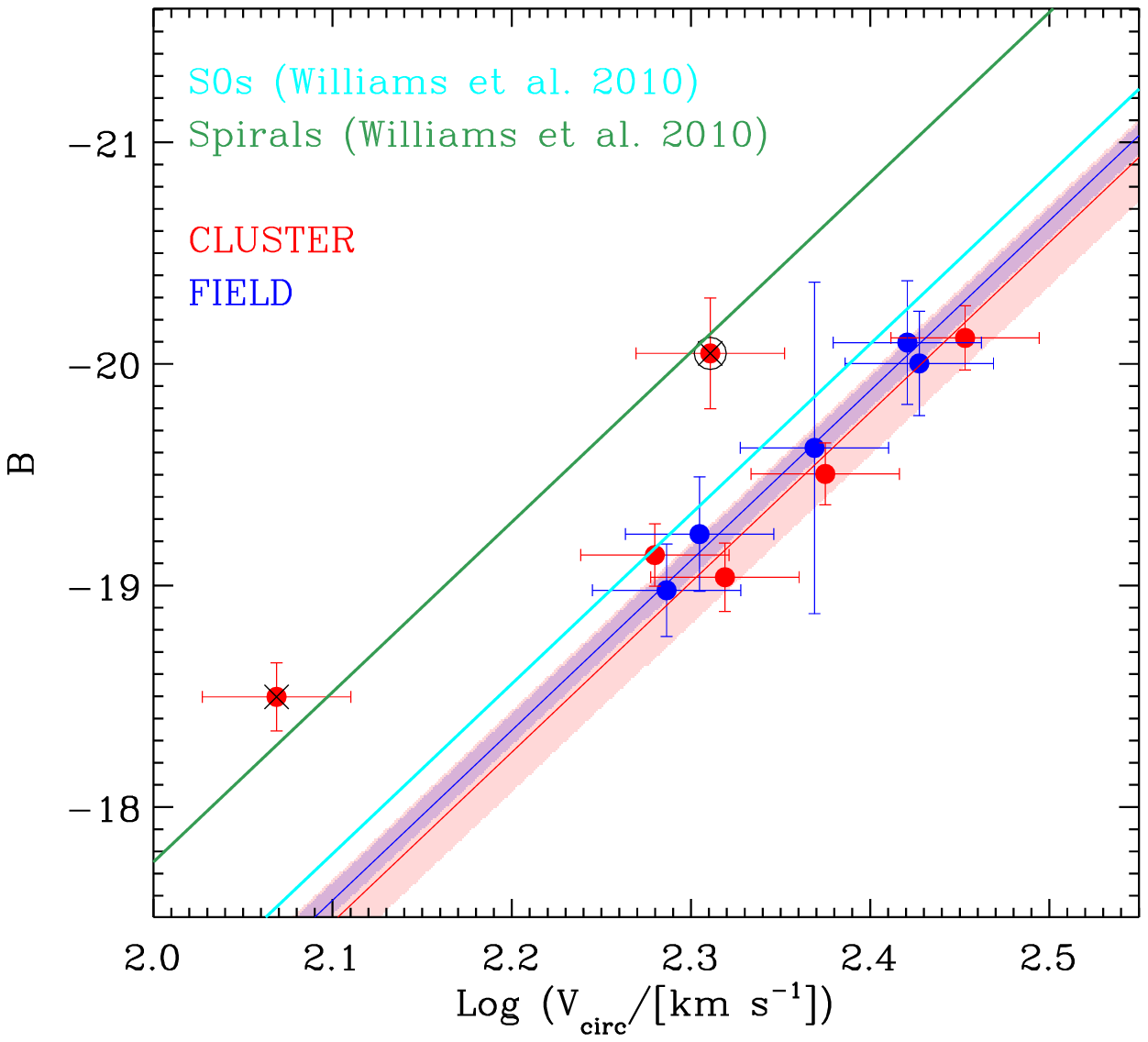,clip=,width=8.cm,bb=28 398 393 687}
\psfig{file=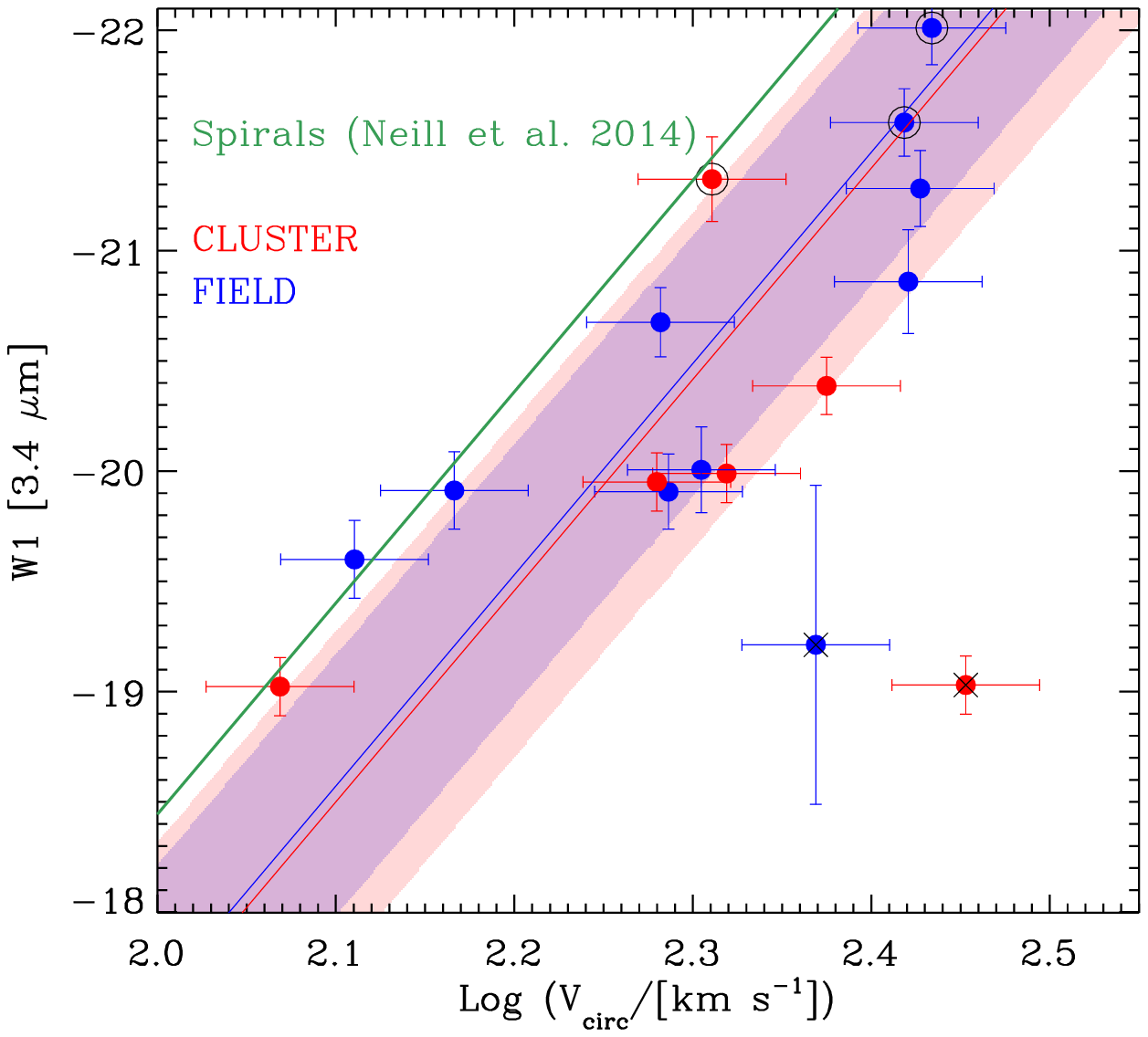,clip=,width=8.cm,bb=28 360 393 687}
\caption{Tully-Fisher relation between circular velocity as determined
  by Jeans axisymmetric models and K-band (upper panel), B-band (middle panel), and 3.6 $\mu$m (lower panel) for field (blue) and cluster (red) galaxies. The thick blue and red lines show the linear fit to the data; shaded area indicate the 1-$\sigma$ error of the fit. The slope of the S0 Tully-Fisher relation is fixed to the one determined for the Spirals, whose reference is given in each panel. MUSE data are identified by a black circle. As comparison, the Tully-Fisher relations for spirals (green) and S0s (cyan) from \citet{Williams+10} and \citet{Neill+14} are shown. Outliers are identified with crosses in the plots and removed from the fit.}
\label{fig:tf}
\end{figure}

\subsubsection{Calculation of circular velocity}
\label{sec:jam}

We constructed axisimmetric two-dimensional maps of $V_{\rm rms}=\sqrt{V^2+\sigma^2}$ by folding observed kinematic quantities. 
For the calculation of $V_{\rm rms}$ we repeated the {\tt ppxf} analysis by fitting only velocity and velocity dispersion, not the higher moments, in order to align our analysis of the MUSE observations with the one for the ATLAS3D data. The two-dimensional maps were rotated to align the kinematic major axis (determined in \ref{sec:vsigma} and listed in
Table \ref{tab:best_fit_parameters}) along the x-axis. 
The fit to the folded $V_{\rm rms}$ maps was performed using the {\tt JAM} software following the same methodology adopted in \citet{Williams+09}, which we summarise here.

Two different {\tt JAM} models are considered. One in which the potential of the dark matter follows the light distribution (model A), and one in which we the dark matter has a spherical distribution (model B). 
In both models, the mass density distribution is parametrised by a sum of two-dimensional Gaussians aligned along the galaxy photometric major axis. 
Each Gaussian has a given surface density (in $M_{\sun}\,{\rm pc}^{-2}$), dispersion along the major axis (in arcsec) and axial ratio. 

In model A, light traces the total mass; we applied the MGE procedure on the reconstructed images obtained by integrating the MUSE datacubes over the SDSS-r bandpass; these images are deeper than other images available in the literature for the same band. Foreground and background sources were masked to avoid biasing the fit. 
The light profiles of non-saturated foreground stars were used for the determination of the point-spread functions. For the galaxies in the ATLAS 3D sample, we used the MGE parameters as determined by \citet{Scott+13} in the same band.
In model B, we added the contribution of a dark matter halo, which was assumed spherical and following the \citet{Navarro+97} profile. We refer to \citep{Williams+09} for further details on the equations used.

The total number of free parameters in the fit is five: stellar mass-to-light ratio ($M/L$, which represents the total mass-to-light ratio in model A, including both luminous and dark matter), inclination ($Incl$), orbital anisotropy ($\beta$), total dark matter content ($M_{\rm DM}$, untied from the light distribution, which is set to 0 for model A) and scale radius $r_s$ (not considered in model A). 
The influence of a central super-massive black hole is negligible given the spatial resolution of the data.

For the majority of the sample galaxies, the data quality was not sufficient to separate the degeneracy between stellar and dark matter distribution. 
We therefore adopted best model A for the galaxies in which the $M_{\rm DM}$ and $r_s$ are consistent with 0, and best model B for the remaining systems. 
For the ATLAS3D sample, in which model A has been chosen, we fixed the inclination as in Table~1 of \citet{Cappellari+13}, which lists values for the same model A.

Once the best-fit models parameters are determined, we computed the radial profile of the circular velocity using the {\tt  mge\_circular\_velocity} tool in the {\tt JAM} software distribution.

We report in Table \ref{tab:best_fit_parameters} the best-fit parameters related to the chosen model (A or B), and the ``flat'' value of $V_{\rm circ}$, which we then use to construct the
Tully-Fisher relation.  
We classify a circular velocity curve as ``flat'' if the variation in the last half part of the curve is less than 15\%. We then compute $<V_{\rm circ}>$  as the average of the values of $V_{\rm circ}(R)$ in the outermost 75\% of the radial range.  

\subsection{Mass-weighted stellar populations}
\label{sec:ssp}

We inferred the  mass-weighted ages and metallicities ([Z/H]) of the stellar population in the sample galaxies within $1 R_e$.
For the ATLAS3D sample we used the ages and metallicities determined by \citet{McDermid+15}, which were obtained by integrating the spectra within $1 R_e$ and fitting them with stellar templates from the MILES library \citep{Vazdekis+12}. 
The mass-weighted values of age and metallicity were determined as weighted sum of the age and metallicity of the templates used (Equations 1 and 2 in their paper). The weights were determined by the fitting routine {\tt ppxf} executed with regularization. Emission lines where fitted and removed from the spectra using {\tt gandalf} \citep{Sarzi+06}. Errors were computed by means of Monte Carlo simulations.
For the MUSE sample, we adopted the same strategy as in \citet{McDermid+15}, and limited our spectral interval to theirs ($4750 - 5400$ \AA) to avoid systematic differences in the analysis between the two datasets. From the mass weights of the individual templates, we determined also $t_{\rm 50}$, the time in Gyr needed to form 50\% of the current-day stellar mass within $1R_e$, following the prescriptions of Section 5.1 in \citet{McDermid+15}. The quantity $t_{\rm 50}$ offers a useful proxy for the star formation time-scale.
 
The inferred mass-weighted values of age, [Z/H], and $t_{\rm 50}$ are listed in Table \ref{tab:indices}
and shown in Figure \ref{fig:ssp}. There is no apparent difference between the properties of field and cluster S0s. However, there is a tendency for more massive galaxies to have higher ages and metallicities, and lower $t_{\rm 50}$. 
The stronger dependency of stellar population parameters on mass suggested by Figure~\ref{fig:ssp} is not new, and is particularly evident in larger samples from the literature.
\citet{McDermid+15} (their Figure~13) showed that the mass-weighted stellar populations in early-type galaxies strongly depend on mass, whereas the correlation with galaxy density is weak though present (galaxies in less dense environments are younger, more metal poor, and have a longer star formation time-scale). In \citet{Fraser+18} only the dependency of stellar populations with mass is evident (their Figure~5), and no dependency on environment is seen (their Figure~9, although they show only the results for the bulge-dominated region).

In a forthcoming paper (Johnston et al, in preparation), we will investigate the two-dimensional maps of the stellar populations, exploiting the full wavelength range of the MUSE sample, and separating the contribution of young and old components,  and of the disk and the bulge by means of spectral decomposition analysis (e.g., \citealt{Coccato+11, Coccato+18, Johnston+12, Johnston+17}), to determine if there are any subtler dependencies on environment.

\begin{table}   
\begin{center}
\caption{Mass weighted values of Age, [Z/H], and $t_{\rm 50}$ within $1 R_e$ of the sample galaxies.}
\begin{tabular}{lccc}
\hline
GALAXY     &      Age          &      [Z/H]                 & $t_{\rm 50}$\\
           &    [Gry]          & [$\log_{10}(Z/Z_{\odot})$] &   [Gyr]     \\
 (1)       &     (2)           &       (3)                  &    (4)      \\
\hline
field       &                   &                           &             \\
 PGC 004187 & $13.61 \pm 0.54$  &  $  0.02 \pm 0.02 $       & $1.70 \pm 0.54$  \\
 IC 1989    & $ 7.04 \pm 0.31$  &  $  0.12 \pm 0.02 $       & $4.51 \pm 0.31$  \\ 
 NGC 3546   & $12.35 \pm 0.49$  &  $ -0.11 \pm 0.03 $       & $2.02 \pm 0.49$  \\ 
 PGC 045474 & $13.57 \pm 0.21$  &  $ -0.00 \pm 0.03 $       & $1.72 \pm 0.21$  \\
 NGC 2880    & $11.37 \pm 0.68$  &  $ -0.21 \pm 0.03 $      & $2.59 \pm 0.68$  \\ 
 NGC 3098    & $ 9.46 \pm 0.54$  &  $ -0.34 \pm 0.02 $      & $4.51 \pm 0.54$  \\ 
 NGC 6149    & $10.50 \pm 0.78$  &  $ -0.40 \pm 0.02 $      & $3.56 \pm 0.78$  \\ 
 NGC 6278    & $11.29 \pm 0.80$  &  $  0.03 \pm 0.05 $      & $2.58 \pm 0.80$  \\ 
 NGC 6548    & $ 9.99 \pm 0.91$  &  $ -0.04 \pm 0.02 $      & $3.50 \pm 0.91$  \\ 
 NGC 6703    & $12.50 \pm 0.75$  &  $ -0.17 \pm 0.03 $      & $1.61 \pm 0.76$  \\ 
 NGC 6798    & $10.58 \pm 0.76$  &  $ -0.22 \pm 0.02 $      & $3.37 \pm 0.76$  \\ 
 PGC 056772  & $ 6.39 \pm 1.31$  &  $ -0.48 \pm 0.08 $      & $6.81 \pm 1.31$   \\ 
cluster     & 		         	&	                        &                  \\   
 NGC 4696D  & $ 9.82 \pm 0.56$  &  $ -0.16 \pm 0.04 $       & $4.59 \pm 0.76$  \\
 NGC 4706   & $13.06 \pm 0.15$  &  $ -0.20 \pm 0.01 $       & $0.77 \pm 0.72$  \\
 PGC 043435 & $10.51 \pm 0.39$  &  $  0.06 \pm 0.04 $       & $1.75 \pm 0.71$  \\
 PGC 043466 & $ 7.12 \pm 0.46$  &  $ -0.26 \pm 0.03 $       & $2.06 \pm 0.71$  \\
 NGC 4425    & $ 8.97 \pm 0.76$  &  $ -0.18 \pm 0.03 $      & $1.00 \pm 0.70$  \\  
 NGC 4429    & $13.60 \pm 0.72$  &  $ -0.04 \pm 0.02 $      & $2.82 \pm 0.56$  \\  
 NGC 4435    & $12.36 \pm 0.71$  &  $ -0.18 \pm 0.03 $      & $2.01 \pm 0.14$  \\  
 NGC 4461    & $11.96 \pm 0.71$  &  $ -0.10 \pm 0.02 $      & $1.96 \pm 0.40$  \\  
 NGC 4503    & $13.28 \pm 0.70$  &  $ -0.11 \pm 0.01 $      & $5.44 \pm 0.46$  \\   
\hline
\end{tabular}
\label{tab:indices}
\begin{minipage}{8cm}
  Notes: Values in Columns 2-4 for the ATLAS3D sample are from \citet{McDermid+15}.
\end{minipage}
\end{center}
\end{table}

\begin{figure}
\begin{center}
\hspace{-0.3cm}
\psfig{file=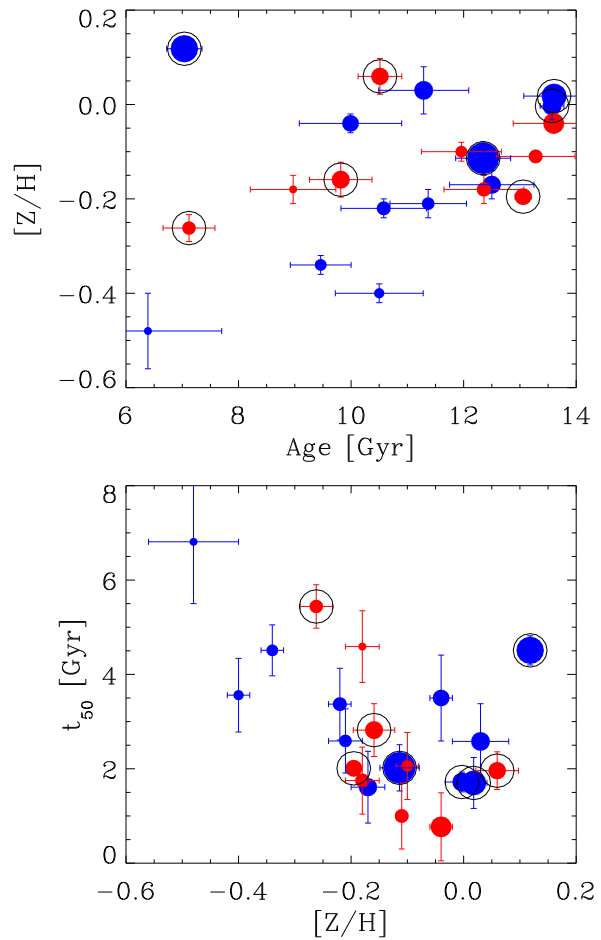,clip=,width=8.cm}
\caption{Mass-weighted values of Age, metallicity, and $t_{\rm 50}$ within $1 R_e$. Colours indicate the field (blue) and cluster (red) galaxies; the size of the symbol is proportional to the total mass indicated in Table \ref{tab:best_fit_parameters}. Circles identify galaxies from the MUSE sample.}
\label{fig:ssp}
\end{center}
\end{figure}

\section{Discussion and conclusions}
\label{sec:conclusions}

We have investigated the kinematics and stellar population properties in a sample of 21 lenticular galaxies that reside in two extremely different environments (field and cluster) with the aim of detecting any dependence on environment in their formation and evolution. In particular, we compared the radial profile of the ratio between stellar rotation and velocity dispersion ($V_{rot} / \sigma$), and the $\lambda$ parameter (a proxy for the specific angular momentum), the $M_k$, $B$ and $3.4 \mu m$ Tully-Fisher relations, and the integrated mass-weighted age, metallicity, and half-mass formation time scale $t_{\rm 50}$ at $1 R_e$ in 12 field and 9 cluster S0s.

The kinematic analysis revealed (at a $\sim 1.5\sigma$ level) that S0s in the field tend to have lower $V/\sigma(R_e)$ and $\lambda(R_e)$ than their cluster counter parts. 
In the Tully--Fisher relations however we do not find any quantifiable difference with environment.

\subsection{Effects of observational biases}
\label{sec:observational_biases}
Before embarking on the interpretation of these results, we investigate the effects of observational biases in our sample selection.

\subsubsection{Morphological mis-classification}
\label{sec:morph_bias}

The morphological distinction between elliptical and lenticular galaxies could be ambiguous, especially if the object is nearly face-on. 
As discussed in Section~\ref{sec:sample}, 8 galaxies in the ATLAS 3D sample, {\bf classified} as S0 according to LEDA, were indicated as ``pure'' Sersic objects by \citet{Krajnovic+13}. They are marked with an asterisk in the last column of Table~\ref{tab:best_fit_parameters}. 
However, our photometric decomposition of their 2MASS images revealed the presence of a prominent disk component in all cases. 
In order to test our results against this possible contamination, we repeated the kinematic analysis after removing these 8 galaxies from the sample. 
We found that this test does not substantially change the results: the largest distinction between the kinematics is still observed when splitting the sample into field and cluster, although the significance moderately decreases due to the smaller sample statistics (see Figures \ref{fig:vsigma_lambda_small_sample} and \ref{fig:parameter_separation_small_sample}).
Regarding the Tully--Fisher relation, the removal of these 8 galaxies would leave in only 2 galaxies with a flat $V_{\rm circ}$ profile in the field sample, so a sensible comparison with cluster galaxies is not possible. 
We also explored the extreme and unlikely case of 100\% contamination, in which {\it all} the sample galaxies are either spirals or ellipticals. From the measured values of $\lambda(<R_e)$ of spirals and ellipticals in \citet{Graham+18} and \citet{Emsellem+11}, we created 1000 mock random catalogues with 12 randomly selected galaxies to represent the field population and 9 randomly selected galaxies to represent the cluster population. We made sure not to have duplicates in each mock catalogue, and that all the mock catalogues are unique. We selected only galaxies with $\lambda(<R_e) < 0.7$ as in our observations. We then computed the mean $\lambda(<R_e)$ and its error for each of the field and cluster mock catalogues and check the significance in the difference between $\lambda$ of field and cluster galaxies. We found that in only $\lesssim 10\%$ of the simulations the difference between the average values of $\lambda_{\rm FIELD}$ and  $\lambda_{\rm CLUSTER}$ is at least as significant as our results (i.e., at $1.43\sigma$ level, in the sense that the $1\sigma$ error bars associated to $\lambda_{\rm FIELD}$ and  $\lambda_{\rm CLUSTER}$ touch each other if multiplied by 1.43). This fraction is small, but not insignificant; therefore a larger sample is desirable to confirm the results of this pilot study. 

\begin{figure*}
\hbox{
  \psfig{file=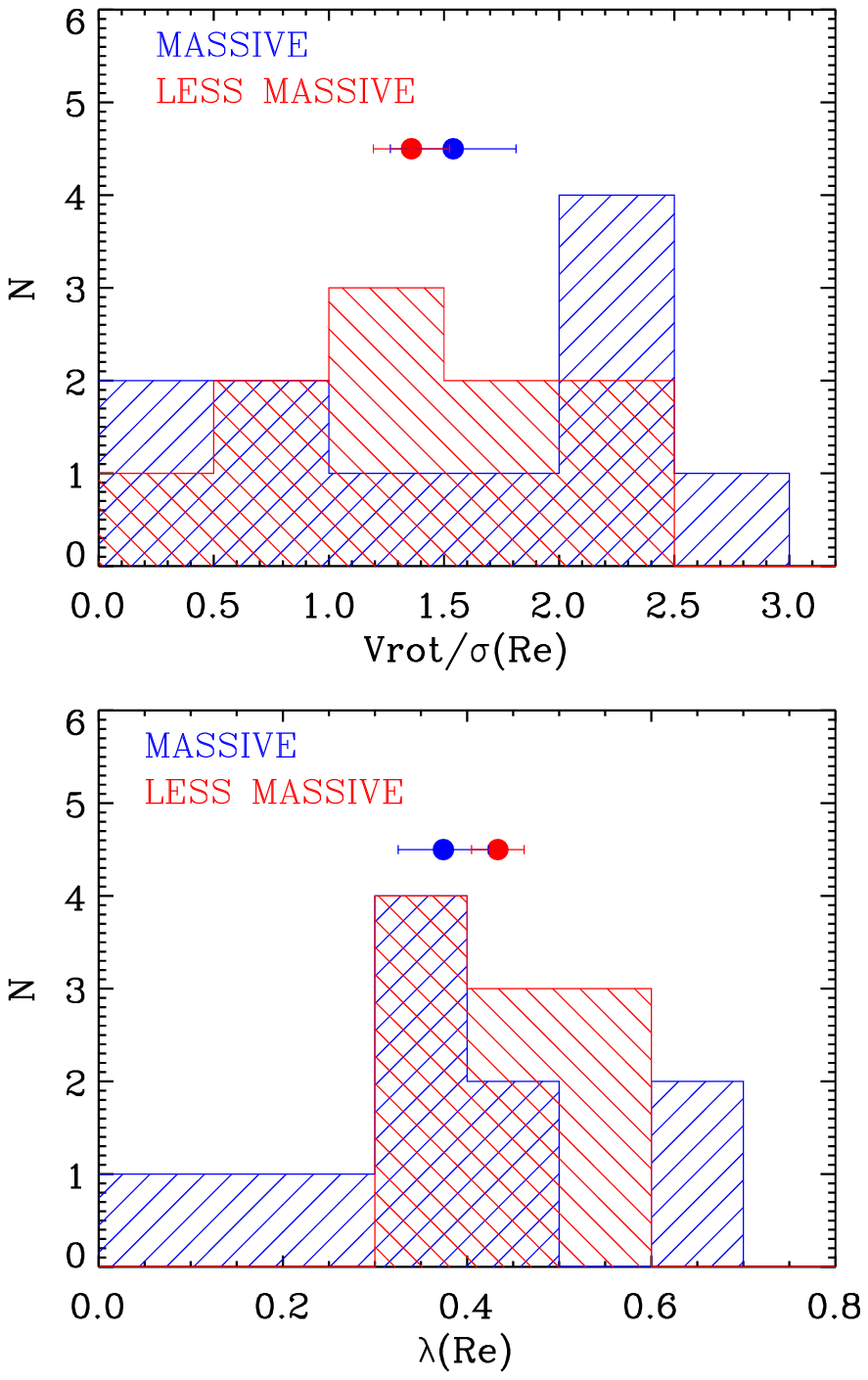,clip=,width=5.8cm}
  \psfig{file=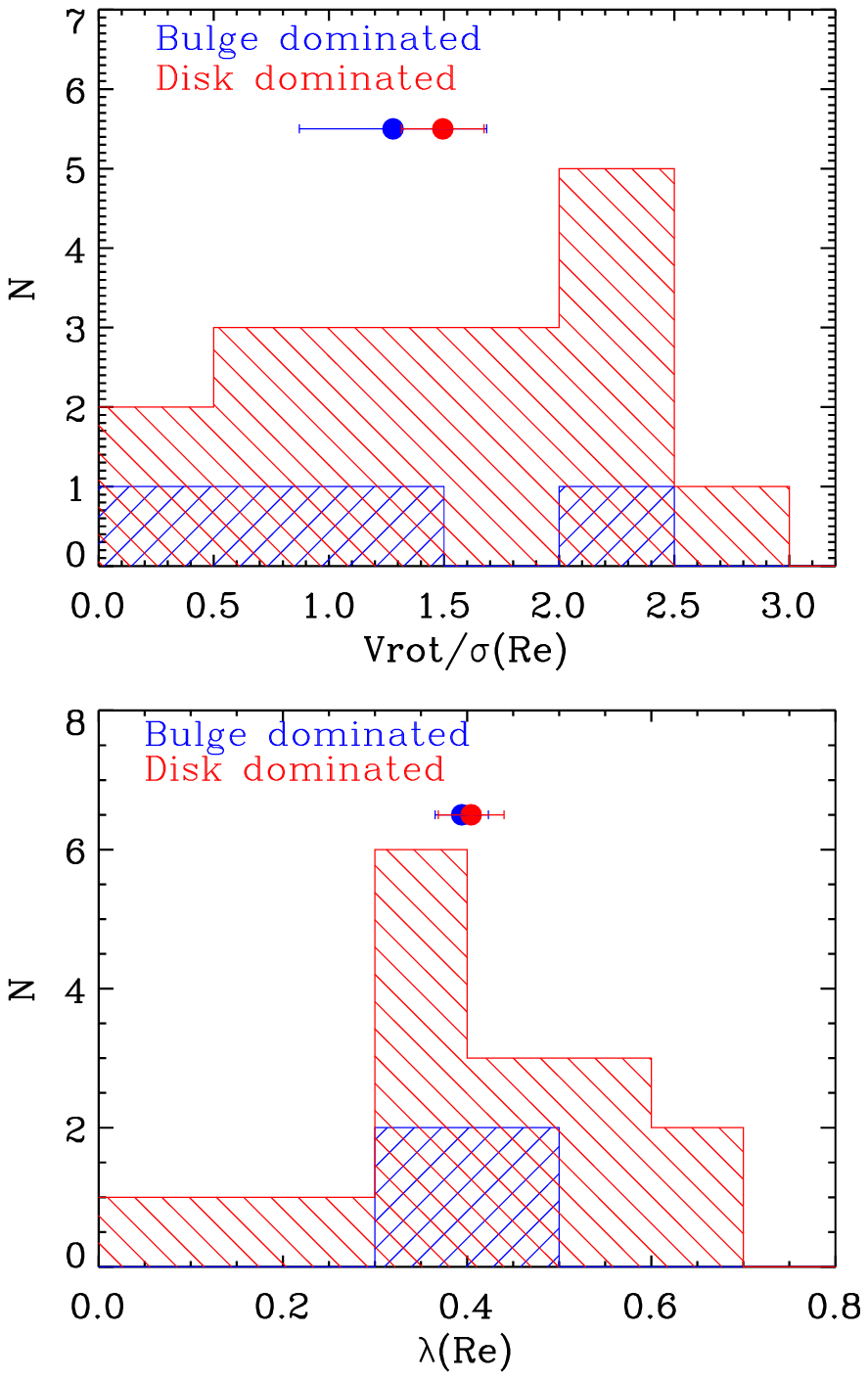,clip=,width=5.8cm}
  \psfig{file=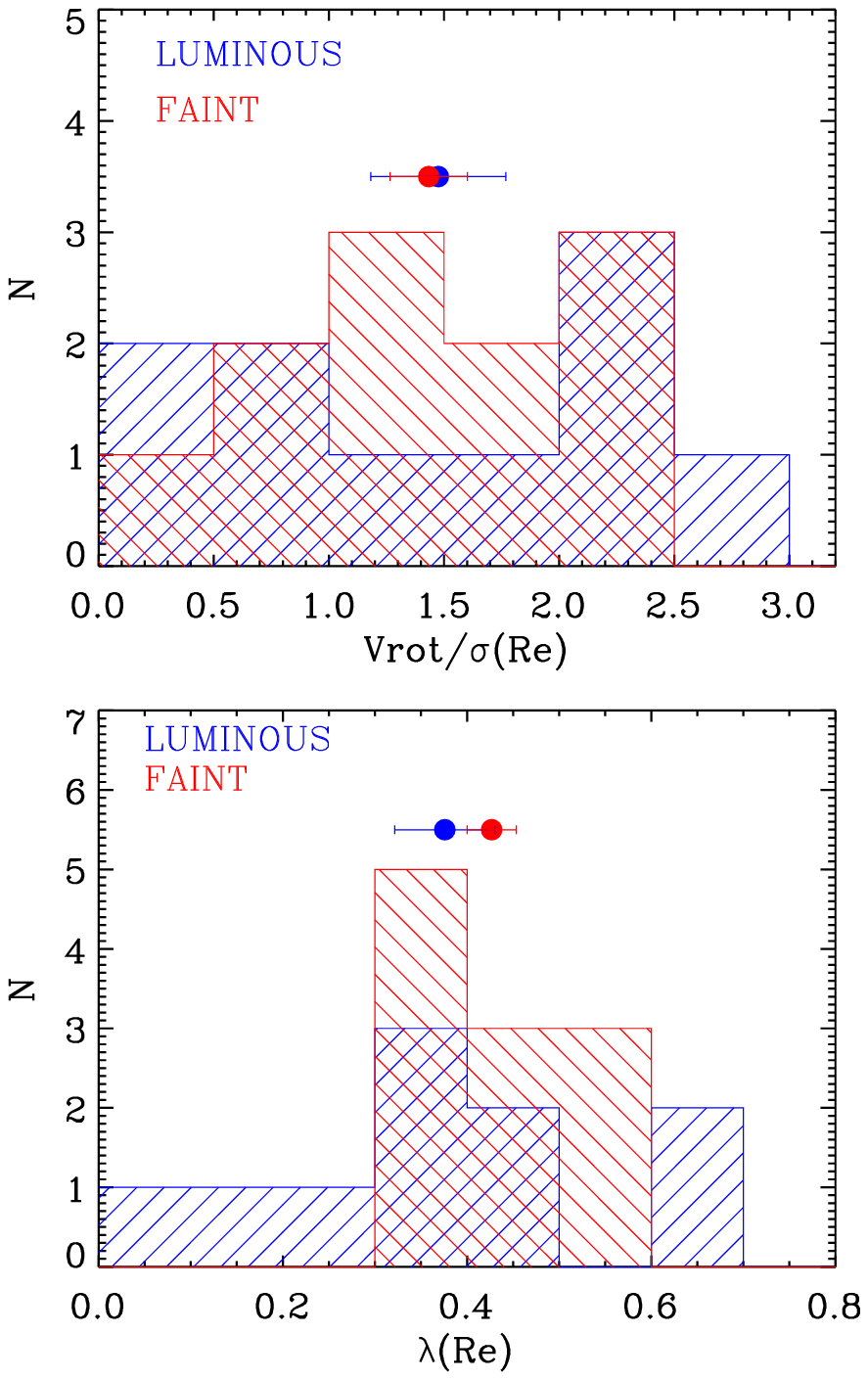,clip=,width=5.8cm}
  }
  \caption{Comparison between the values of profiles of $V/\sigma$ (upper  panel) and the proxy for specific angular momentum $\lambda$  (lower panel) at 1 effective radius of the sample galaxies. Galaxies are divided into massive and less massive (left panels), bulge dominated and disk dominated (central panels), and luminous and faint (right panels). The threshold values used to separate the sample are: total mass  = $6 \times 10^{10}$ M$_{\odot}$, bulge-to-total ratio = 0.5, $M_k$=-23.68 mag, i.e. the median of the $M_k$ magnitudes. Same results are obtained using W1= -20.48, i.e. the median of the W1 magnitudes.}
  \label{fig:parameter_separation}
\end{figure*}

\begin{figure}
  \psfig{file=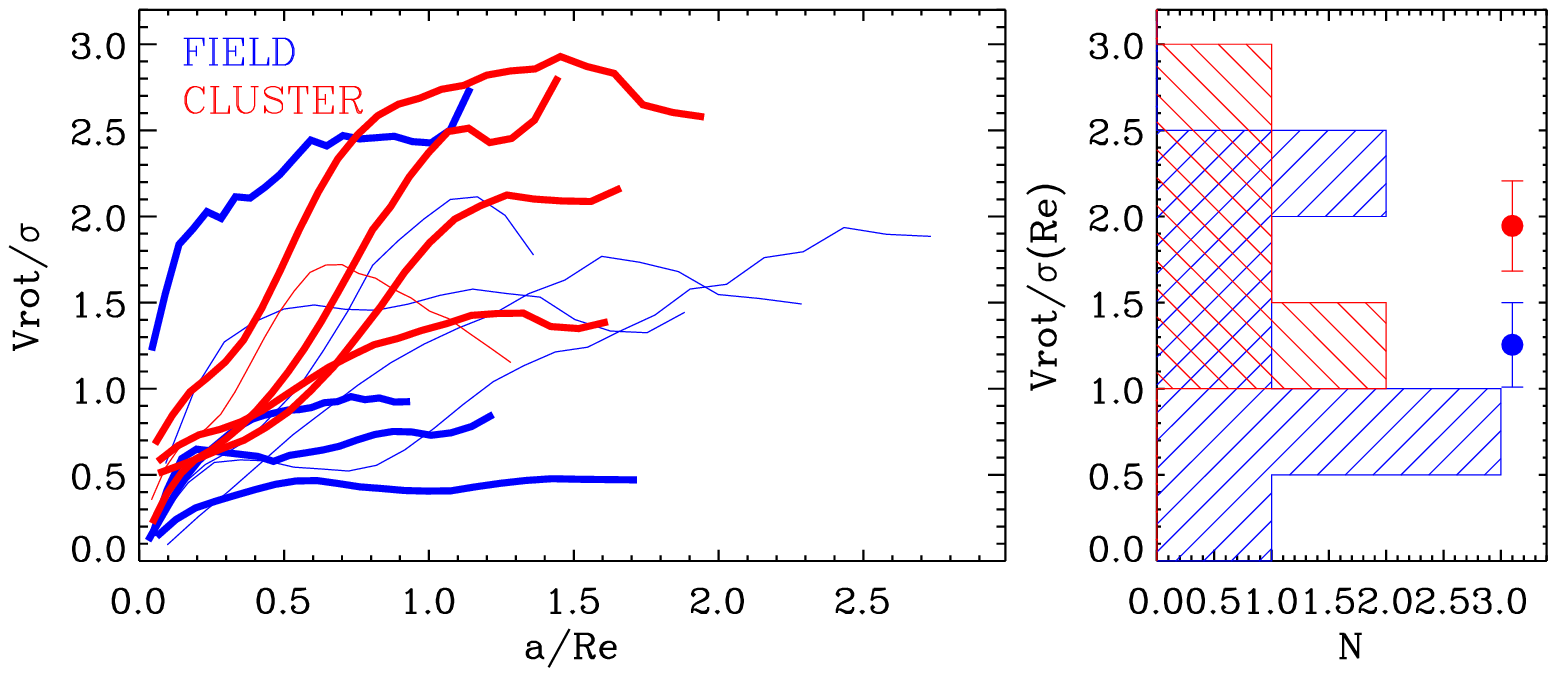,clip=,width=8.5cm,bb=10 360 460 559}
  \psfig{file=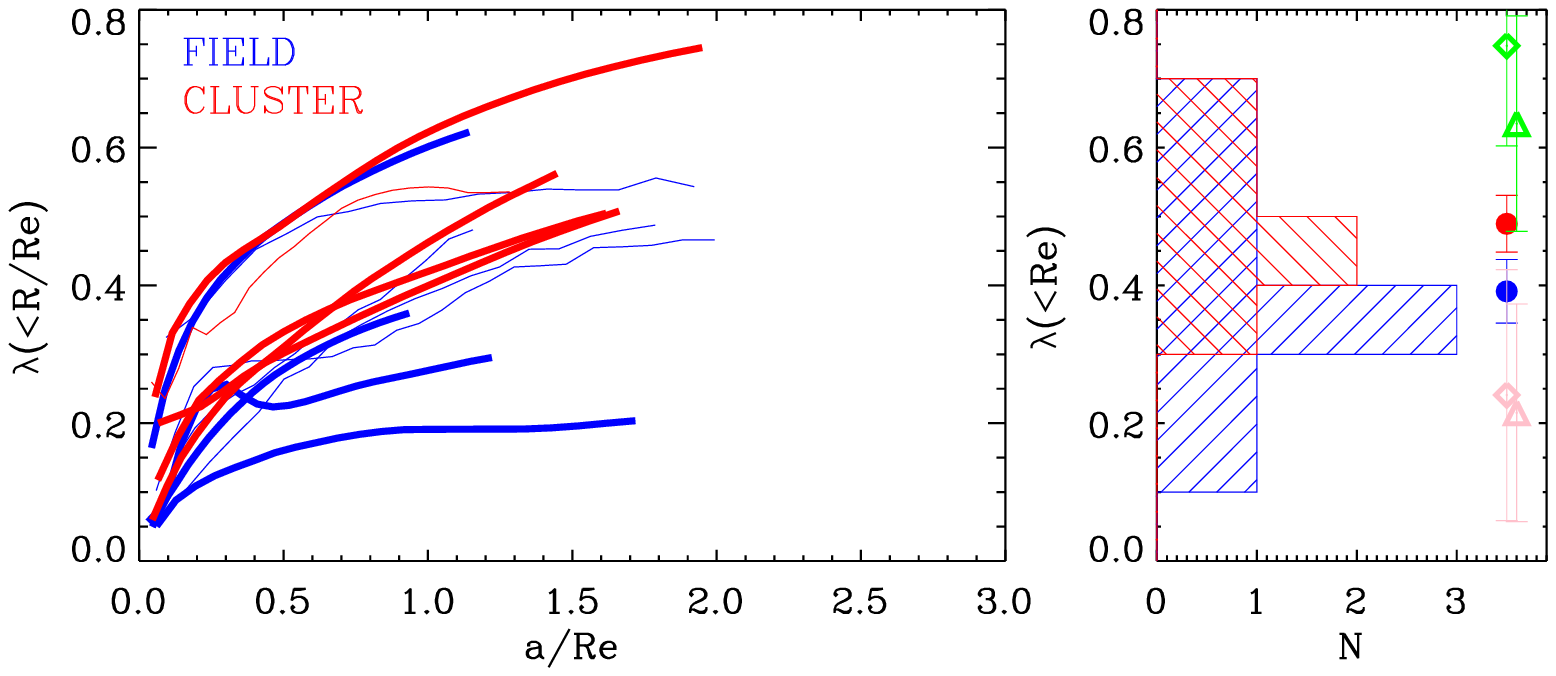,clip=,width=8.5cm,bb=10 360 460 559}
  \caption{As Figs. \ref{fig:vsigma} (upper panels) and \ref{fig:lambda} (lower panels) but after having removed the 8 galaxies marked with asterisks in Table \ref{tab:best_fit_parameters}, which have uncertain morphological classification. Green and pink symbols represent the mean values of spirals and ellipticals, respectively, as in Figs. \ref{fig:vsigma} and \ref{fig:lambda}.} 
  \label{fig:vsigma_lambda_small_sample}
\end{figure}

\begin{figure*}
 \hbox{
   \psfig{file=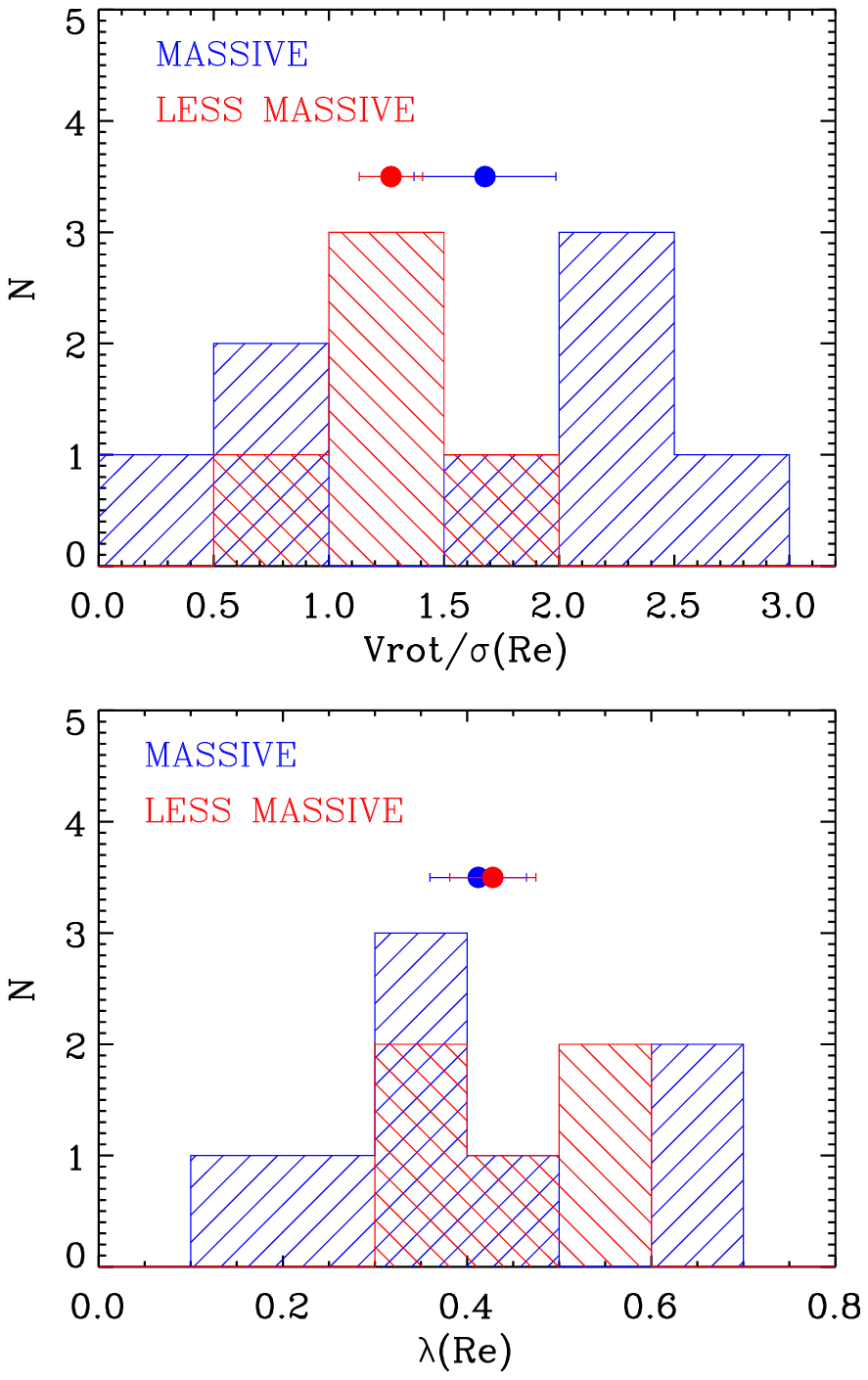,clip=,width=5.8cm}
   \psfig{file=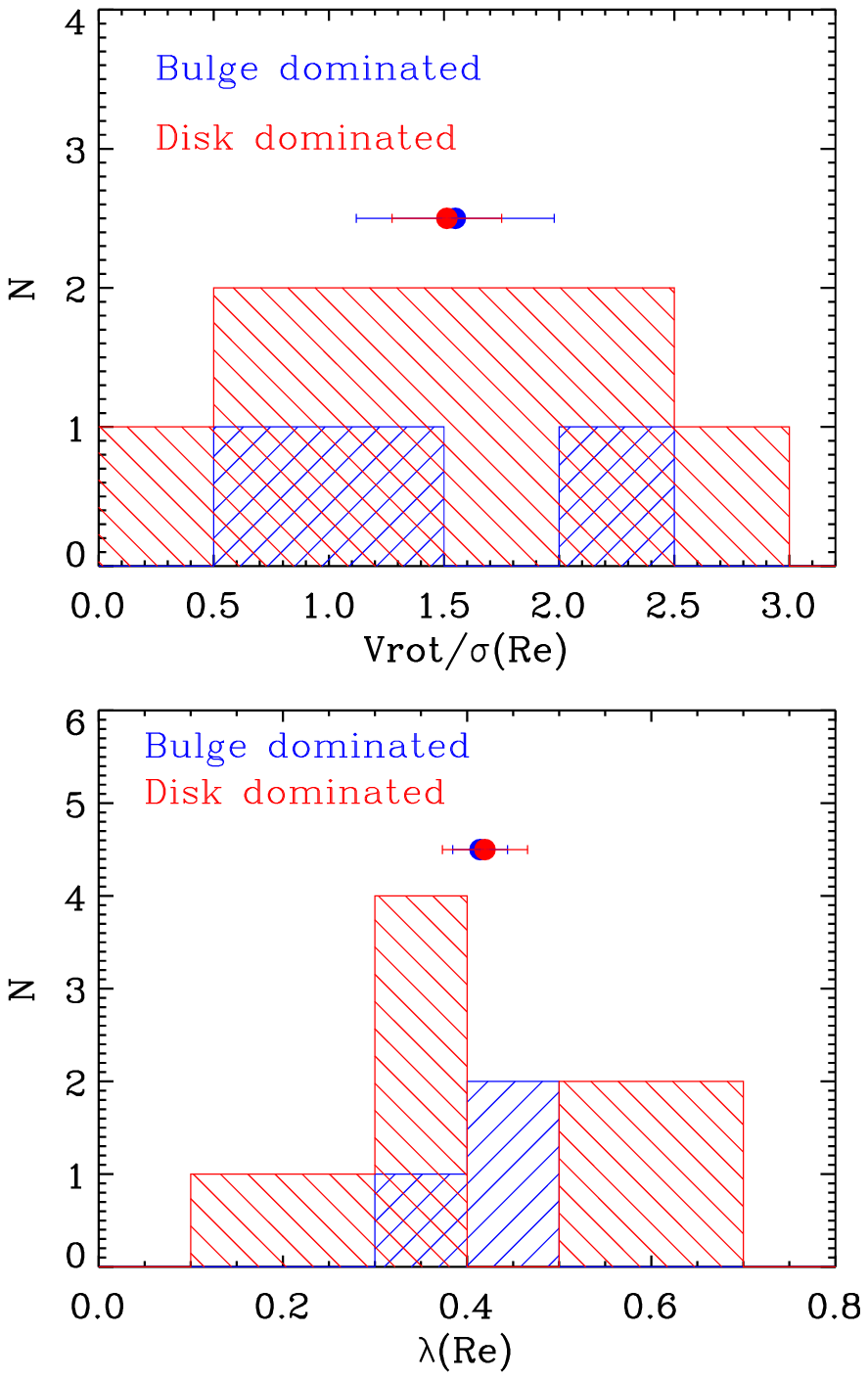,clip=,width=5.8cm}
   \psfig{file=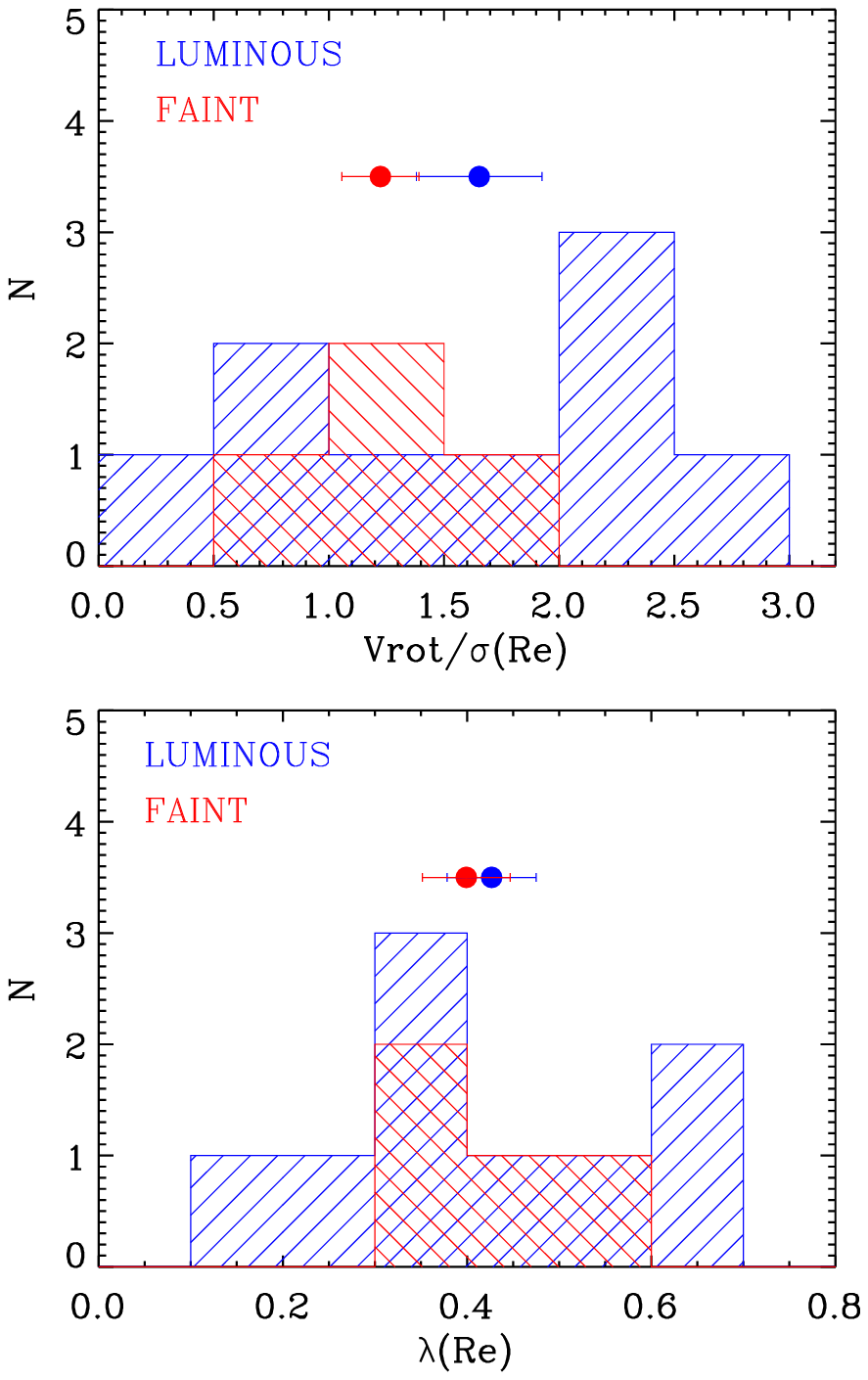,clip=,width=5.8cm}
 }
 \caption{As Fig \ref{fig:parameter_separation}, but after having removed the 8 galaxies marked with asterisks in Table \ref{tab:best_fit_parameters}, which have uncertain morphological classification.}
 \label{fig:parameter_separation_small_sample}
\end{figure*}

\begin{figure}
 \hbox{
   \psfig{file=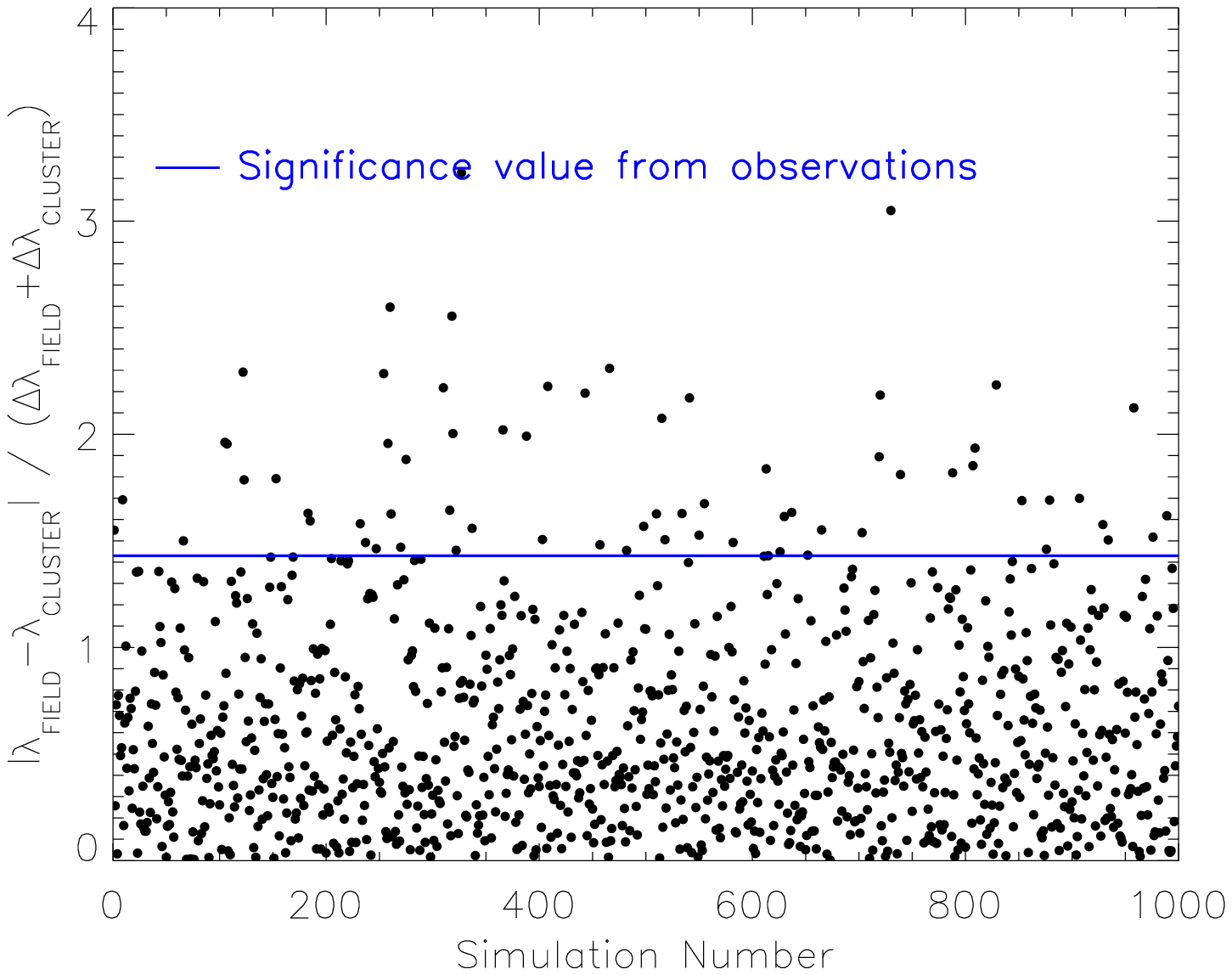,clip=,width=8.5cm,}
 }
 \caption{Significance of the measured difference between $\lambda(<R_e)$ for field and cluster galaxies, in the extreme assumption that all our sample is totally contaminated by spirals and ellipticals. The blue line at 1.43 represent the significance of our pilot project, in the sense that our measurements of $\lambda(<R_e)$ for the field and cluster sample agree with each other if we multiply the error bars by 1.43. 7\% of the simulations are above this line.}
 \label{fig:simulations}
\end{figure}

\subsubsection{Other parameters}
As evident from Figure~\ref{fig:sample}, there is not a fully homogeneous mass distribution, luminosity, or bulge-to-total light ratio between field and cluster galaxies. In addition, galaxy properties may depend on inclination and distance from the cluster centre, which would require a much larger sample to marginalize out.  Therefore, we might be seeing the effects of the contamination from other parameters rather than the environment. 

For example, a mass dependency would be indeed consistent with the recent results of \citet{Fraser+18}, in which they  clearly show a bimodal dependency of stellar population properties with mass. 
Their results indicate that  mass is the main driver for the formation of lenticular galaxies. 
According to their interpretation, massive lenticulars mostly form by morphological or inside-out quenching (stellar mass higher than $6 \times 10^{10}$ M$_{\odot}$), whereas less massive lenticulars mostly form by the fading of a progenitor spiral galaxy.  

In order to evaluate possible selection biases in our results, we repeated the analysis of Sections \ref{sec:lambda} and \ref{sec:vsigma} splitting the sample into massive versus less massive (with the division placed at  $6 \times 10^{10} M\odot$),  into bulge dominated vs disk dominated (split at a disk-to-total luminosity ratio $D/T = 0.5$), and into luminous versus faint galaxies (either side of the median value of K-band magnitudes $\langle M_k \rangle =-23.83$).
Results are shown in Figure \ref{fig:parameter_separation}.
It is evident that the kinematic properties of the galaxies do not differ when split in these ways.  Differences in the distribution of radial profiles and their values at $1 R_e$ are less significant than when the sample is divided by environment.

Thus, this kinematic analysis indicates that mass is not the main driver for different formation scenarios; this may seem at a first glance to be at odds with the recent findings by \citet{Fraser+18}. 
However, one should note that we cover a significantly different range of masses and environments in this work. Our sample galaxies have masses higher than $\log(M/M\odot) = 10.2$, so all belong to the upper mass range of \citet{Fraser+18}, where mass-driven differences between their properties are expected to be less evident. Contrastingly, by construction our sample spans a much wider range of environments than \citet{Fraser+18}, so should be more sensitive to any dependency there. Moreover, their study was mainly focused on stellar population, so did not explore the kinematic differences uncovered here.

Interestingly, if we consider the results of the stellar population analysis, we reach a somewhat different conclusion. 
The size of the symbols in Fig. \ref{fig:ssp} are proportional to the log of the total mass. 
There does seem to be at least some correlation with mass between points in Fig. \ref{fig:ssp}, but not with environment. The massive galaxies tend to concentrate in the region of higher [Z/H] and age, and low $t_{\rm 50}$, consistent with the findings of \citet{McDermid+15} and \citet{Fraser+18}, and also as expected from the broader mass--metallicity relation in galaxies (e.g., \citealt{Gallazzi+06}). 

Although our preliminary results suggest that environment shapes the kinematic properties of lenticular galaxies, whereas mass shapes the stellar populations, one would not expect the effects of mass and environment to be totally independent of each other. Indeed, merging and stripping processes can change the total mass of a galaxy and its morphology.  Moreover, the time-scale of star formation (which is known to correlate with mass; e.g. \citealt{Thomas+05, McDermid+15}) or quenching mechanisms (that can be related to the environment; e.g., \citealt{Davies+19}) also play a role, and suggest that the reality must be a more complex interplay between mass and environment in determining the global properties of a galaxy.

\subsection{Formation scenarios}

A number of scenarios have been proposed to explain the formation of S0s. 
As mentioned in the introduction, one can group them into two main categories: the first category involves galaxy interaction (e.g., mergers, tidal interactions); the second involves changes in the gas content (e.g., ram-pressure stripping, starvation).

Interactions in general tend to destroy the stellar disk leading to less rotationally-supported systems or to an elliptical galaxy. 
We might therefore expect that S0s formed through this channel would have significantly lower $V_{\rm rot}/\sigma$ and $\lambda$ than their spiral counterparts of similar mass \citep[e.g.,][]{Bournaud+05, Querejeta+15, Bekki+11, Lagos+17}. 
Also, merger remnants would be systematically shifted to lower $V_{\rm circ}$ for the same mass right after their formation, but they would then move to the local Tully--Fisher relation after 4--7~Gyr of passive evolution.  
On the other hand, if a large amount of gas is involved in the merger, then new stars in rapidly-rotating structures can boost $V_{\rm rot}/\sigma$. However, simulations shows that such enhancement affects only the very central regions in the majority of cases \citep[e.g.,][]{Lagos+18}.
In general, the modification of the gas content has less effect on the dynamical structure of the galaxy, so the resulting S0s would be expected to have the dynamical structure of their spiral progenitors. 

Our results point in the direction that the field environment is dominated by the merger processes that decrease $V_{\rm rot}/\sigma$ and $\lambda$. The fact that we do not observe systematic differences between the Tully--Fisher relations of S0s in different environments (at least within the scatter of our measurements) suggests that such mergers occurred at high redshift, and then the merger remnant has evolved passively and in isolation for 4-7 Gyrs \citep{Tapia+17} moving back to the local Tully--Fisher S0s relation. On the other hand, dense environments seem to be dominated by the stripping processes that change $V_{\rm rot}/\sigma$ and $\lambda$ less dramatically. 
We clarify that here we consider a process to be "dominating" only with respect to the other mechanisms that contribute {\it to form} S0s, not with respect to all the mechanisms that can occur in a given environment and that can change, for example, the morphological type of an S0 galaxy.

From a qualitative point of view, the kinematic results are consistent with the idea that i) isolated S0 galaxies are the end product of past mergers resulting in a stellar component ``dynamically hotter'' than cluster S0s, and ii) cluster S0s are formed through processes that involve the rapid consumption or removal of gas resulting in a stellar component "dynamically colder" than field S0s. S0s in cluster could therefore be "faded" spirals, although a small contribution from minor mergers cannot be rule out (at least in some cases) to explain why the mean $\lambda(R_e)$ of cluster S0 is not as high as those of spirals. Also the effect of the cluster environment is more directed to the gas content (e.g. ram pressure stripping), consistent with  the fact that the majority of our cluster S0s have have even less gas in their central regions than S0s in the field (see Table \ref{tab:sample}).

\subsection{Summary and future perspective}
\label{sec:future}
We have studied the properties of a sample of lenticular galaxies in the field and in clusters, with the aim of investigating the influence of the environment on their formation processes. 
Even with the relatively small number of galaxies in our sample, we find clear indications that galaxies in the field are generally more pressure supported than galaxies in the cluster, suggesting that galaxy interactions (mergers, tidal interactions) play a more major role in the field in shaping thee systems. 
This scenario is also reflected in the fact that the specific angular momentum of field S0s is closer to the range of values of elliptical galaxies, which are believed to form mainly via mergers.
On the other hand, the kinematic properties of S0s in the cluster are closer to their spiral counterparts, indicating that processes involving the modification of the gas content (stripping, starvation, rapid star formation) are dominant in the cluster environment . 

Mass or luminosity do not seem to be the main drivers of these kinematic properties; however in our study we have only considered galaxies with total mass $\log(M/M_\odot)>10.2$, so we do not cover the faint end of the S0s population.  By contrast, the properties of the stellar populations (age, metallicity, $t_{\rm 50}$) are more correlated with mass than with environment, consistent with previous studies that note the correlation between, for example, mass and metallicity \citep[e.g., ][]{Thomas+05}. 

One of the main cautionary aspects of this pilot study is the sample size, which is not entirely representative of the S0 galaxy population nor is it uniform in parameter distribution. A larger sample is therefore highly recommended to obtain unequivocal conclusions that either confirm or negate the findings and interpretation presented here.
Notwithstanding this limitation, this work highlights that environment does seem to be an important parameter in shaping the kinematic properties of lenticular galaxies.
While the previous study by \citet{Fraser+18} indicated stellar mass as the main driver for dictating the formation mechanism of S0s, a direct comparison between their work and this analysis is difficult to make at this stage because of the differences in mass range end environment explored, and because the previous work focused less on the kinematic signatures.

In summary, the results and limitations of this pilot study argue strongly for a much larger survey of S0s that spans a wide range of both masses and environments, and investigates the signatures of both kinematics and stellar populations in seeking to establish formation mechanisms. This approach will be the key to disentangling the contributions of mass {\it and} environment to dictating the channels by which lenticular galaxies form.

\section*{Acknowledgements}
This research is based on observations collected at the European Southern Observatory under ESO programme 096.B-0325 and it has made use of the services of the ESO Science Archive Facility.
This publication makes use of data products from the Wide-field Infrared Survey Explorer, which is a joint project of the University of California, Los Angeles, and the Jet Propulsion Laboratory/California Institute of Technology, funded by the National Aeronautics and Space Administration.
This publication makes use of data products from the NASA/IPAC Extragalactic Database (NED), which is operated by the Jet Propulsion Laboratory, California Institute of Technology, under contract with the National Aeronautics and Space Administration.
This publication makes use of data products from the Two Micron All Sky Survey, which is a joint project of the University of Massachusetts and the Infrared Processing and Analysis Center/California Institute of Technology, funded by the National Aeronautics and Space Administration and the National Science Foundation.

YJ acknowledges financial support from CONICYT PAI (Concurso Nacional de Inserci\'{o}n en la Academia 2017), No. 79170132 and FONDECYT Iniciaci\'{o}n 2018 No. 11180558. 
EJ acknowledges support from FONDECYT Postdoctoral Fellowship Project No. 3180557.
BRP acknowledges support from the Spanish Ministry of Economy and Competitiveness through grant ESP2017-83197.
AC acknowledges support from PNPD/CAPES.
YKS acknowledges support from the National Research Foundation of Korea (NRF) grant funded by the Ministry of Science, ICT \& Future Planning (No. NRF-2019R1C1C1010279).
ACS acknowledges funding from the brazilian agencies \textit {Conselho Nacional de Desenvolvimento Cient\'ifico e Tecnol\'ogico} (CNPq) and the Rio Grande do Sul Research Foundation (FAPERGS) through grants CNPq-403580/2016-1, CNPq-310845/2015-7, PqG/FAPERGS-17/2551-0001.

We wish to thank the referee, R. Peletier, for constructive feedback.

\bibliography{s0-kinematics}

\begin{thebibliography}{}
\makeatletter
\relax
\def\mn@urlcharsother{\let\do\@makeother \do\$\do\&\do\#\do\^\do\_\do\%\do\~}
\def\mn@doi{\begingroup\mn@urlcharsother \@ifnextchar [ {\mn@doi@}
  {\mn@doi@[]}}
\def\mn@doi@[#1]#2{\def\@tempa{#1}\ifx\@tempa\@empty \href
  {http://dx.doi.org/#2} {doi:#2}\else \href {http://dx.doi.org/#2} {#1}\fi
  \endgroup}
\def\mn@eprint#1#2{\mn@eprint@#1:#2::\@nil}
\def\mn@eprint@arXiv#1{\href {http://arxiv.org/abs/#1} {{\tt arXiv:#1}}}
\def\mn@eprint@dblp#1{\href {http://dblp.uni-trier.de/rec/bibtex/#1.xml}
  {dblp:#1}}
\def\mn@eprint@#1:#2:#3:#4\@nil{\def\@tempa {#1}\def\@tempb {#2}\def\@tempc
  {#3}\ifx \@tempc \@empty \let \@tempc \@tempb \let \@tempb \@tempa \fi \ifx
  \@tempb \@empty \def\@tempb {arXiv}\fi \@ifundefined
  {mn@eprint@\@tempb}{\@tempb:\@tempc}{\expandafter \expandafter \csname
  mn@eprint@\@tempb\endcsname \expandafter{\@tempc}}}

\bibitem[\protect\citeauthoryear{{Aguerri}}{{Aguerri}}{2012}]{Aguerri12}
{Aguerri} J. A.~L.,  2012, \mn@doi [Advances in Astronomy]
  {10.1155/2012/382674}, \href
  {https://ui.adsabs.harvard.edu/abs/2012AdAst2012E..28A} {2012, 382674}

\bibitem[\protect\citeauthoryear{{Arnold}, {Romanowsky}, {Brodie}, {Chomiuk},
  {Spitler}, {Strader}, {Benson}  \& {Forbes}}{{Arnold}
  et~al.}{2011}]{Arnold+11}
{Arnold} J.~A.,  {Romanowsky} A.~J.,  {Brodie} J.~P.,  {Chomiuk} L.,  {Spitler}
  L.~R.,  {Strader} J.,  {Benson} A.~J.,   {Forbes} D.~A.,  2011, \mn@doi
  [\apjl] {10.1088/2041-8205/736/2/L26}, \href
  {http://adsabs.harvard.edu/abs/2011ApJ...736L..26A} {736, L26}

\bibitem[\protect\citeauthoryear{{Bacon} et~al.,}{{Bacon}
  et~al.}{2010}]{Bacon+10}
{Bacon} R.,  et~al., 2010, in Ground-based and Airborne Instrumentation for
  Astronomy III. p. 773508, \mn@doi{10.1117/12.856027}

\bibitem[\protect\citeauthoryear{{Bekki} \& {Couch}}{{Bekki} \&
  {Couch}}{2011}]{Bekki+11}
{Bekki} K.,  {Couch} W.~J.,  2011, \mn@doi [\mnras]
  {10.1111/j.1365-2966.2011.18821.x}, \href
  {http://adsabs.harvard.edu/abs/2011MNRAS.415.1783B} {415, 1783}

\bibitem[\protect\citeauthoryear{{Bellstedt}, {Forbes}, {Foster}, {Romanowsky},
  {Brodie}, {Pastorello}, {Alabi}  \& {Villaume}}{{Bellstedt}
  et~al.}{2017}]{Bellstedt+17}
{Bellstedt} S.,  {Forbes} D.~A.,  {Foster} C.,  {Romanowsky} A.~J.,  {Brodie}
  J.~P.,  {Pastorello} N.,  {Alabi} A.,   {Villaume} A.,  2017, \mn@doi
  [\mnras] {10.1093/mnras/stx418}, \href
  {http://adsabs.harvard.edu/abs/2017MNRAS.467.4540B} {467, 4540}

\bibitem[\protect\citeauthoryear{{Bernardi}, {Shankar}, {Hyde}, {Mei},
  {Marulli}  \& {Sheth}}{{Bernardi} et~al.}{2010}]{Bernardi+10}
{Bernardi} M.,  {Shankar} F.,  {Hyde} J.~B.,  {Mei} S.,  {Marulli} F.,
  {Sheth} R.~K.,  2010, \mn@doi [\mnras] {10.1111/j.1365-2966.2010.16425.x},
  \href {http://adsabs.harvard.edu/abs/2010MNRAS.404.2087B} {404, 2087}

\bibitem[\protect\citeauthoryear{{Binney} \& {Tremaine}}{{Binney} \&
  {Tremaine}}{2008}]{Binney+08}
{Binney} J.,  {Tremaine} S.,  2008, {Galactic Dynamics: Second Edition}.
Princeton University Press

\bibitem[\protect\citeauthoryear{{Bournaud}, {Jog}  \& {Combes}}{{Bournaud}
  et~al.}{2005}]{Bournaud+05}
{Bournaud} F.,  {Jog} C.~J.,   {Combes} F.,  2005, \mn@doi [\aap]
  {10.1051/0004-6361:20042036}, \href
  {http://adsabs.harvard.edu/abs/2005A\%26A...437...69B} {437, 69}

\bibitem[\protect\citeauthoryear{{Cappellari}}{{Cappellari}}{2002}]{Cappellari02}
{Cappellari} M.,  2002, \mn@doi [\mnras] {10.1046/j.1365-8711.2002.05412.x},
  \href {http://adsabs.harvard.edu/abs/2002MNRAS.333..400C} {333, 400}

\bibitem[\protect\citeauthoryear{{Cappellari}}{{Cappellari}}{2008}]{Cappellari+08}
{Cappellari} M.,  2008, \mn@doi [\mnras] {10.1111/j.1365-2966.2008.13754.x},
  \href {http://adsabs.harvard.edu/abs/2008MNRAS.390...71C} {390, 71}

\bibitem[\protect\citeauthoryear{{Cappellari} \& {Copin}}{{Cappellari} \&
  {Copin}}{2003}]{Cappellari+03}
{Cappellari} M.,  {Copin} Y.,  2003, \mn@doi [\mnras]
  {10.1046/j.1365-8711.2003.06541.x}, \href
  {http://adsabs.harvard.edu/abs/2003MNRAS.342..345C} {342, 345}

\bibitem[\protect\citeauthoryear{{Cappellari} \& {Emsellem}}{{Cappellari} \&
  {Emsellem}}{2004}]{Cappellari+04}
{Cappellari} M.,  {Emsellem} E.,  2004, \mn@doi [\pasp] {10.1086/381875}, \href
  {http://adsabs.harvard.edu/abs/2004PASP..116..138C} {116, 138}

\bibitem[\protect\citeauthoryear{{Cappellari} et~al.,}{{Cappellari}
  et~al.}{2011a}]{Cappellari+11a}
{Cappellari} M.,  et~al., 2011a, \mn@doi [\mnras]
  {10.1111/j.1365-2966.2010.18174.x}, \href
  {http://adsabs.harvard.edu/abs/2011MNRAS.413..813C} {413, 813}

\bibitem[\protect\citeauthoryear{{Cappellari} et~al.,}{{Cappellari}
  et~al.}{2011b}]{Cappellari+11b}
{Cappellari} M.,  et~al., 2011b, \mn@doi [\mnras]
  {10.1111/j.1365-2966.2011.18600.x}, \href
  {http://adsabs.harvard.edu/abs/2011MNRAS.416.1680C} {416, 1680}

\bibitem[\protect\citeauthoryear{{Cappellari} et~al.,}{{Cappellari}
  et~al.}{2013}]{Cappellari+13}
{Cappellari} M.,  et~al., 2013, \mn@doi [\mnras] {10.1093/mnras/stt562}, \href
  {http://adsabs.harvard.edu/abs/2013MNRAS.432.1709C} {432, 1709}

\bibitem[\protect\citeauthoryear{{Chies-Santos}, {Larsen}, {Kuntschner},
  {Anders}, {Wehner}, {Strader}, {Brodie}  \& {Santos}}{{Chies-Santos}
  et~al.}{2011}]{Chies-Santos+11}
{Chies-Santos} A.~L.,  {Larsen} S.~S.,  {Kuntschner} H.,  {Anders} P.,
  {Wehner} E.~M.,  {Strader} J.,  {Brodie} J.~P.,   {Santos} J.~F.~C.,  2011,
  \mn@doi [\aap] {10.1051/0004-6361/201015683}, \href
  {http://adsabs.harvard.edu/abs/2011A\%26A...525A..20C} {525, A20}

\bibitem[\protect\citeauthoryear{{Christlein} \& {Zabludoff}}{{Christlein} \&
  {Zabludoff}}{2004}]{Christlein+04}
{Christlein} D.,  {Zabludoff} A.~I.,  2004, \mn@doi [\apj] {10.1086/424909},
  \href {http://adsabs.harvard.edu/abs/2004ApJ...616..192C} {616, 192}

\bibitem[\protect\citeauthoryear{{Coccato}, {Morelli}, {Corsini}, {Buson},
  {Pizzella}, {Vergani}  \& {Bertola}}{{Coccato} et~al.}{2011}]{Coccato+11}
{Coccato} L.,  {Morelli} L.,  {Corsini} E.~M.,  {Buson} L.,  {Pizzella} A.,
  {Vergani} D.,   {Bertola} F.,  2011, \mn@doi [\mnras]
  {10.1111/j.1745-3933.2011.01016.x}, \href
  {https://ui.adsabs.harvard.edu/abs/2011MNRAS.412L.113C} {412, L113}

\bibitem[\protect\citeauthoryear{{Coccato}, {Fabricius}, {Saglia}, {Bender},
  {Erwin}, {Drory}  \& {Morelli}}{{Coccato} et~al.}{2018}]{Coccato+18}
{Coccato} L.,  {Fabricius} M.~H.,  {Saglia} R.~P.,  {Bender} R.,  {Erwin} P.,
  {Drory} N.,   {Morelli} L.,  2018, \mn@doi [\mnras] {10.1093/mnras/sty705},
  \href {https://ui.adsabs.harvard.edu/abs/2018MNRAS.477.1958C} {477, 1958}

\bibitem[\protect\citeauthoryear{{Cortesi} et~al.,}{{Cortesi}
  et~al.}{2013}]{Cortesi+13}
{Cortesi} A.,  et~al., 2013, \mn@doi [\mnras] {10.1093/mnras/stt529}, \href
  {http://adsabs.harvard.edu/abs/2013MNRAS.432.1010C} {432, 1010}

\bibitem[\protect\citeauthoryear{{Courtois}, {Tully}  \&
  {H{\'e}raudeau}}{{Courtois} et~al.}{2011}]{Courtois+11}
{Courtois} H.~M.,  {Tully} R.~B.,   {H{\'e}raudeau} P.,  2011, \mn@doi [\mnras]
  {10.1111/j.1365-2966.2011.18839.x}, \href
  {http://cdsads.u-strasbg.fr/abs/2011MNRAS.415.1935C} {415, 1935}

\bibitem[\protect\citeauthoryear{{Davies} et~al.,}{{Davies}
  et~al.}{2019}]{Davies+19}
{Davies} L.~J.~M.,  et~al., 2019, \mn@doi [\mnras] {10.1093/mnras/sty3393},
  \href {http://adsabs.harvard.edu/abs/2019MNRAS.483.5444D} {483, 5444}

\bibitem[\protect\citeauthoryear{{Dressler}}{{Dressler}}{1980}]{Dressler80}
{Dressler} A.,  1980, \mn@doi [\apj] {10.1086/157753}, \href
  {http://adsabs.harvard.edu/abs/1980ApJ...236..351D} {236, 351}

\bibitem[\protect\citeauthoryear{{Dressler} et~al.,}{{Dressler}
  et~al.}{1997}]{Dressler+97}
{Dressler} A.,  et~al., 1997, \mn@doi [\apj] {10.1086/304890}, \href
  {http://adsabs.harvard.edu/abs/1997ApJ...490..577D} {490, 577}

\bibitem[\protect\citeauthoryear{{Eliche-Moral}, {Gonz{\'a}lez-Garc{\'{\i}}a},
  {Aguerri}, {Gallego}, {Zamorano}, {Balcells}  \& {Prieto}}{{Eliche-Moral}
  et~al.}{2012}]{Eliche-Moral+12}
{Eliche-Moral} M.~C.,  {Gonz{\'a}lez-Garc{\'{\i}}a} A.~C.,  {Aguerri} J.~A.~L.,
   {Gallego} J.,  {Zamorano} J.,  {Balcells} M.,   {Prieto} M.,  2012, \mn@doi
  [\aap] {10.1051/0004-6361/201118711}, \href
  {http://adsabs.harvard.edu/abs/2012A\%26A...547A..48E} {547, A48}

\bibitem[\protect\citeauthoryear{{Eliche-Moral}, {Gonz{\'a}lez-Garc{\'{\i}}a},
  {Aguerri}, {Gallego}, {Zamorano}, {Balcells}  \& {Prieto}}{{Eliche-Moral}
  et~al.}{2013}]{Eliche-Moral+13}
{Eliche-Moral} M.~C.,  {Gonz{\'a}lez-Garc{\'{\i}}a} A.~C.,  {Aguerri} J.~A.~L.,
   {Gallego} J.,  {Zamorano} J.,  {Balcells} M.,   {Prieto} M.,  2013, \mn@doi
  [\aap] {10.1051/0004-6361/201220841}, \href
  {http://adsabs.harvard.edu/abs/2013A\%26A...552A..67E} {552, A67}

\bibitem[\protect\citeauthoryear{{Emsellem} et~al.,}{{Emsellem}
  et~al.}{2007}]{Emsellem+07}
{Emsellem} E.,  et~al., 2007, \mn@doi [\mnras]
  {10.1111/j.1365-2966.2007.11752.x}, \href
  {http://adsabs.harvard.edu/abs/2007MNRAS.379..401E} {379, 401}

\bibitem[\protect\citeauthoryear{{Emsellem} et~al.,}{{Emsellem}
  et~al.}{2011}]{Emsellem+11}
{Emsellem} E.,  et~al., 2011, \mn@doi [\mnras]
  {10.1111/j.1365-2966.2011.18496.x}, \href
  {https://ui.adsabs.harvard.edu/abs/2011MNRAS.414..888E} {414, 888}

\bibitem[\protect\citeauthoryear{{Falc{\'o}n-Barroso}
  et~al.,}{{Falc{\'o}n-Barroso} et~al.}{2019}]{barroso+19}
{Falc{\'o}n-Barroso} J.,  et~al., 2019, \mn@doi [\aap]
  {10.1051/0004-6361/201936413}, \href
  {https://ui.adsabs.harvard.edu/abs/2019A&A...632A..59F} {632, A59}

\bibitem[\protect\citeauthoryear{{Fraser-McKelvie}, {Arag{\'o}n-Salamanca},
  {Merrifield}, {Tabor}, {Bernardi}, {Drory}, {Parikh}  \&
  {Argudo-Fern{\'a}ndez}}{{Fraser-McKelvie} et~al.}{2018}]{Fraser+18}
{Fraser-McKelvie} A.,  {Arag{\'o}n-Salamanca} A.,  {Merrifield} M.,  {Tabor}
  M.,  {Bernardi} M.,  {Drory} N.,  {Parikh} T.,   {Argudo-Fern{\'a}ndez} M.,
  2018, \mn@doi [\mnras] {10.1093/mnras/sty2563}, \href
  {http://adsabs.harvard.edu/abs/2018MNRAS.481.5580F} {481, 5580}

\bibitem[\protect\citeauthoryear{{Freudling}, {Romaniello}, {Bramich},
  {Ballester}, {Forchi}, {Garc{\'{\i}}a-Dabl{\'o}}, {Moehler}  \&
  {Neeser}}{{Freudling} et~al.}{2013}]{Freudling+13}
{Freudling} W.,  {Romaniello} M.,  {Bramich} D.~M.,  {Ballester} P.,  {Forchi}
  V.,  {Garc{\'{\i}}a-Dabl{\'o}} C.~E.,  {Moehler} S.,   {Neeser} M.~J.,  2013,
  \mn@doi [\aap] {10.1051/0004-6361/201322494}, \href
  {http://adsabs.harvard.edu/abs/2013A\%26A...559A..96F} {559, A96}

\bibitem[\protect\citeauthoryear{{Gallazzi}, {Charlot}, {Brinchmann}  \&
  {White}}{{Gallazzi} et~al.}{2006}]{Gallazzi+06}
{Gallazzi} A.,  {Charlot} S.,  {Brinchmann} J.,   {White} S.~D.~M.,  2006,
  \mn@doi [\mnras] {10.1111/j.1365-2966.2006.10548.x}, \href
  {http://adsabs.harvard.edu/abs/2006MNRAS.370.1106G} {370, 1106}

\bibitem[\protect\citeauthoryear{{Graham} et~al.,}{{Graham}
  et~al.}{2018}]{Graham+18}
{Graham} M.~T.,  et~al., 2018, \mn@doi [\mnras] {10.1093/mnras/sty504}, \href
  {http://adsabs.harvard.edu/abs/2018MNRAS.477.4711G} {477, 4711}

\bibitem[\protect\citeauthoryear{{Gunn} \& {Gott}}{{Gunn} \&
  {Gott}}{1972}]{Gunn+72}
{Gunn} J.~E.,  {Gott} III J.~R.,  1972, \mn@doi [\apj] {10.1086/151605}, \href
  {http://adsabs.harvard.edu/abs/1972ApJ...176....1G} {176, 1}

\bibitem[\protect\citeauthoryear{{Jaff{\'e}} et~al.,}{{Jaff{\'e}}
  et~al.}{2011}]{jaffe+11b}
{Jaff{\'e}} Y.~L.,  et~al., 2011, \mn@doi [\mnras]
  {10.1111/j.1365-2966.2011.19384.x}, \href
  {http://adsabs.harvard.edu/abs/2011MNRAS.417.1996J} {417, 1996}

\bibitem[\protect\citeauthoryear{{Jaff{\'e}} et~al.,}{{Jaff{\'e}}
  et~al.}{2014}]{Jaffe+14}
{Jaff{\'e}} Y.~L.,  et~al., 2014, \mn@doi [\mnras] {10.1093/mnras/stu507},
  \href {http://adsabs.harvard.edu/abs/2014MNRAS.440.3491J} {440, 3491}

\bibitem[\protect\citeauthoryear{{Jerjen} \& {Dressler}}{{Jerjen} \&
  {Dressler}}{1997}]{Jerien+97}
{Jerjen} H.,  {Dressler} A.,  1997, \mn@doi [\aaps] {10.1051/aas:1997355},
  \href {https://ui.adsabs.harvard.edu/abs/1997A&AS..124....1J} {124, 1}

\bibitem[\protect\citeauthoryear{{Johnston}, {Arag{\'o}n-Salamanca},
  {Merrifield}  \& {Bedregal}}{{Johnston} et~al.}{2012}]{Johnston+12}
{Johnston} E.~J.,  {Arag{\'o}n-Salamanca} A.,  {Merrifield} M.~R.,   {Bedregal}
  A.~G.,  2012, \mn@doi [\mnras] {10.1111/j.1365-2966.2012.20813.x}, \href
  {http://adsabs.harvard.edu/abs/2012MNRAS.422.2590J} {422, 2590}

\bibitem[\protect\citeauthoryear{{Johnston}, {Arag{\'o}n-Salamanca}  \&
  {Merrifield}}{{Johnston} et~al.}{2014}]{Johnston+14}
{Johnston} E.~J.,  {Arag{\'o}n-Salamanca} A.,   {Merrifield} M.~R.,  2014,
  \mn@doi [\mnras] {10.1093/mnras/stu582}, \href
  {http://adsabs.harvard.edu/abs/2014MNRAS.441..333J} {441, 333}

\bibitem[\protect\citeauthoryear{{Johnston} et~al.,}{{Johnston}
  et~al.}{2017}]{Johnston+17}
{Johnston} E.~J.,  et~al., 2017, \mn@doi [\mnras] {10.1093/mnras/stw2823},
  \href {https://ui.adsabs.harvard.edu/abs/2017MNRAS.465.2317J} {465, 2317}

\bibitem[\protect\citeauthoryear{{Karachentseva}, {Mitronova}, {Melnyk}  \&
  {Karachentsev}}{{Karachentseva} et~al.}{2010}]{Karachentseva+10}
{Karachentseva} V.~E.,  {Mitronova} S.~N.,  {Melnyk} O.~V.,   {Karachentsev}
  I.~D.,  2010, \mn@doi [Astrophysical Bulletin] {10.1134/S1990341310010013},
  \href {http://adsabs.harvard.edu/abs/2010AstBu..65....1K} {65, 1}

\bibitem[\protect\citeauthoryear{{Katkov}, {Sil'chenko}  \&
  {Afanasiev}}{{Katkov} et~al.}{2014}]{Katkov+14}
{Katkov} I.~Y.,  {Sil'chenko} O.~K.,   {Afanasiev} V.~L.,  2014, \mn@doi
  [\mnras] {10.1093/mnras/stt2365}, \href
  {http://adsabs.harvard.edu/abs/2014MNRAS.438.2798K} {438, 2798}

\bibitem[\protect\citeauthoryear{{Krajnovi{\'c}}, {Cappellari}, {de Zeeuw}  \&
  {Copin}}{{Krajnovi{\'c}} et~al.}{2006}]{Krajnovic+06}
{Krajnovi{\'c}} D.,  {Cappellari} M.,  {de Zeeuw} P.~T.,   {Copin} Y.,  2006,
  \mn@doi [\mnras] {10.1111/j.1365-2966.2005.09902.x}, \href
  {http://adsabs.harvard.edu/abs/2006MNRAS.366..787K} {366, 787}

\bibitem[\protect\citeauthoryear{{Krajnovi{\'c}} et~al.,}{{Krajnovi{\'c}}
  et~al.}{2013}]{Krajnovic+13}
{Krajnovi{\'c}} D.,  et~al., 2013, \mn@doi [\mnras] {10.1093/mnras/sts315},
  \href {http://adsabs.harvard.edu/abs/2013MNRAS.432.1768K} {432, 1768}

\bibitem[\protect\citeauthoryear{{Lagos}, {Theuns}, {Stevens}, {Cortese},
  {Padilla}, {Davis}, {Contreras}  \& {Croton}}{{Lagos}
  et~al.}{2017}]{Lagos+17}
{Lagos} C.~d.~P.,  {Theuns} T.,  {Stevens} A.~R.~H.,  {Cortese} L.,  {Padilla}
  N.~D.,  {Davis} T.~A.,  {Contreras} S.,   {Croton} D.,  2017, \mn@doi
  [\mnras] {10.1093/mnras/stw2610}, \href
  {http://adsabs.harvard.edu/abs/2017MNRAS.464.3850L} {464, 3850}

\bibitem[\protect\citeauthoryear{{Lagos} et~al.,}{{Lagos}
  et~al.}{2018}]{Lagos+18}
{Lagos} C.~d.~P.,  et~al., 2018, \mn@doi [\mnras] {10.1093/mnras/stx2667},
  \href {http://adsabs.harvard.edu/abs/2018MNRAS.473.4956L} {473, 4956}

\bibitem[\protect\citeauthoryear{{Lee}, {Chung}  \& {Yoon}}{{Lee}
  et~al.}{2019}]{Lee+19}
{Lee} S.-Y.,  {Chung} C.,   {Yoon} S.-J.,  2019, \mn@doi [\apjs]
  {10.3847/1538-4365/aaecd4}, \href
  {https://ui.adsabs.harvard.edu/abs/2019ApJS..240....2L} {240, 2}

\bibitem[\protect\citeauthoryear{{Leung} et~al.,}{{Leung}
  et~al.}{2018}]{leung+18}
{Leung} G.~Y.~C.,  et~al., 2018, \mn@doi [\mnras] {10.1093/mnras/sty288}, \href
  {http://adsabs.harvard.edu/abs/2018MNRAS.477..254L} {477, 254}

\bibitem[\protect\citeauthoryear{{McDermid} et~al.,}{{McDermid}
  et~al.}{2015}]{McDermid+15}
{McDermid} R.~M.,  et~al., 2015, \mn@doi [\mnras] {10.1093/mnras/stv105}, \href
  {http://adsabs.harvard.edu/abs/2015MNRAS.448.3484M} {448, 3484}

\bibitem[\protect\citeauthoryear{{Merluzzi}, {Busarello}, {Dopita}, {Haines},
  {Steinhauser}, {Bourdin}  \& {Mazzotta}}{{Merluzzi}
  et~al.}{2016}]{Merluzzi+16}
{Merluzzi} P.,  {Busarello} G.,  {Dopita} M.~A.,  {Haines} C.~P.,
  {Steinhauser} D.,  {Bourdin} H.,   {Mazzotta} P.,  2016, \mn@doi [\mnras]
  {10.1093/mnras/stw1198}, \href
  {http://adsabs.harvard.edu/abs/2016MNRAS.460.3345M} {460, 3345}

\bibitem[\protect\citeauthoryear{{Mishra}, {Wadadekar}  \& {Barway}}{{Mishra}
  et~al.}{2018}]{Mishra+18}
{Mishra} P.~K.,  {Wadadekar} Y.,   {Barway} S.,  2018, \mn@doi [\mnras]
  {10.1093/mnras/sty1107}, \href
  {https://ui.adsabs.harvard.edu/abs/2018MNRAS.478..351M} {478, 351}

\bibitem[\protect\citeauthoryear{{Navarro}, {Frenk}  \& {White}}{{Navarro}
  et~al.}{1997}]{Navarro+97}
{Navarro} J.~F.,  {Frenk} C.~S.,   {White} S.~D.~M.,  1997, \mn@doi [\apj]
  {10.1086/304888}, \href {http://adsabs.harvard.edu/abs/1997ApJ...490..493N}
  {490, 493}

\bibitem[\protect\citeauthoryear{{Neill}, {Seibert}, {Tully}, {Courtois},
  {Sorce}, {Jarrett}, {Scowcroft}  \& {Masci}}{{Neill} et~al.}{2014}]{Neill+14}
{Neill} J.~D.,  {Seibert} M.,  {Tully} R.~B.,  {Courtois} H.,  {Sorce} J.~G.,
  {Jarrett} T.~H.,  {Scowcroft} V.,   {Masci} F.~J.,  2014, \mn@doi [\apj]
  {10.1088/0004-637X/792/2/129}, \href
  {http://adsabs.harvard.edu/abs/2014ApJ...792..129N} {792, 129}

\bibitem[\protect\citeauthoryear{{Peng}, {Ho}, {Impey}  \& {Rix}}{{Peng}
  et~al.}{2002}]{Peng+02}
{Peng} C.~Y.,  {Ho} L.~C.,  {Impey} C.~D.,   {Rix} H.-W.,  2002, \mn@doi [\aj]
  {10.1086/340952}, \href
  {https://ui.adsabs.harvard.edu/abs/2002AJ....124..266P} {124, 266}

\bibitem[\protect\citeauthoryear{{Poggianti} et~al.,}{{Poggianti}
  et~al.}{2009}]{Poggianti+09}
{Poggianti} B.~M.,  et~al., 2009, \mn@doi [\apjl]
  {10.1088/0004-637X/697/2/L137}, \href
  {http://adsabs.harvard.edu/abs/2009ApJ...697L.137P} {697, L137}

\bibitem[\protect\citeauthoryear{{Ponman}, {Allan}, {Jones}, {Merrifield},
  {McHardy}, {Lehto}  \& {Luppino}}{{Ponman} et~al.}{1994}]{Ponman+94}
{Ponman} T.~J.,  {Allan} D.~J.,  {Jones} L.~R.,  {Merrifield} M.,  {McHardy}
  I.~M.,  {Lehto} H.~J.,   {Luppino} G.~A.,  1994, \mn@doi [\nat]
  {10.1038/369462a0}, \href
  {https://ui.adsabs.harvard.edu/abs/1994Natur.369..462P} {369, 462}

\bibitem[\protect\citeauthoryear{{Querejeta} et~al.,}{{Querejeta}
  et~al.}{2015}]{Querejeta+15}
{Querejeta} M.,  et~al., 2015, \mn@doi [\aap] {10.1051/0004-6361/201526354},
  \href {http://adsabs.harvard.edu/abs/2015A%26A...579L...2Q} {579, L2}

\bibitem[\protect\citeauthoryear{{Rizzo}, {Fraternali}  \& {Iorio}}{{Rizzo}
  et~al.}{2018}]{Rizzo+18}
{Rizzo} F.,  {Fraternali} F.,   {Iorio} G.,  2018, \mn@doi [\mnras]
  {10.1093/mnras/sty347}, \href
  {http://adsabs.harvard.edu/abs/2018MNRAS.476.2137R} {476, 2137}

\bibitem[\protect\citeauthoryear{{Rodr{\'\i}guez Del Pino}, {Bamford},
  {Arag{\'o}n-Salamanca}, {Milvang-Jensen}, {Merrifield}  \&
  {Balcells}}{{Rodr{\'\i}guez Del Pino} et~al.}{2014}]{DelPino+14}
{Rodr{\'\i}guez Del Pino} B.,  {Bamford} S.~P.,  {Arag{\'o}n-Salamanca} A.,
  {Milvang-Jensen} B.,  {Merrifield} M.~R.,   {Balcells} M.,  2014, \mn@doi
  [\mnras] {10.1093/mnras/stt2202}, \href
  {https://ui.adsabs.harvard.edu/abs/2014MNRAS.438.1038R} {438, 1038}

\bibitem[\protect\citeauthoryear{{Saha} \& {Cortesi}}{{Saha} \&
  {Cortesi}}{2018}]{Saha+18}
{Saha} K.,  {Cortesi} A.,  2018, \mn@doi [\apj] {10.3847/2041-8213/aad23a},
  \href {https://ui.adsabs.harvard.edu/abs/2018ApJ...862L..12S} {862, L12}

\bibitem[\protect\citeauthoryear{{Sarzi} et~al.,}{{Sarzi}
  et~al.}{2006}]{Sarzi+06}
{Sarzi} M.,  et~al., 2006, \mn@doi [\mnras] {10.1111/j.1365-2966.2005.09839.x},
  \href {http://adsabs.harvard.edu/abs/2006MNRAS.366.1151S} {366, 1151}

\bibitem[\protect\citeauthoryear{{Scott} et~al.,}{{Scott}
  et~al.}{2013}]{Scott+13}
{Scott} N.,  et~al., 2013, \mn@doi [\mnras] {10.1093/mnras/sts422}, \href
  {http://adsabs.harvard.edu/abs/2013MNRAS.432.1894S} {432, 1894}

\bibitem[\protect\citeauthoryear{{Skrutskie} et~al.,}{{Skrutskie}
  et~al.}{2006}]{Skrutskie+06}
{Skrutskie} M.~F.,  et~al., 2006, \mn@doi [\aj] {10.1086/498708}, \href
  {https://ui.adsabs.harvard.edu/abs/2006AJ....131.1163S} {131, 1163}

\bibitem[\protect\citeauthoryear{{Sorce}, {Courtois}  \& {Tully}}{{Sorce}
  et~al.}{2012}]{Sorce+12}
{Sorce} J.~G.,  {Courtois} H.~M.,   {Tully} R.~B.,  2012, \mn@doi [\aj]
  {10.1088/0004-6256/144/5/133}, \href
  {http://adsabs.harvard.edu/abs/2012AJ....144..133S} {144, 133}

\bibitem[\protect\citeauthoryear{{Sorce} et~al.,}{{Sorce}
  et~al.}{2013}]{Sorce+13}
{Sorce} J.~G.,  et~al., 2013, \mn@doi [\apj] {10.1088/0004-637X/765/2/94},
  \href {http://adsabs.harvard.edu/abs/2013ApJ...765...94S} {765, 94}

\bibitem[\protect\citeauthoryear{{Soto}, {Lilly}, {Bacon}, {Richard}  \&
  {Conseil}}{{Soto} et~al.}{2016}]{Soto+16}
{Soto} K.~T.,  {Lilly} S.~J.,  {Bacon} R.,  {Richard} J.,   {Conseil} S.,
  2016, \mn@doi [\mnras] {10.1093/mnras/stw474}, \href
  {http://adsabs.harvard.edu/abs/2016MNRAS.458.3210S} {458, 3210}

\bibitem[\protect\citeauthoryear{{Steinhauser}, {Haider}, {Kapferer}  \&
  {Schindler}}{{Steinhauser} et~al.}{2012}]{Steinhauser+12}
{Steinhauser} D.,  {Haider} M.,  {Kapferer} W.,   {Schindler} S.,  2012,
  \mn@doi [\aap] {10.1051/0004-6361/201118311}, \href
  {http://adsabs.harvard.edu/abs/2012A\%26A...544A..54S} {544, A54}

\bibitem[\protect\citeauthoryear{{Tapia}, {Eliche-Moral}, {Aceves},
  {Rodr{\'{\i}}guez-P{\'e}rez}, {Borlaff}  \& {Querejeta}}{{Tapia}
  et~al.}{2017}]{Tapia+17}
{Tapia} T.,  {Eliche-Moral} M.~C.,  {Aceves} H.,  {Rodr{\'{\i}}guez-P{\'e}rez}
  C.,  {Borlaff} A.,   {Querejeta} M.,  2017, \mn@doi [\aap]
  {10.1051/0004-6361/201628821}, \href
  {http://adsabs.harvard.edu/abs/2017A%26A...604A.105T} {604, A105}

\bibitem[\protect\citeauthoryear{{Thomas}, {Maraston}, {Bender}  \& {Mendes de
  Oliveira}}{{Thomas} et~al.}{2005}]{Thomas+05}
{Thomas} D.,  {Maraston} C.,  {Bender} R.,   {Mendes de Oliveira} C.,  2005,
  \mn@doi [\apj] {10.1086/426932}, \href
  {http://adsabs.harvard.edu/abs/2005ApJ...621..673T} {621, 673}

\bibitem[\protect\citeauthoryear{{Tully} \& {Fisher}}{{Tully} \&
  {Fisher}}{1977}]{Tully+77}
{Tully} R.~B.,  {Fisher} J.~R.,  1977, \aap, \href
  {http://adsabs.harvard.edu/abs/1977A%26A....54..661T} {54, 661}

\bibitem[\protect\citeauthoryear{{Tully}, {Shaya}  \& {Pierce}}{{Tully}
  et~al.}{1992}]{Tully+92}
{Tully} R.~B.,  {Shaya} E.~J.,   {Pierce} M.~J.,  1992, \mn@doi [\apjs]
  {10.1086/191673}, \href
  {https://ui.adsabs.harvard.edu/abs/1992ApJS...80..479T} {80, 479}

\bibitem[\protect\citeauthoryear{{Vazdekis}, {Ricciardelli}, {Cenarro},
  {Rivero-Gonz{\'a}lez}, {D{\'{\i}}az-Garc{\'{\i}}a}  \&
  {Falc{\'o}n-Barroso}}{{Vazdekis} et~al.}{2012}]{Vazdekis+12}
{Vazdekis} A.,  {Ricciardelli} E.,  {Cenarro} A.~J.,  {Rivero-Gonz{\'a}lez}
  J.~G.,  {D{\'{\i}}az-Garc{\'{\i}}a} L.~A.,   {Falc{\'o}n-Barroso} J.,  2012,
  \mn@doi [\mnras] {10.1111/j.1365-2966.2012.21179.x}, \href
  {http://adsabs.harvard.edu/abs/2012MNRAS.424..157V} {424, 157}

\bibitem[\protect\citeauthoryear{{Vollmer}}{{Vollmer}}{2013}]{Vollmer13}
{Vollmer} B.,  2013, {The Influence of Environment on Galaxy Evolution}.
p.~207, \mn@doi{10.1007/978-94-007-5609-0_5}

\bibitem[\protect\citeauthoryear{{Weilbacher}, {Streicher}, {Urrutia}, {Jarno},
  {P{\'e}contal-Rousset}, {Bacon}  \& {B{\"o}hm}}{{Weilbacher}
  et~al.}{2012}]{Weilbacher+12}
{Weilbacher} P.~M.,  {Streicher} O.,  {Urrutia} T.,  {Jarno} A.,
  {P{\'e}contal-Rousset} A.,  {Bacon} R.,   {B{\"o}hm} P.,  2012, in Software
  and Cyberinfrastructure for Astronomy II. p. 84510B,
  \mn@doi{10.1117/12.925114}

\bibitem[\protect\citeauthoryear{{Williams}, {Bureau}  \&
  {Cappellari}}{{Williams} et~al.}{2009}]{Williams+09}
{Williams} M.~J.,  {Bureau} M.,   {Cappellari} M.,  2009, \mn@doi [\mnras]
  {10.1111/j.1365-2966.2009.15582.x}, \href
  {http://adsabs.harvard.edu/abs/2009MNRAS.400.1665W} {400, 1665}

\bibitem[\protect\citeauthoryear{{Williams}, {Bureau}  \&
  {Cappellari}}{{Williams} et~al.}{2010}]{Williams+10}
{Williams} M.~J.,  {Bureau} M.,   {Cappellari} M.,  2010, \mn@doi [\mnras]
  {10.1111/j.1365-2966.2010.17406.x}, \href
  {http://adsabs.harvard.edu/abs/2010MNRAS.409.1330W} {409, 1330}

\bibitem[\protect\citeauthoryear{{Wright} et~al.,}{{Wright}
  et~al.}{2010}]{Wright+10}
{Wright} E.~L.,  et~al., 2010, \mn@doi [\aj] {10.1088/0004-6256/140/6/1868},
  \href {http://adsabs.harvard.edu/abs/2010AJ....140.1868W} {140, 1868}

\bibitem[\protect\citeauthoryear{{de Vaucouleurs}, {de Vaucouleurs}, {Corwin},
  {Buta}, {Paturel}  \& {Fouqu{\'e}}}{{de Vaucouleurs} et~al.}{1991}]{RC3}
{de Vaucouleurs} G.,  {de Vaucouleurs} A.,  {Corwin} Jr. H.~G.,  {Buta} R.~J.,
  {Paturel} G.,   {Fouqu{\'e}} P.,  1991, {Third Reference Catalogue of Bright
  Galaxies. Volume I: Explanations and references. Volume II: Data for galaxies
  between 0$^{h}$ and 12$^{h}$. Volume III: Data for galaxies between 12$^{h}$
  and 24$^{h}$.}

\bibitem[\protect\citeauthoryear{{van den Berg}, {Hong}  \& {Grindlay}}{{van
  den Berg} et~al.}{2009}]{vandenBerg+09}
{van den Berg} M.,  {Hong} J.~S.,   {Grindlay} J.~E.,  2009, \mn@doi [\apj]
  {10.1088/0004-637X/700/2/1702}, \href
  {http://adsabs.harvard.edu/abs/2009ApJ...700.1702V} {700, 1702}

\makeatother
\end{thebibliography}
\appendix
\section{Measured kinematic maps of the MUSE lenticular sample}
\label{sec:app_a}

In this section we show the two-dimensional stellar kinematic maps of the sample galaxies observed with MUSE and discuss the individual galaxies (Figures \ref{kinemaps_i}--\ref{kinemaps_cc}). We also show the two-dimensional maps of the ionised-gas velocity only for those galaxies that have a significant detection in H$\alpha$ (Figure \ref{kinemaps_gas}). We refer the reader to Section \ref{sec:muse_sample} for further information about the data reduction and the measurements. 
 
 {\it PGC 0044187 --  2MIG 131}. The stellar velocity field is regular and reaches an observed peak-to-peak rotation amplitude of $\sim 300$ \kms. The stellar velocity dispersion has a central peak of $\sim 300$ \kms\ and then it decrease down to $\sim 150$ \kms\ at $\sim 5"$ from the centre and stay relatively flat.
 
 {\it IC 1989 -- 2MIG 445}. The stellar velocity field is regular and reaches an observed peak-to-peak rotation amplitude of $\sim 200$ \kms. The velocity dispersion ranges between 100 and 150 \kms\ except in the central region, which is characterised by an elongated structure aligned along the kinematic major axis. This structure has a counter-part in the rotation map. A discontinuity in the kinematic maps is observed at $\sim 10"$ on the North West side, along the major axis. It is most likely caused by the presence of a dwarf companion in projection along the line of sight. This region, approximately 5" wide, was removed in the derivation of $\lambda$, $V/sigma$, and $V_{\rm circ}$. The fit of ionised-gas emission lines also fails in this contaminated area. 
 
 {\it NGC 3546 -- 2MIG 1546}. The stellar velocity field is regular and reaches an observed peak-to-peak rotation amplitude of $\sim 350$ \kms. The stellar velocity dispersion has a central peak of $\sim 220$ \kms\ and then it decrease down to $\sim 100$ \kms\ at $\sim 10"$ from the centre and stay relatively flat. The ionised-gas component is extended towards the entire field of view and rotates regularly with a peak-to-peak rotation amplitude of $\sim 500$ \kms.

 {\it PGC 045474 -- 2MIG 1814}. The stellar velocity field is regular and reaches an observed peak-to-peak rotation amplitude of $\sim 300$ \kms. The stellar velocity dispersion has a central peak of $\sim 250$ \kms\ and then it decrease down to $\sim 150$ \kms\ at $\sim 30"$ from the centre. The ionised-gas component is extended towards the entire field of view. It shows a disk-like regular rotation in the central $\sim 10"$ and then it shows an irregular, outflow-like field, with line-of-sight velocity of $\sim 200$ \kms\ with respect to the system centre in the South-West region.
 
 {\it NGC 4696D -- CCC 43}. The stellar velocity field is regular and reaches an observed peak-to-peak rotation amplitude of $\sim 200$ \kms. The stellar velocity dispersion has a central peak of $\sim 170$ \kms\ and then it decrease down to $\sim 50$ \kms\ at $\sim 30"$ from the centre.
 
 {\it NGC 4706 -- CCC 122}. The stellar velocity field is regular and reaches an observed peak-to-peak rotation amplitude of $\sim 400$ \kms. The stellar velocity dispersion has a central peak of $\sim 250$ \kms\ and then it rapidly decrease down to $\sim 50$ \kms\ at $\sim 30"$ from the centre. Ionised-gas is detected in the innermost $\sim 10"$ and shows a disk-like rotation of $\sim 400$ \kms (peak-to-peak).
 
 {\it PGC 043435 -- CCC 137}. The stellar velocity field is regular and reaches an observed peak-to-peak rotation amplitude of $\sim 400$ \kms. The stellar velocity dispersion has a central peak of $\sim 220$ \kms\ and then it rapidly decrease down to $\sim 50$ \kms\ at $\sim 30"$ from the centre.

 {\it PGC 043466 -- CCC 158}. The stellar velocity field is regular and reaches an observed peak-to-peak rotation amplitude of $\sim 200$ \kms. The stellar velocity dispersion has a central peak of $\sim 220$ \kms\ and then it rapidly decrease down to $\sim 50$ \kms\ at $\sim 30"$ from the centre. Ionised-gas is detected in the central $\sim 10"$ and shows a skewed and irregular velocity field that is $\sim 30$ degrees misaligned with respect to the stellar velocity field. The degree of gas-stellar misalignment is poorly constrained because of the irregularity of the ionised gas velocity field.
 
\begin{figure*}
\includegraphics[width=0.40\textwidth]{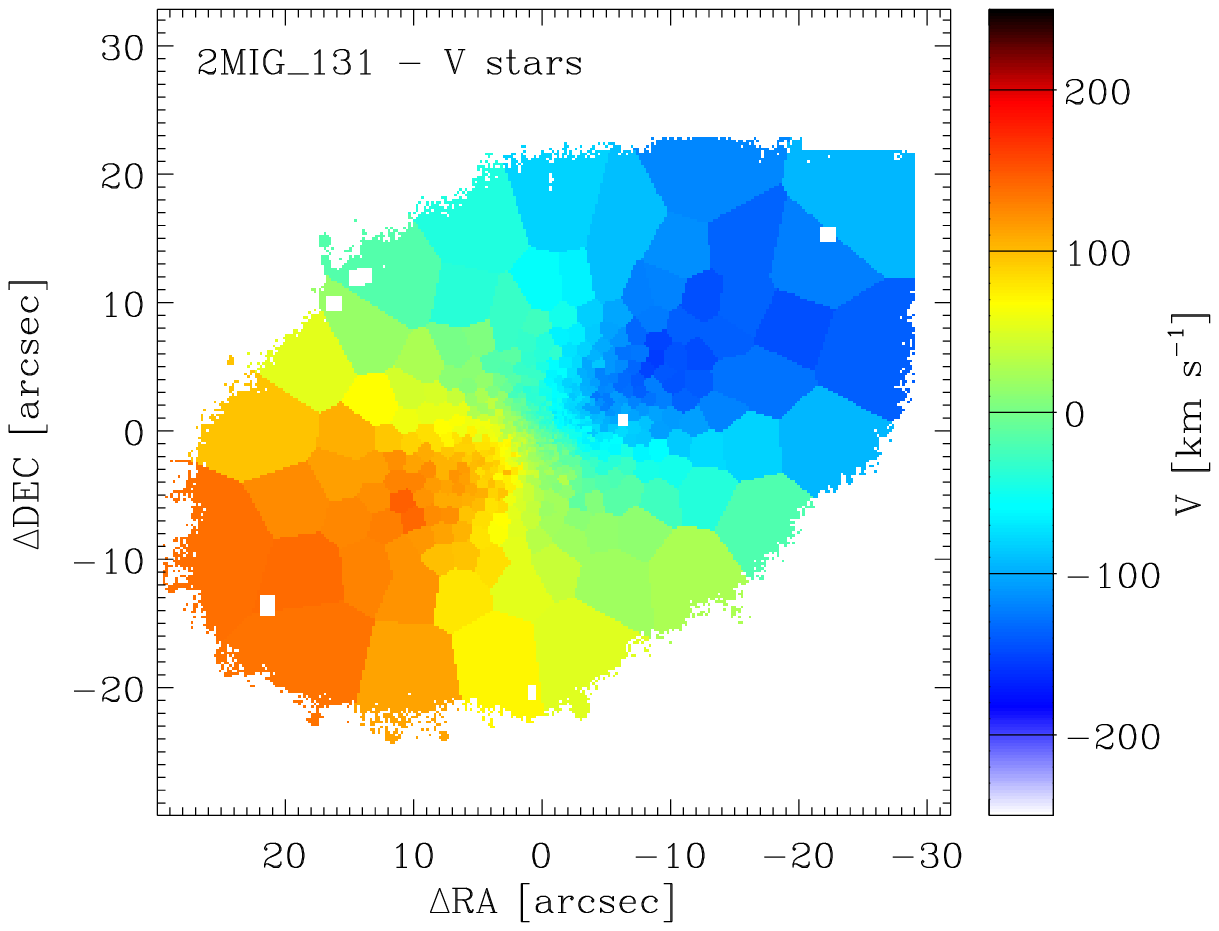}
\includegraphics[width=0.40\textwidth]{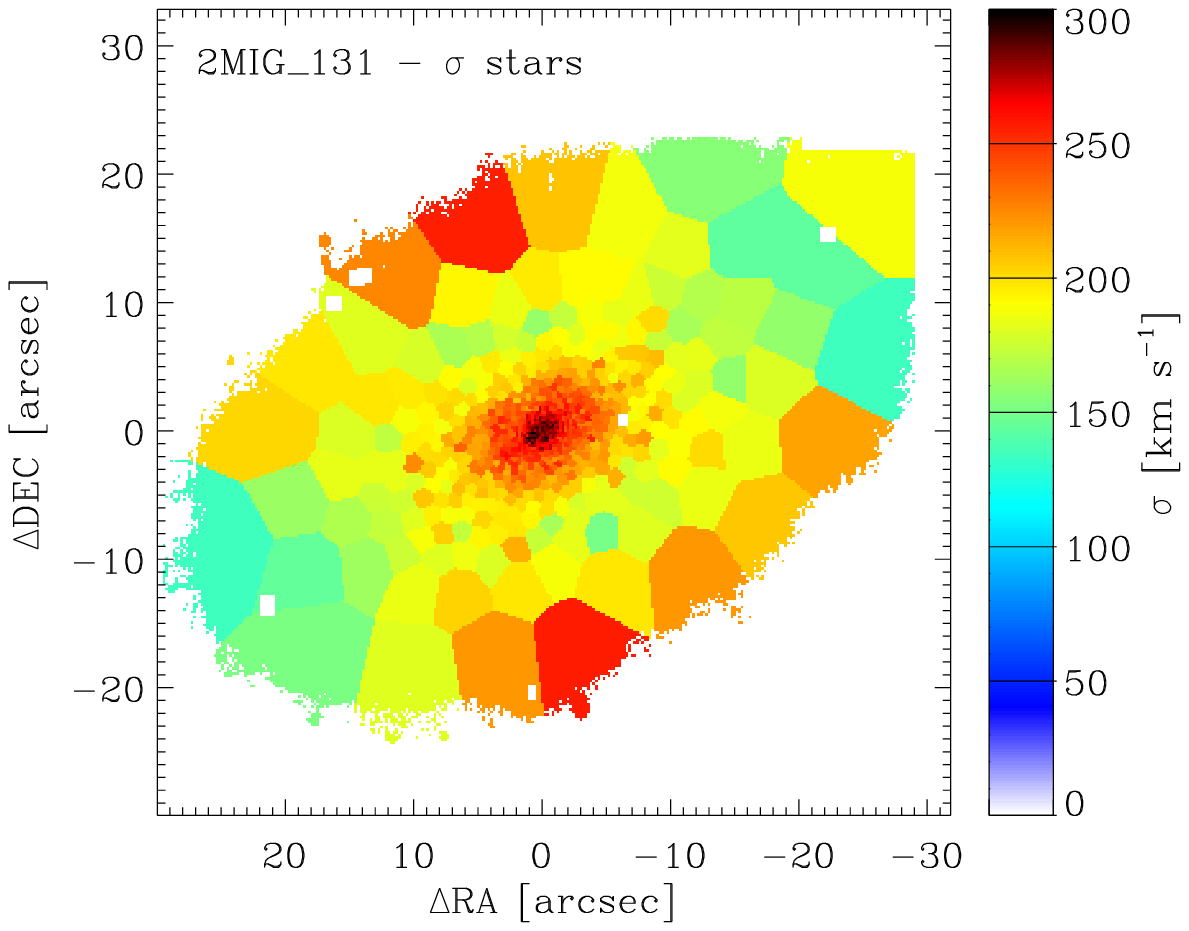}
\includegraphics[width=0.40\textwidth]{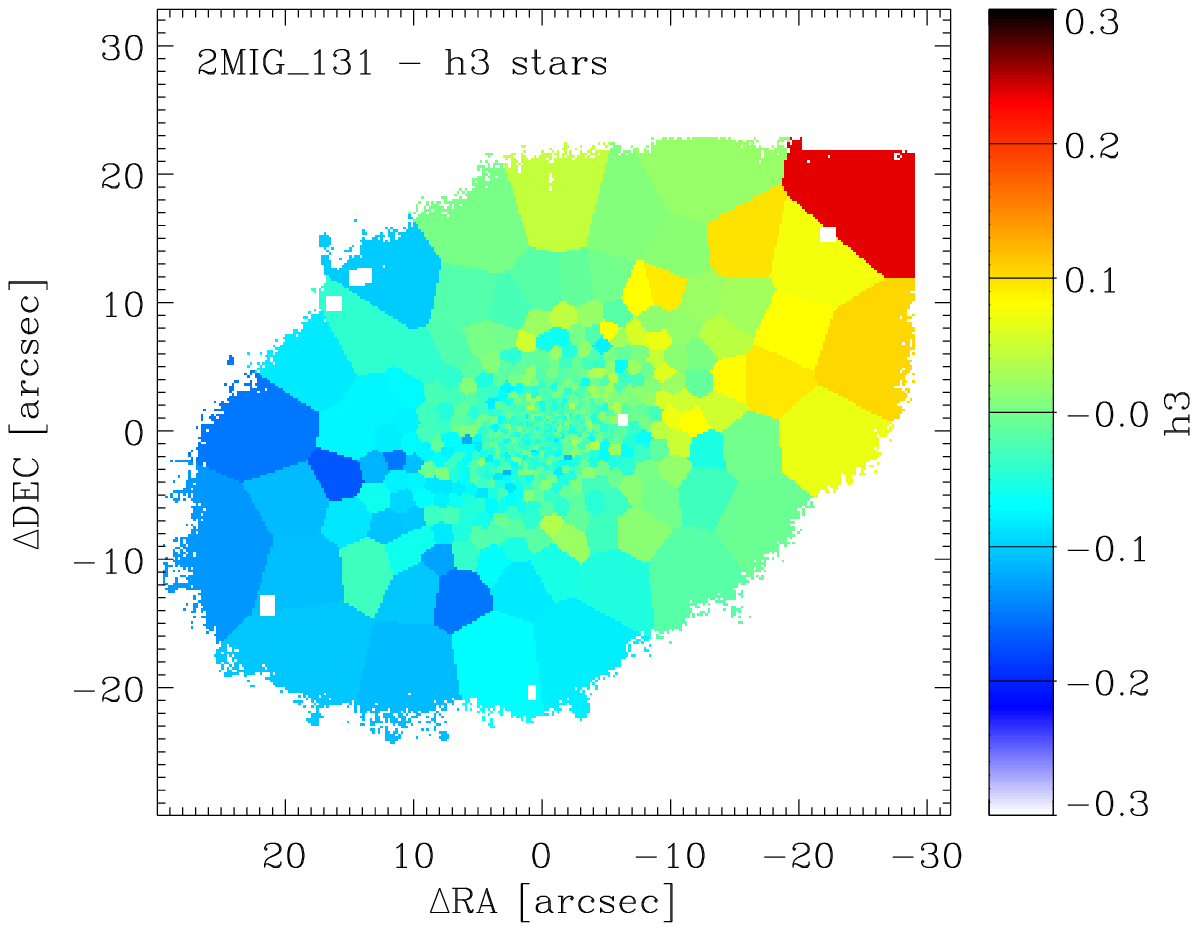}
\includegraphics[width=0.40\textwidth]{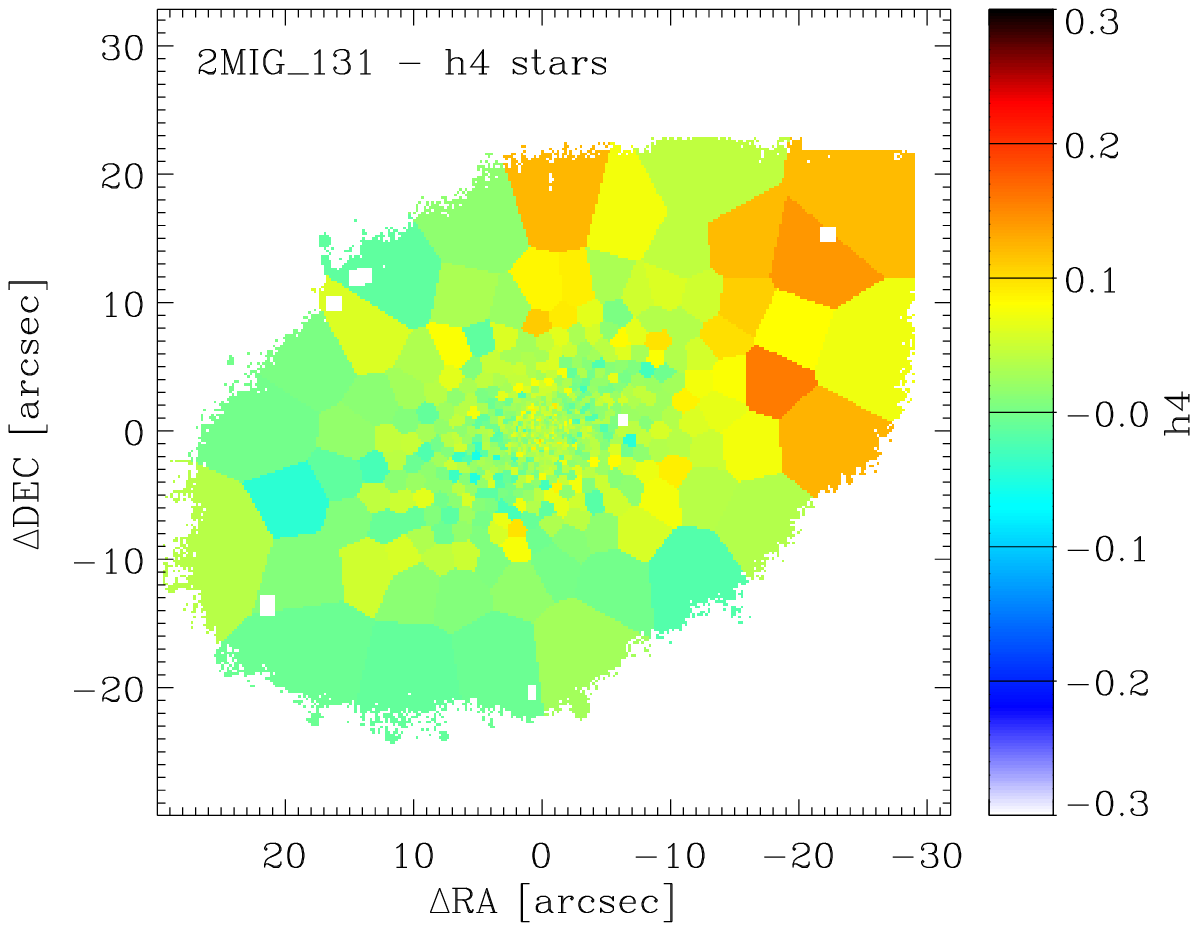}
\includegraphics[width=0.40\textwidth]{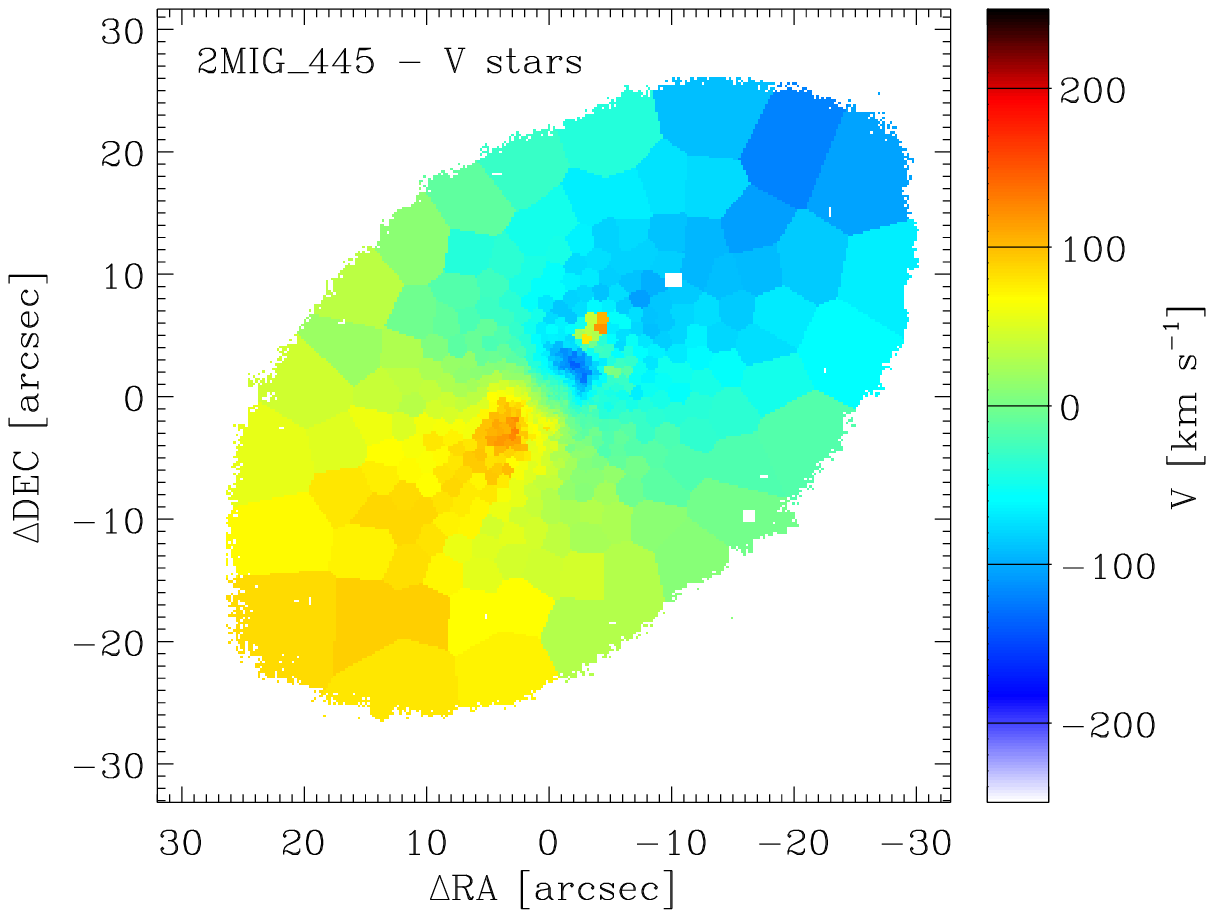}
\includegraphics[width=0.40\textwidth]{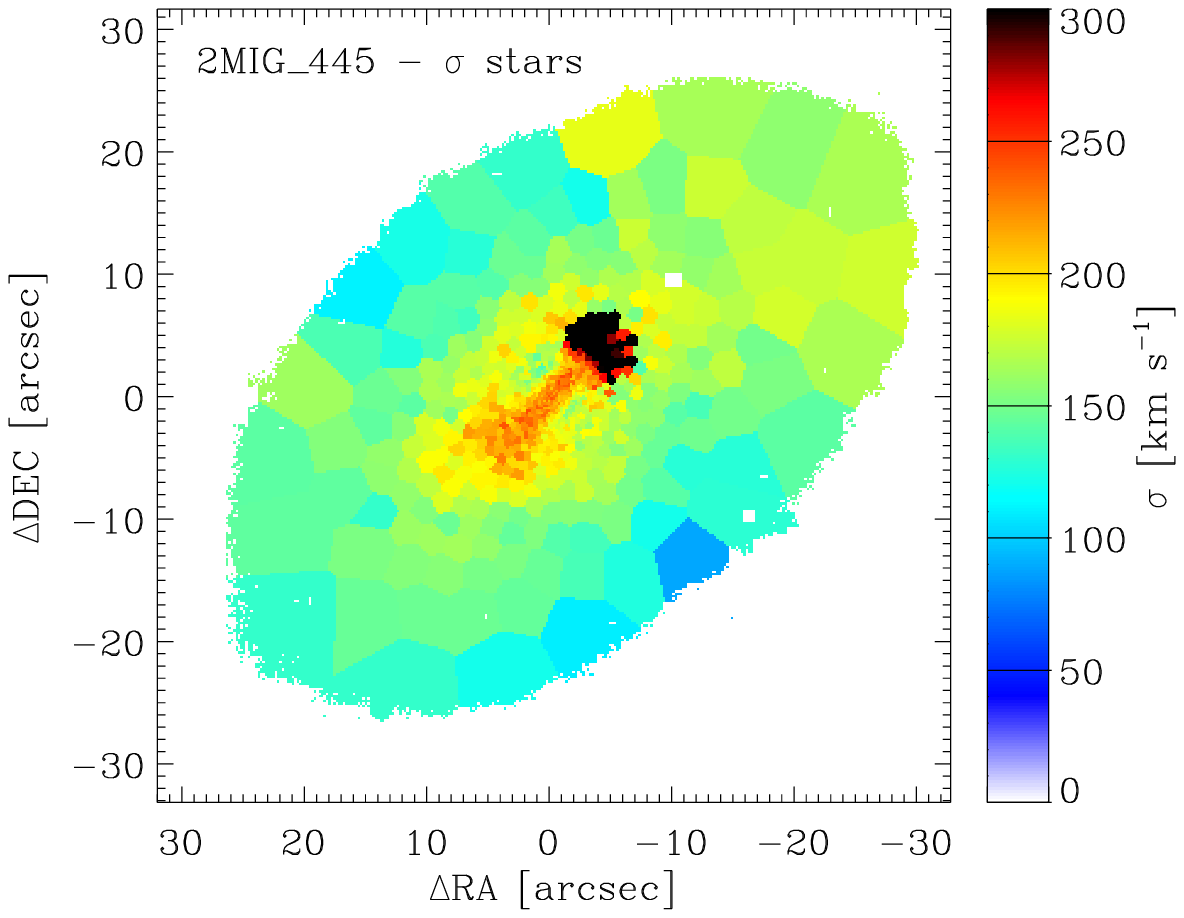}
\includegraphics[width=0.40\textwidth]{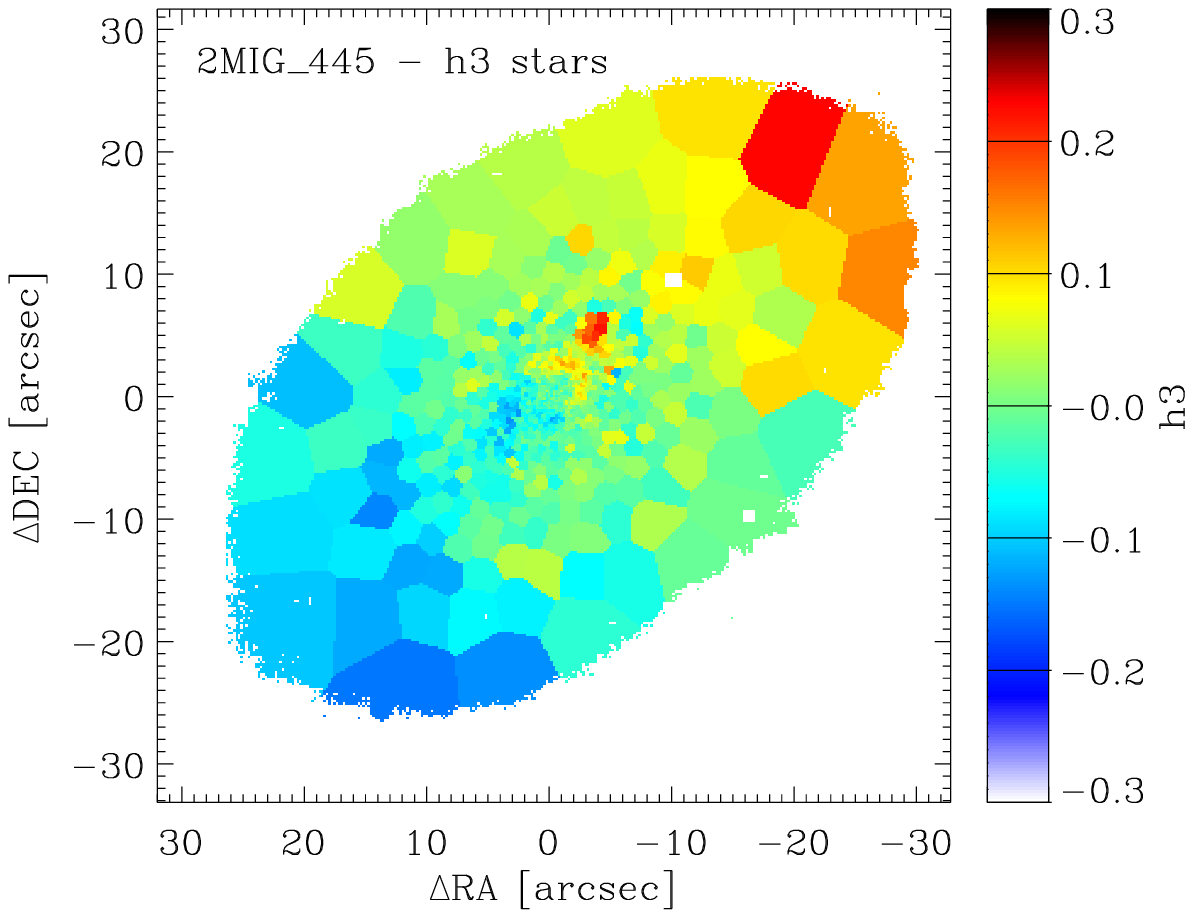}
\includegraphics[width=0.40\textwidth]{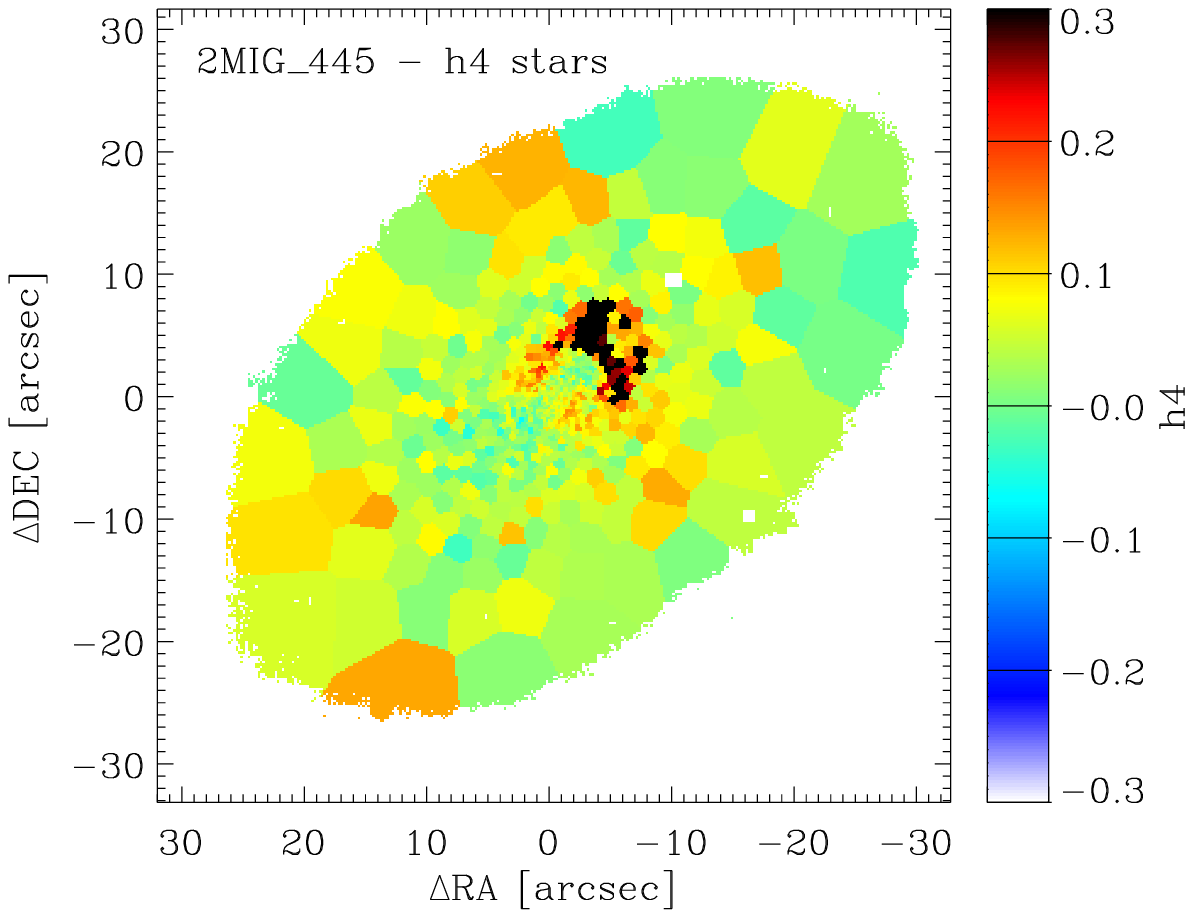}
\caption[]{Velocity, velocity dispersion, $h3$ , and $h4$ maps for the MUSE sample of Isolated galaxies.}
\label{kinemaps_i}
\end{figure*}
\addtocounter{figure}{-1}
\begin{figure*}
\includegraphics[width=0.40\textwidth]{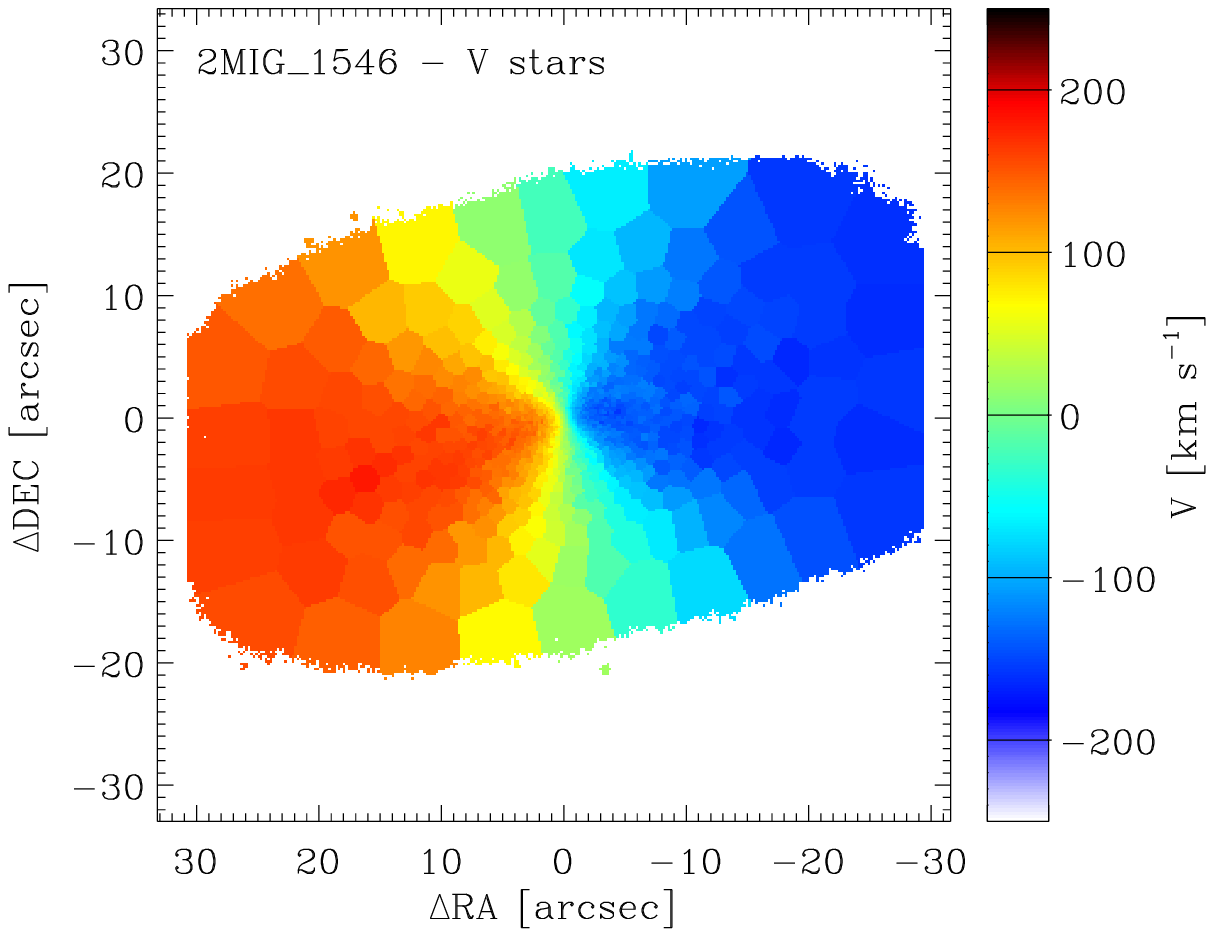}
\includegraphics[width=0.40\textwidth]{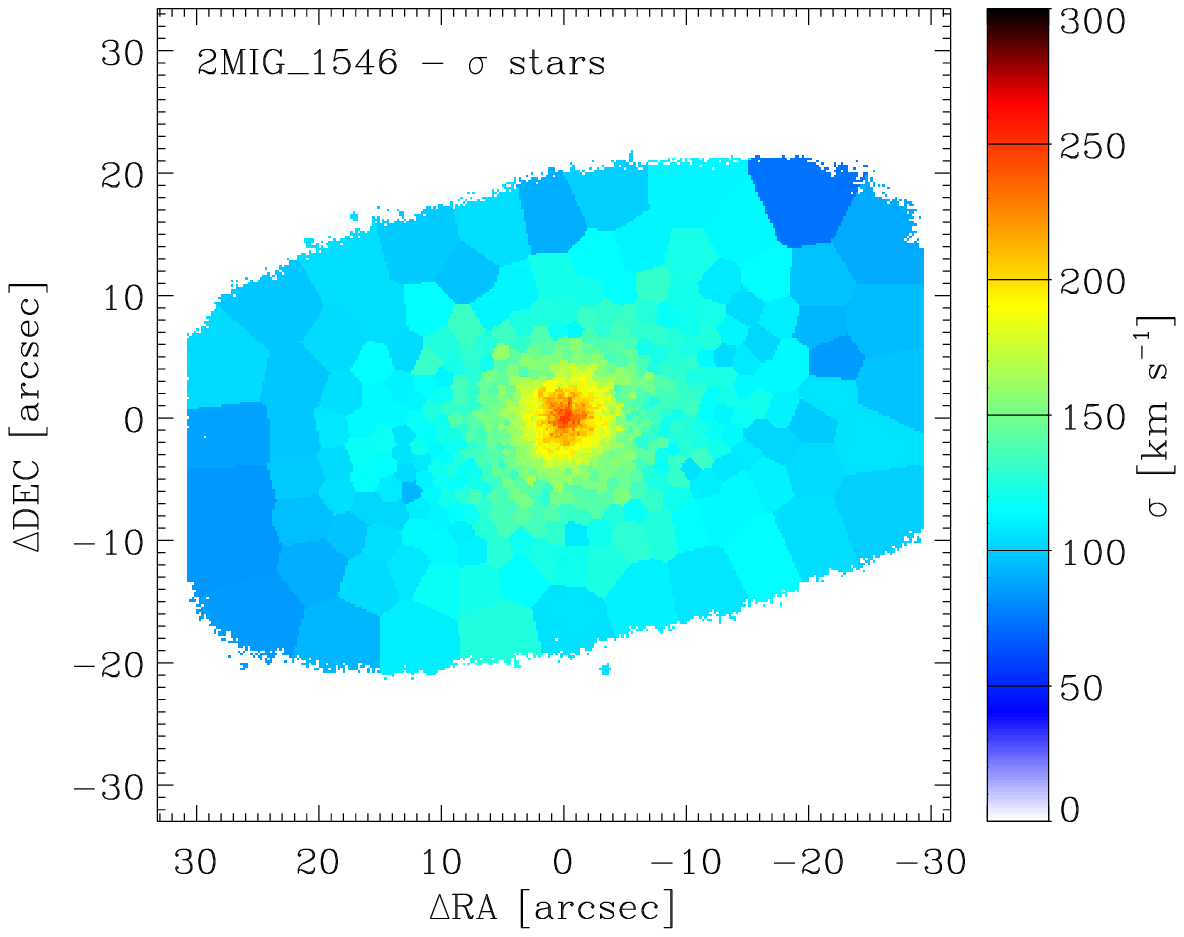}
\includegraphics[width=0.40\textwidth]{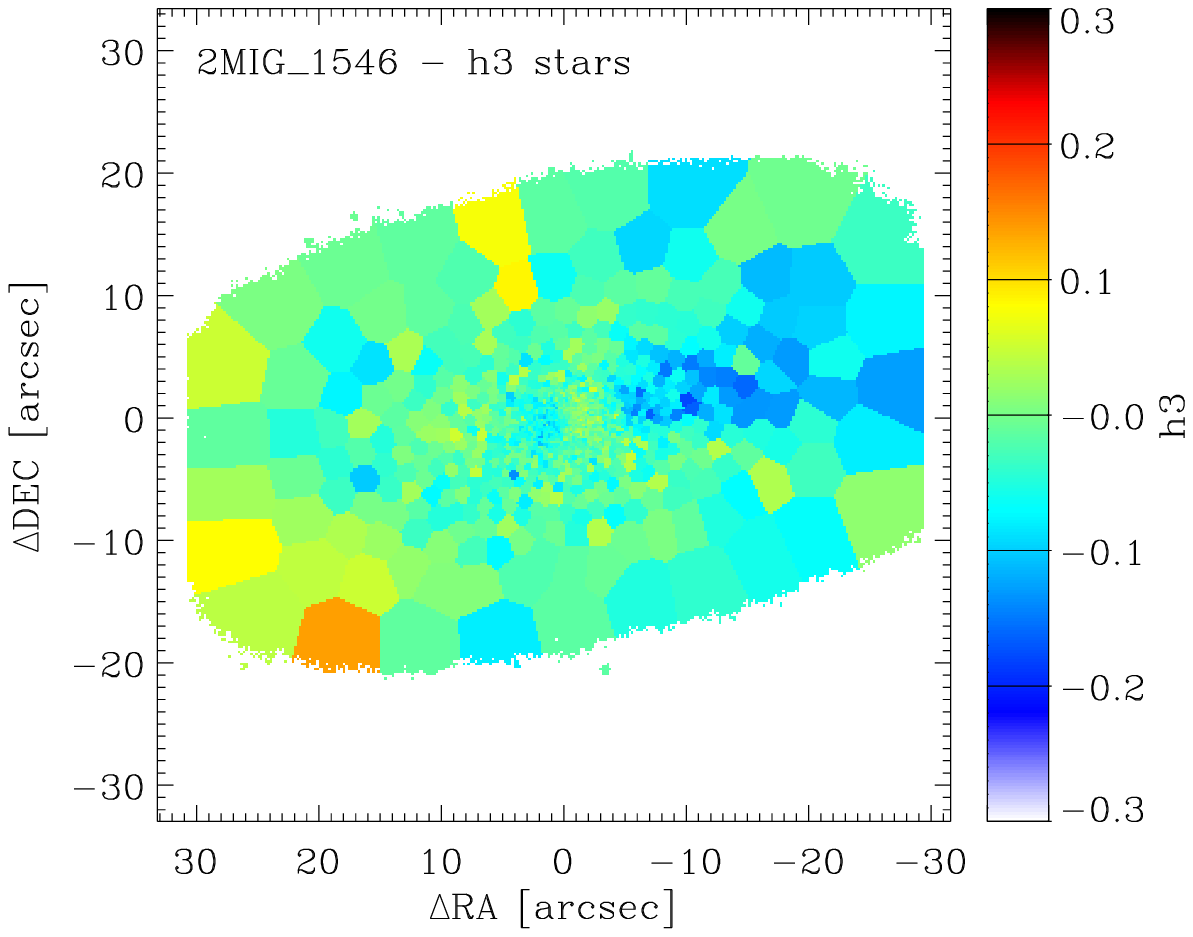}
\includegraphics[width=0.40\textwidth]{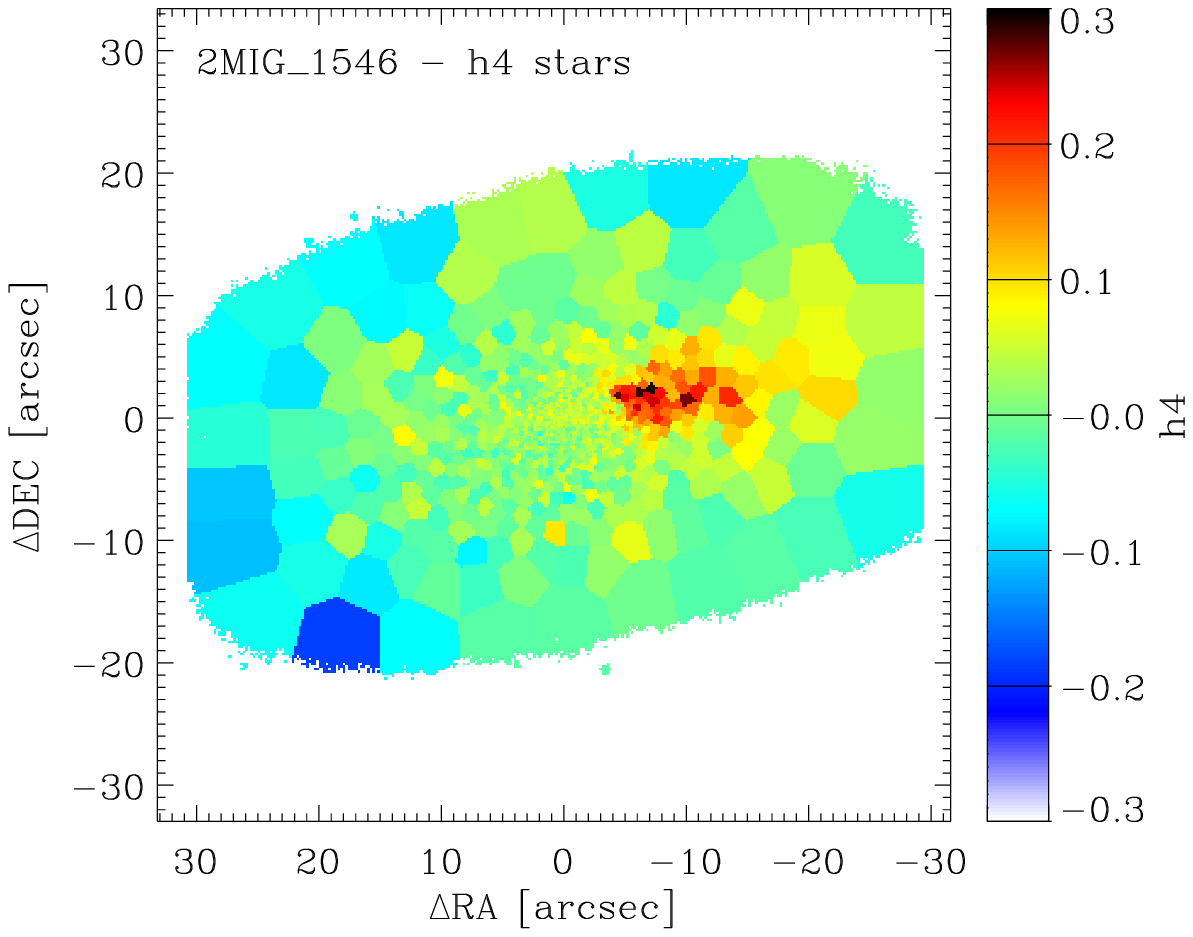}
\includegraphics[width=0.40\textwidth]{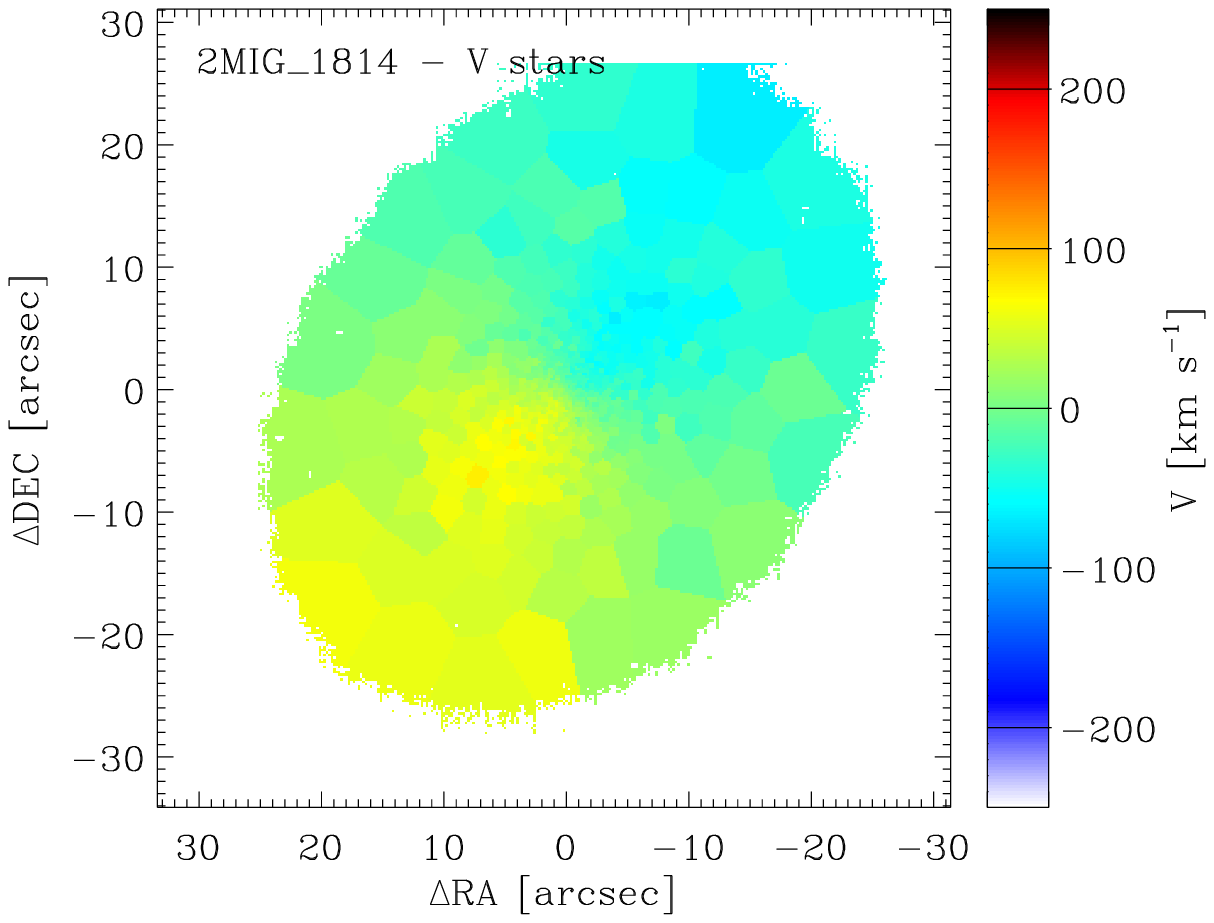}
\includegraphics[width=0.40\textwidth]{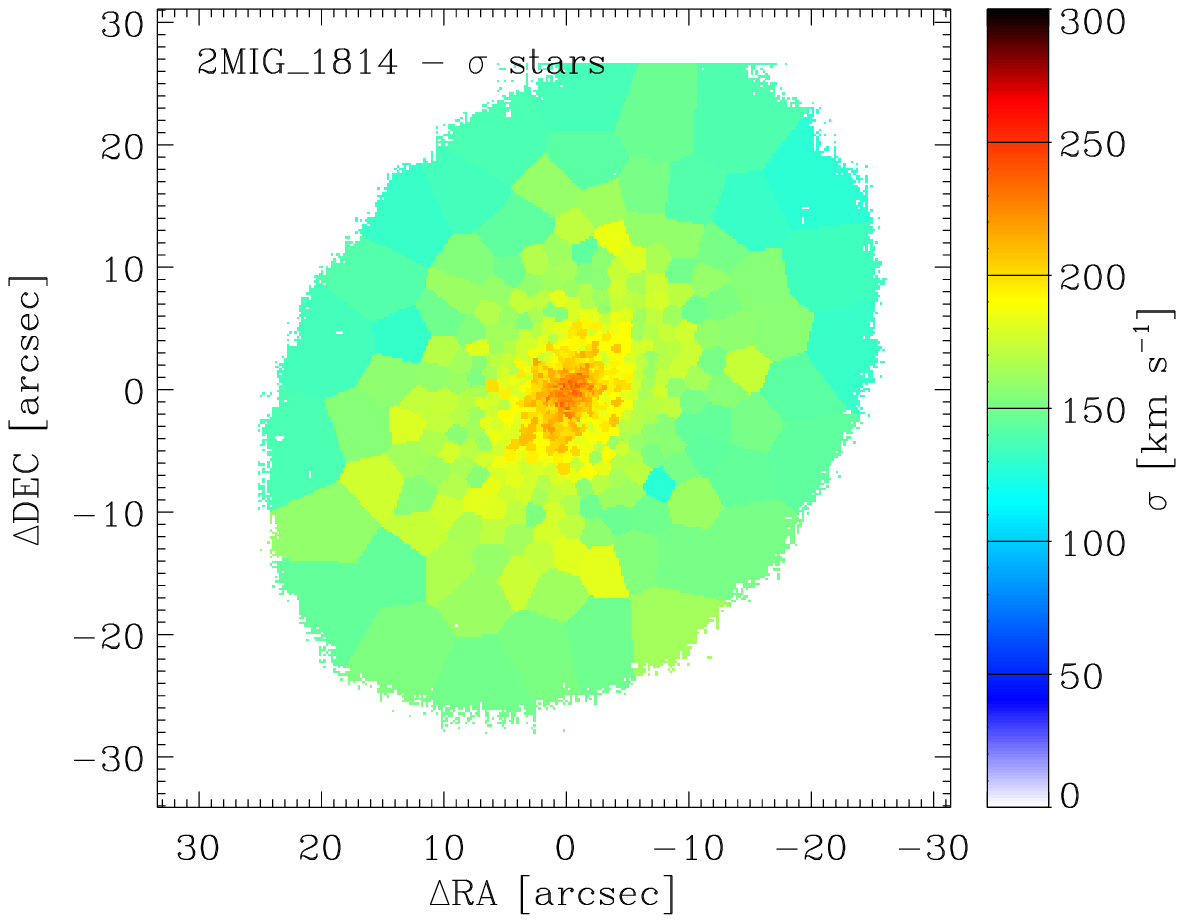}
\includegraphics[width=0.40\textwidth]{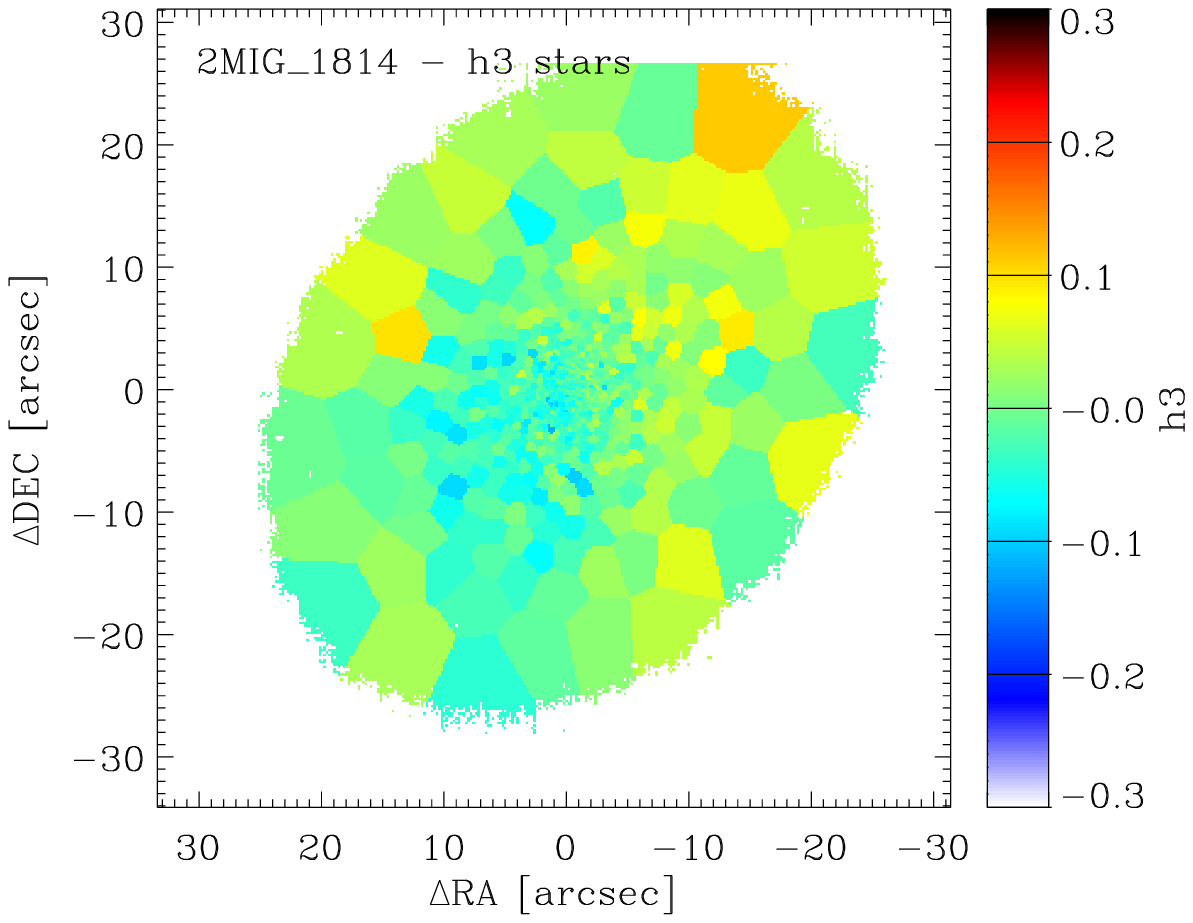}
\includegraphics[width=0.40\textwidth]{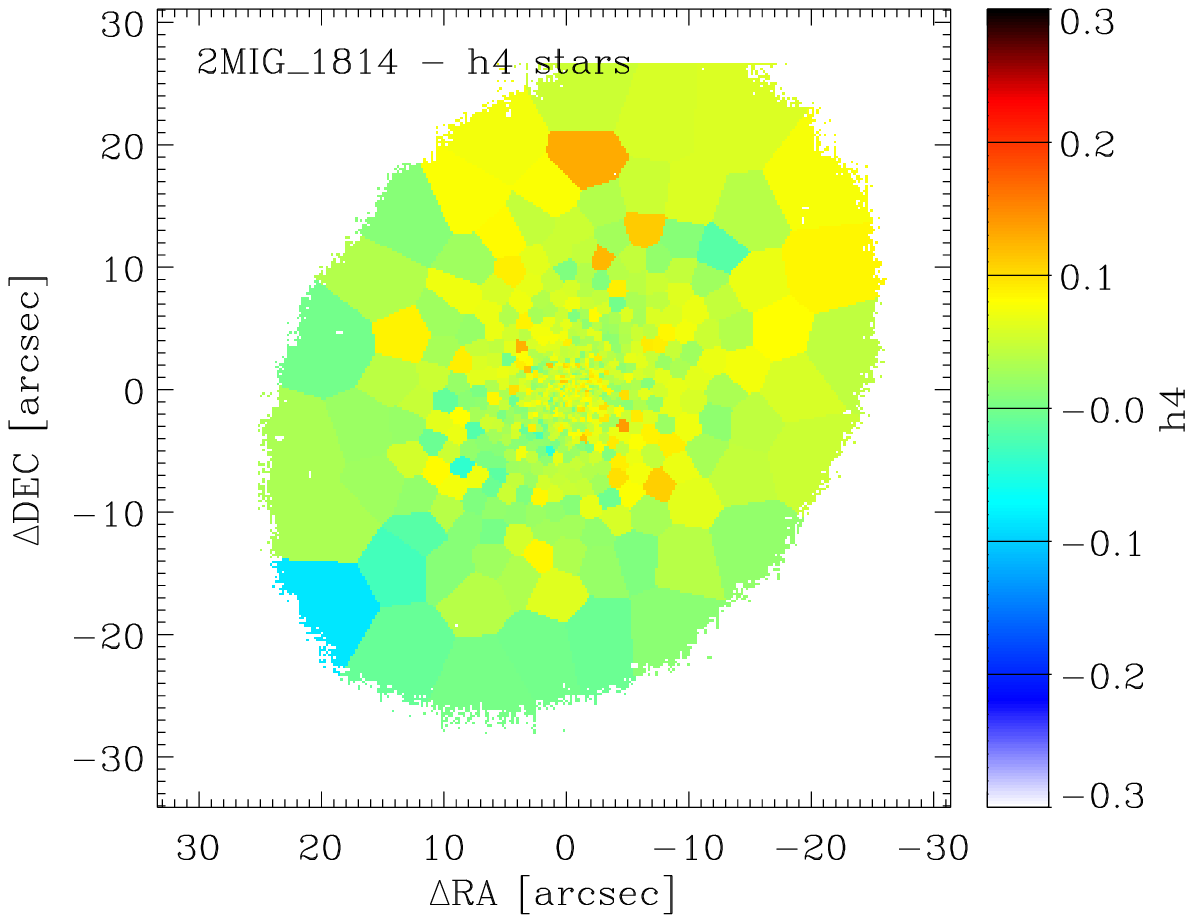}
\caption[]{ -- Continued.}
\label{kinemaps_ii}
\end{figure*}

\begin{figure*}
\includegraphics[width=0.40\textwidth]{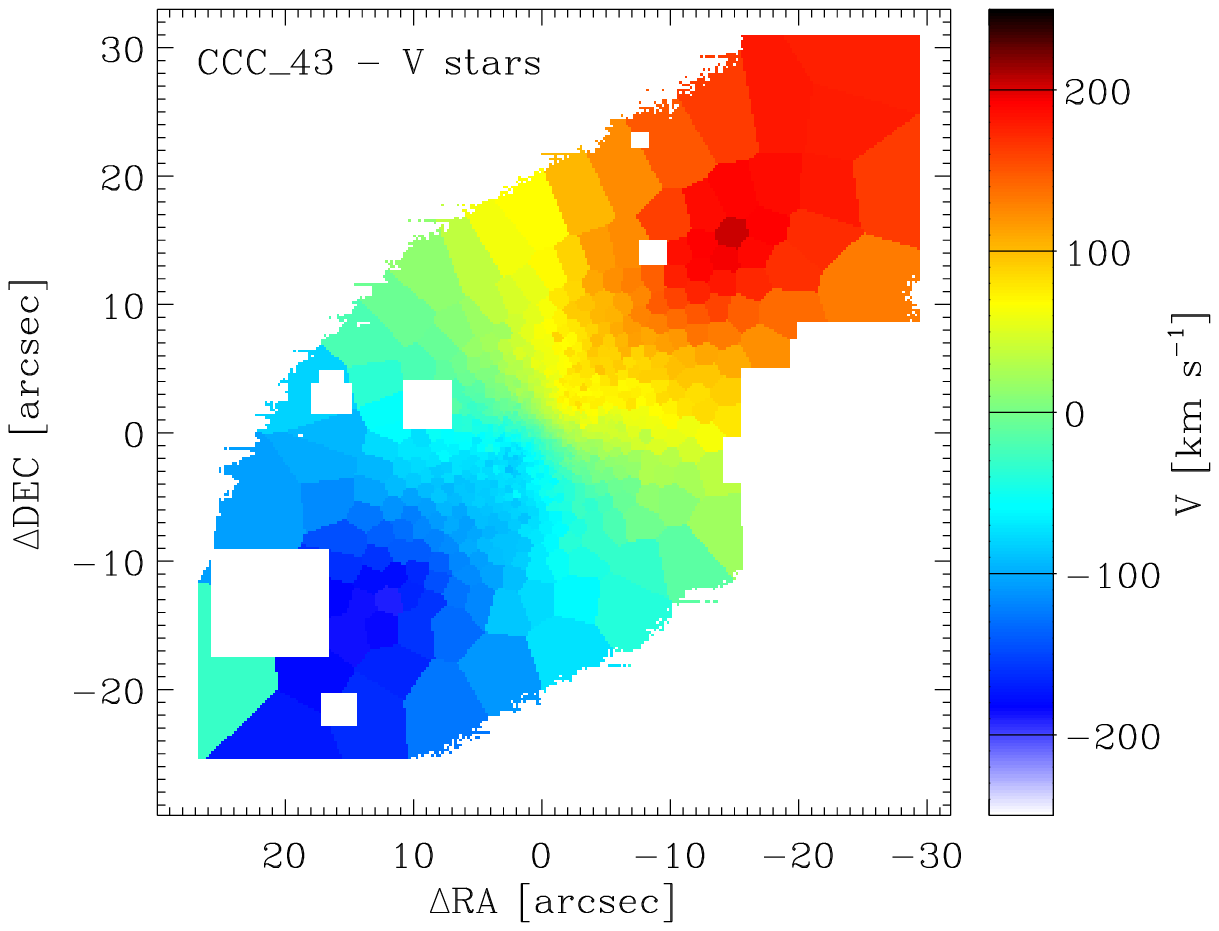}
\includegraphics[width=0.40\textwidth]{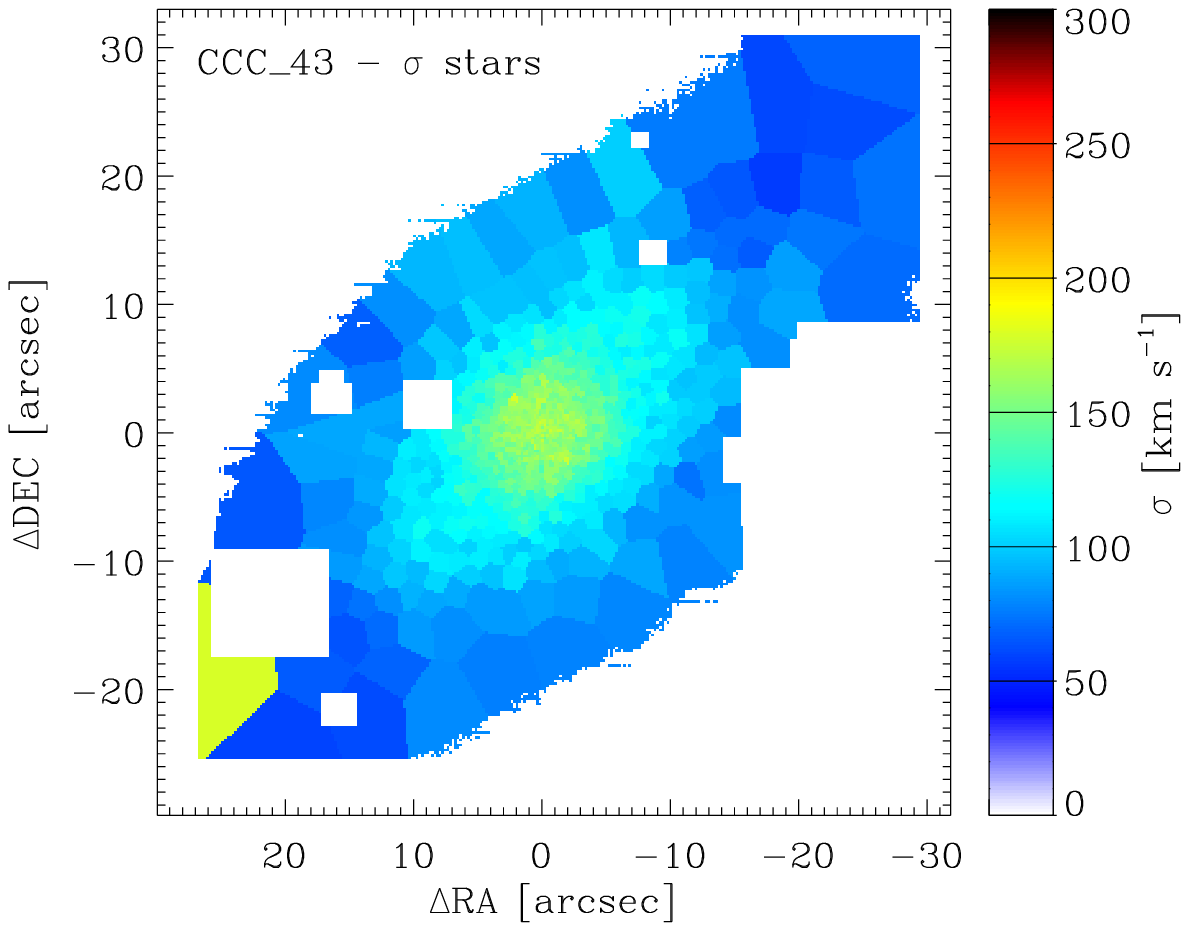}
\includegraphics[width=0.40\textwidth]{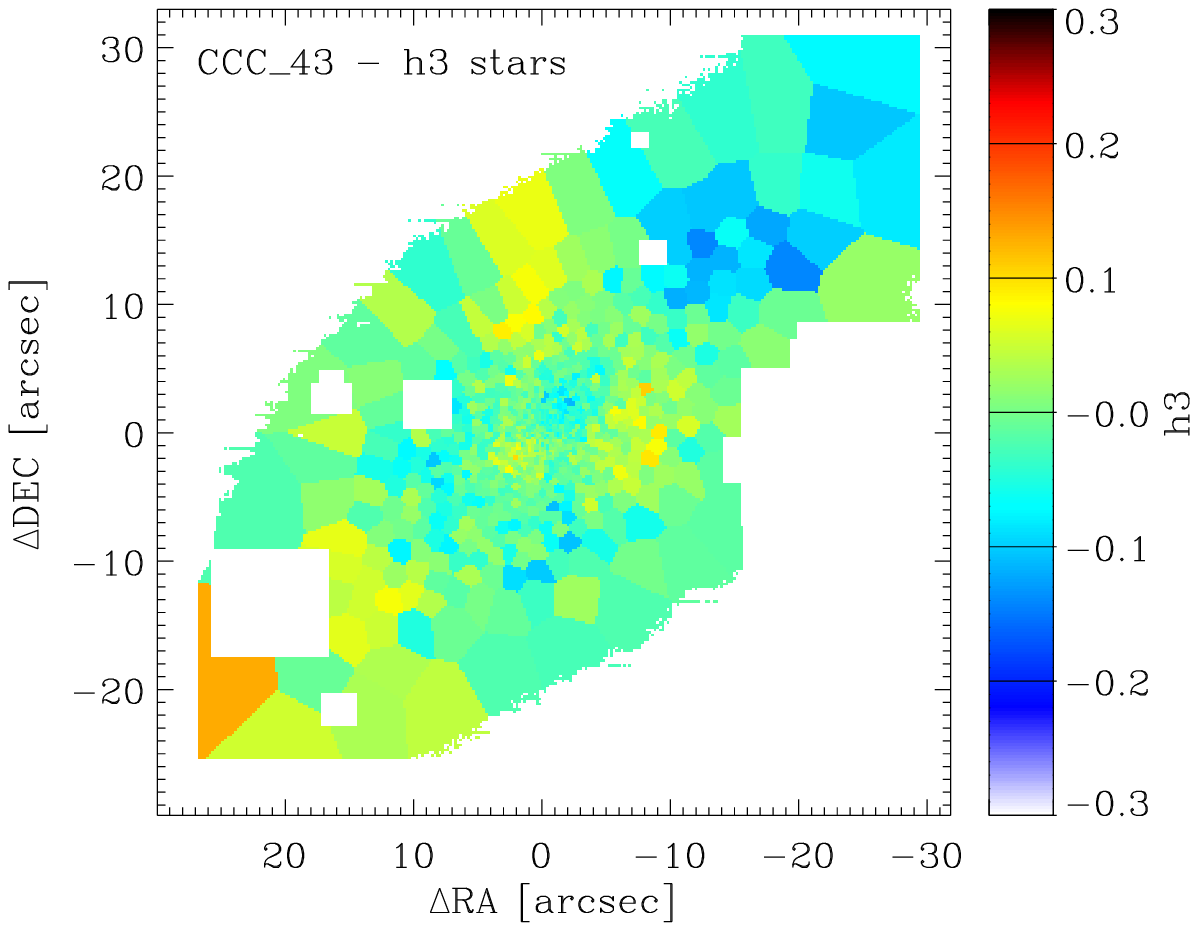}
\includegraphics[width=0.40\textwidth]{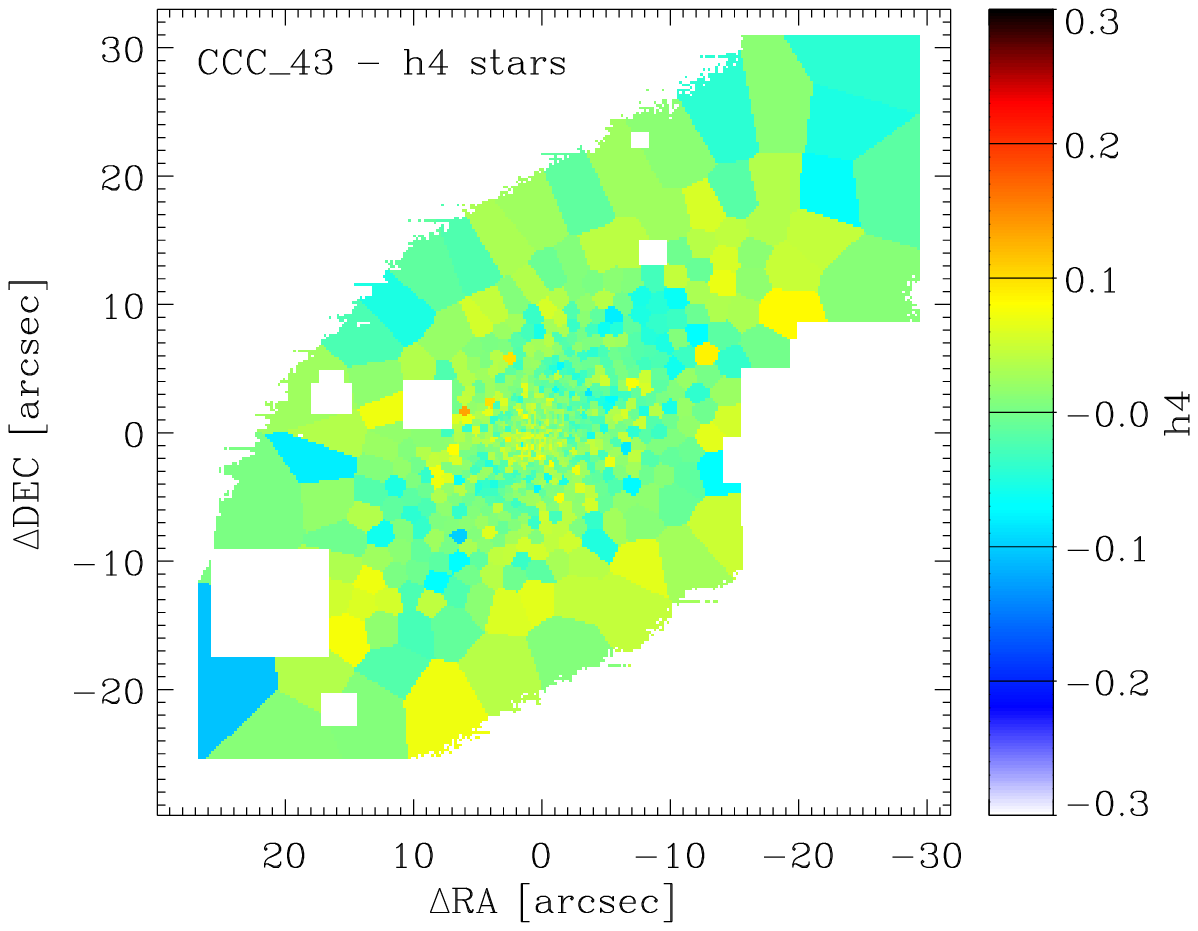}
\includegraphics[width=0.40\textwidth]{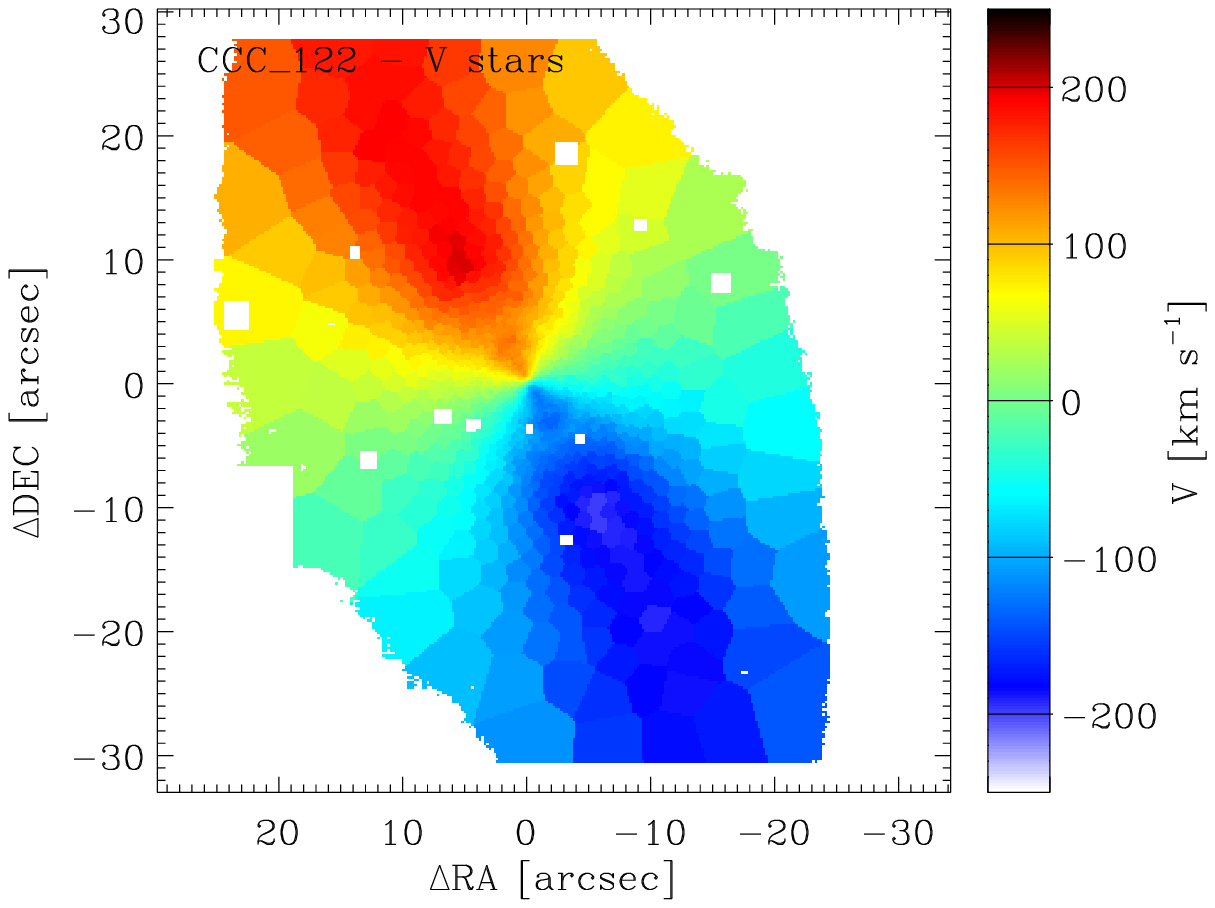}
\includegraphics[width=0.40\textwidth]{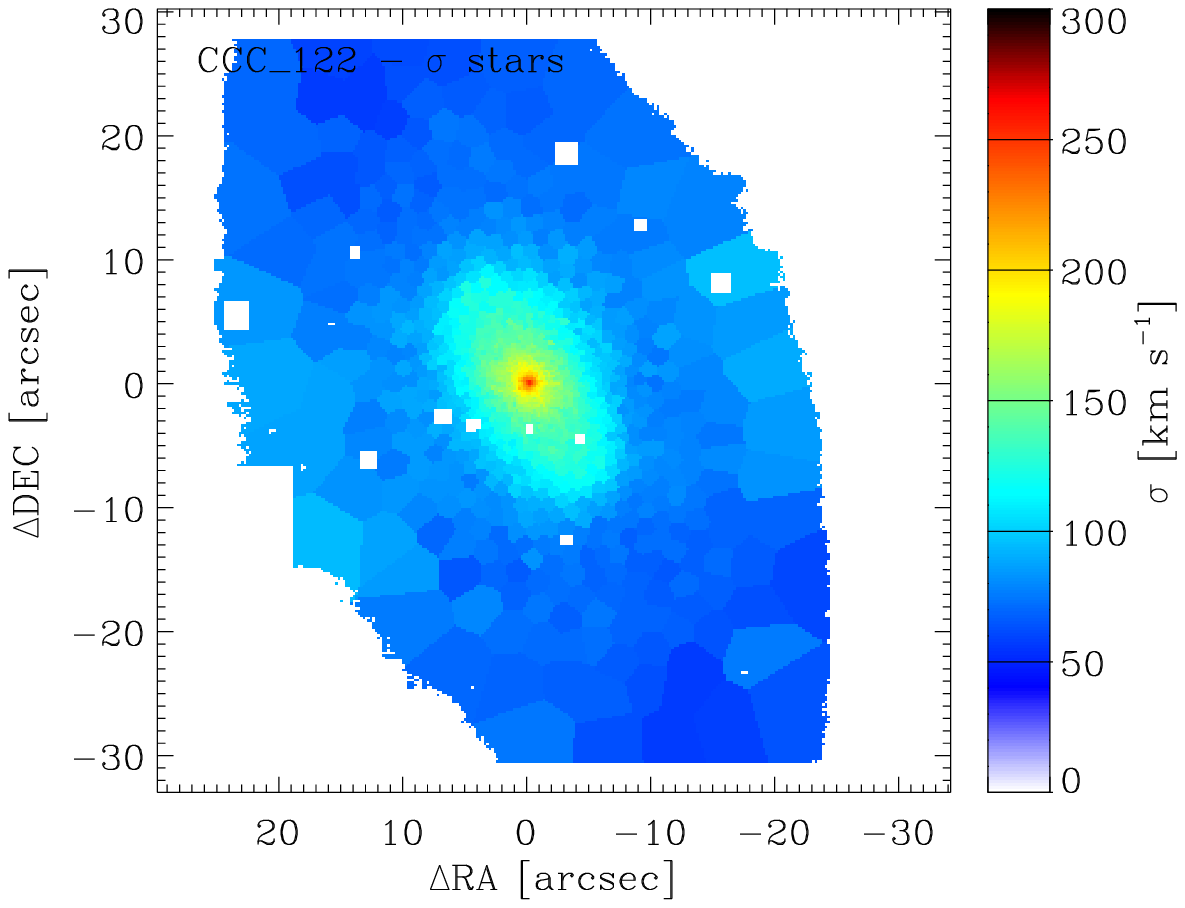}
\includegraphics[width=0.40\textwidth]{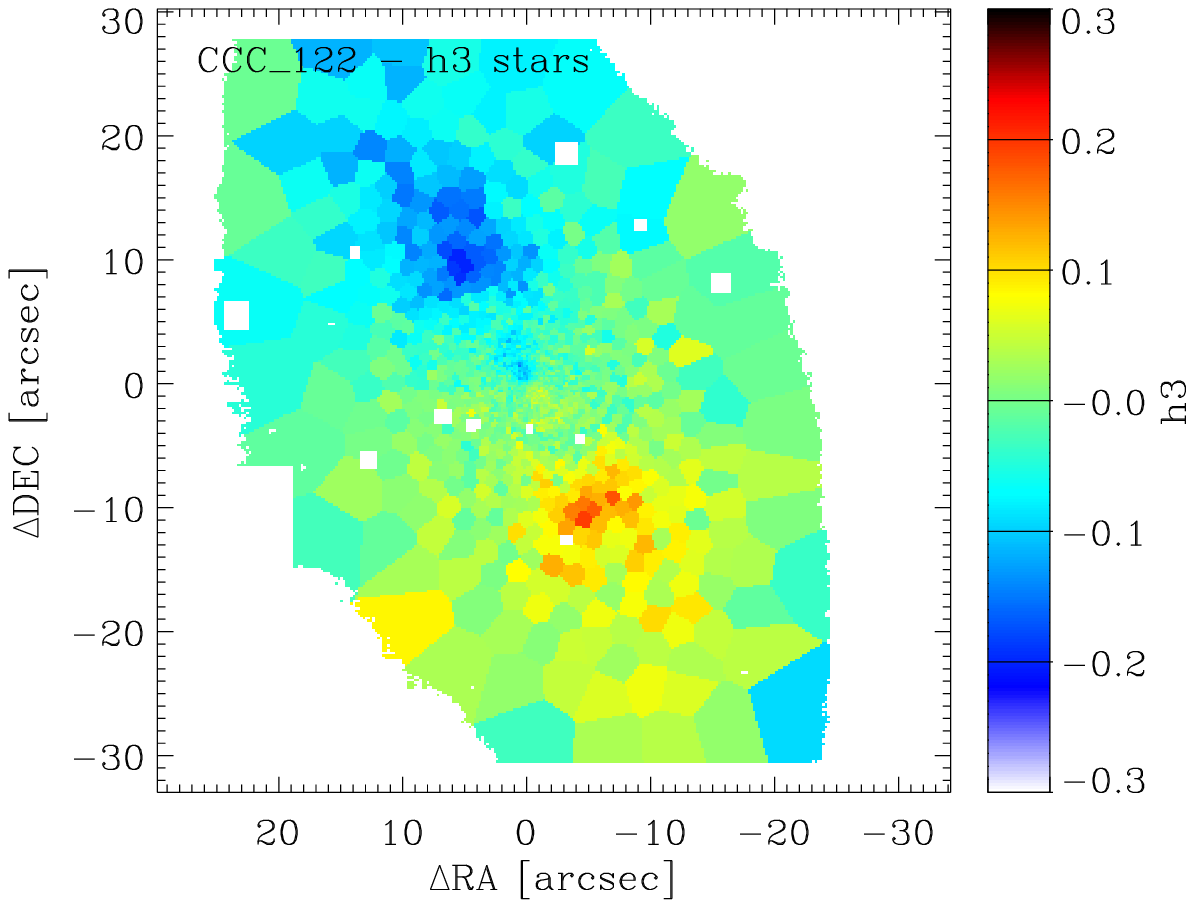}
\includegraphics[width=0.40\textwidth]{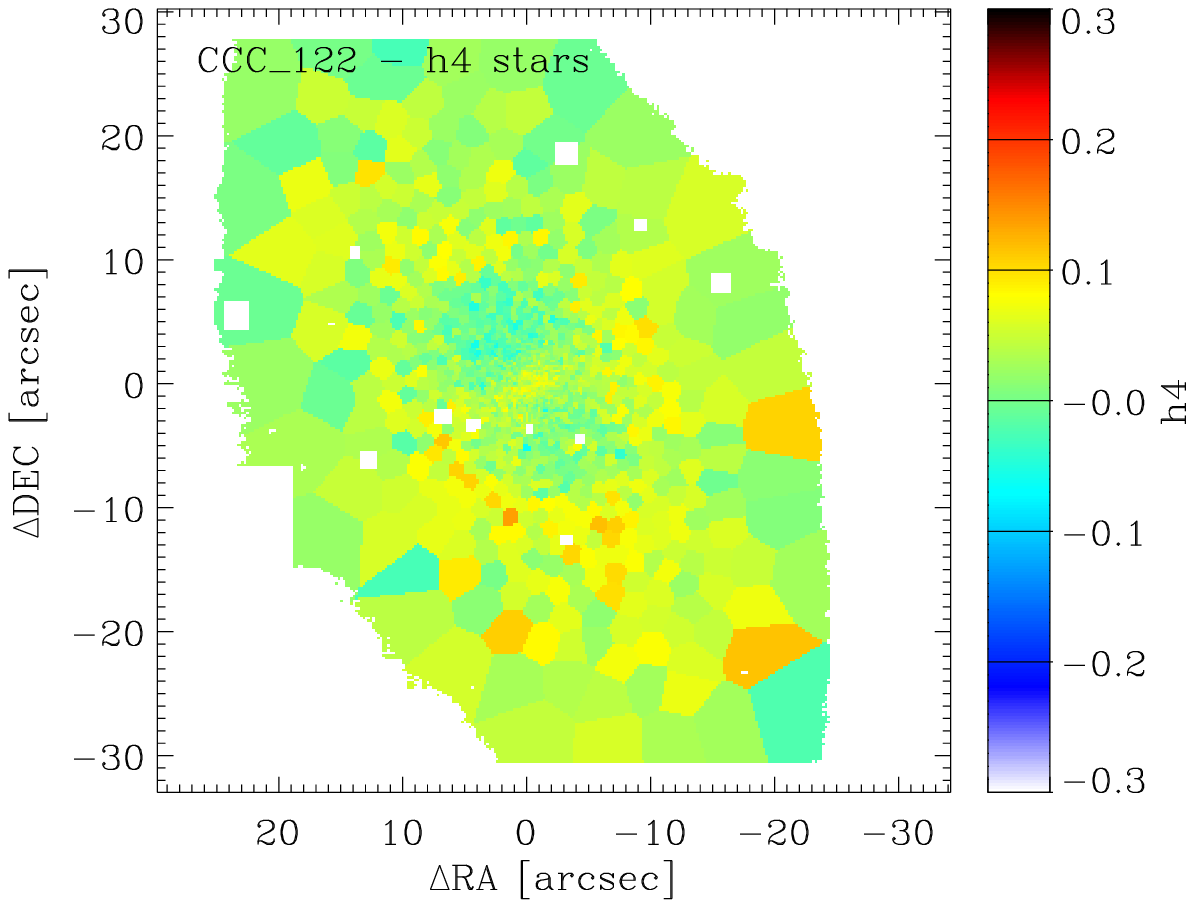}
\caption[]{ Stellar velocity, velocity dispersion, $h3$ , and $h4$ maps for the MUSE sample of Cluster galaxies.}
\label{kinemaps_c}
\end{figure*}
\addtocounter{figure}{-1}
\begin{figure*}
\includegraphics[width=0.40\textwidth]{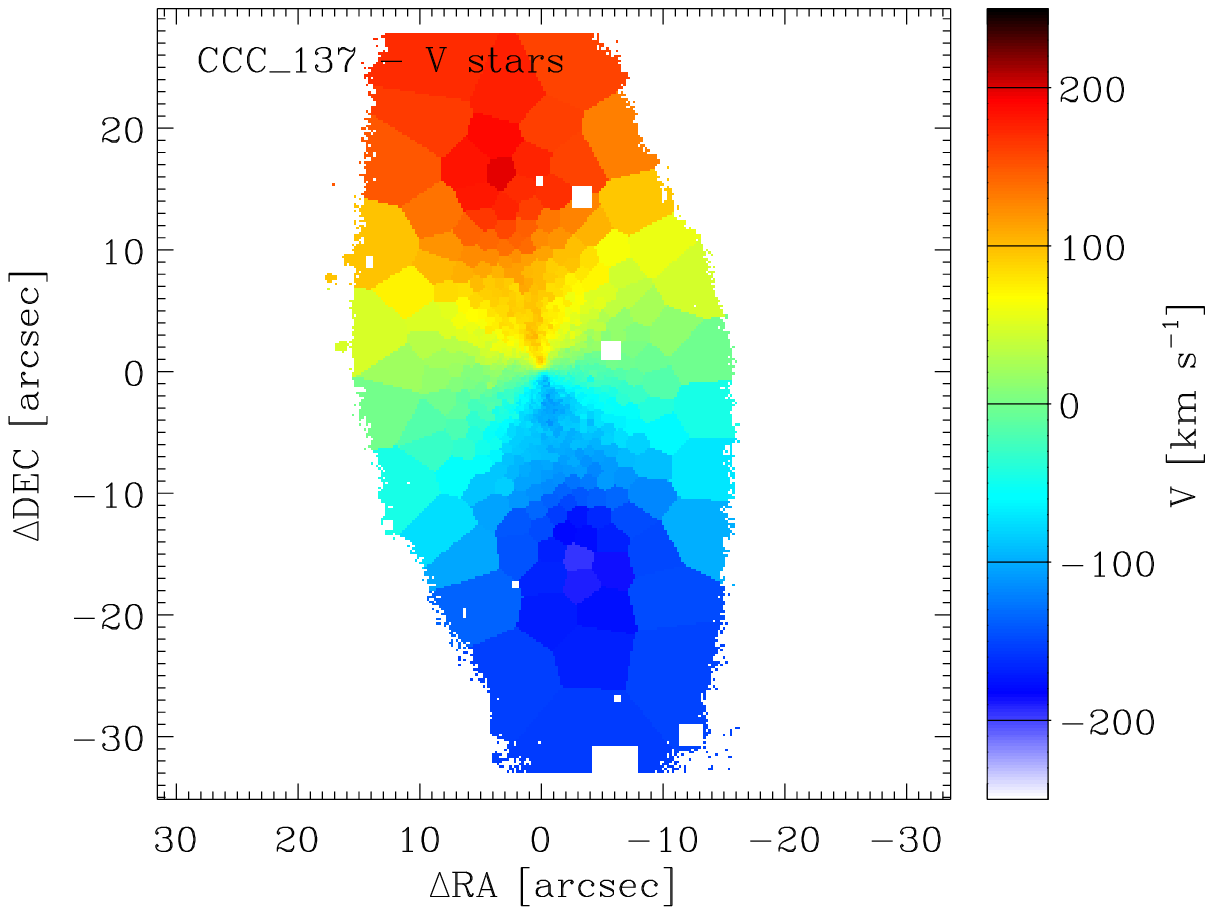}
\includegraphics[width=0.40\textwidth]{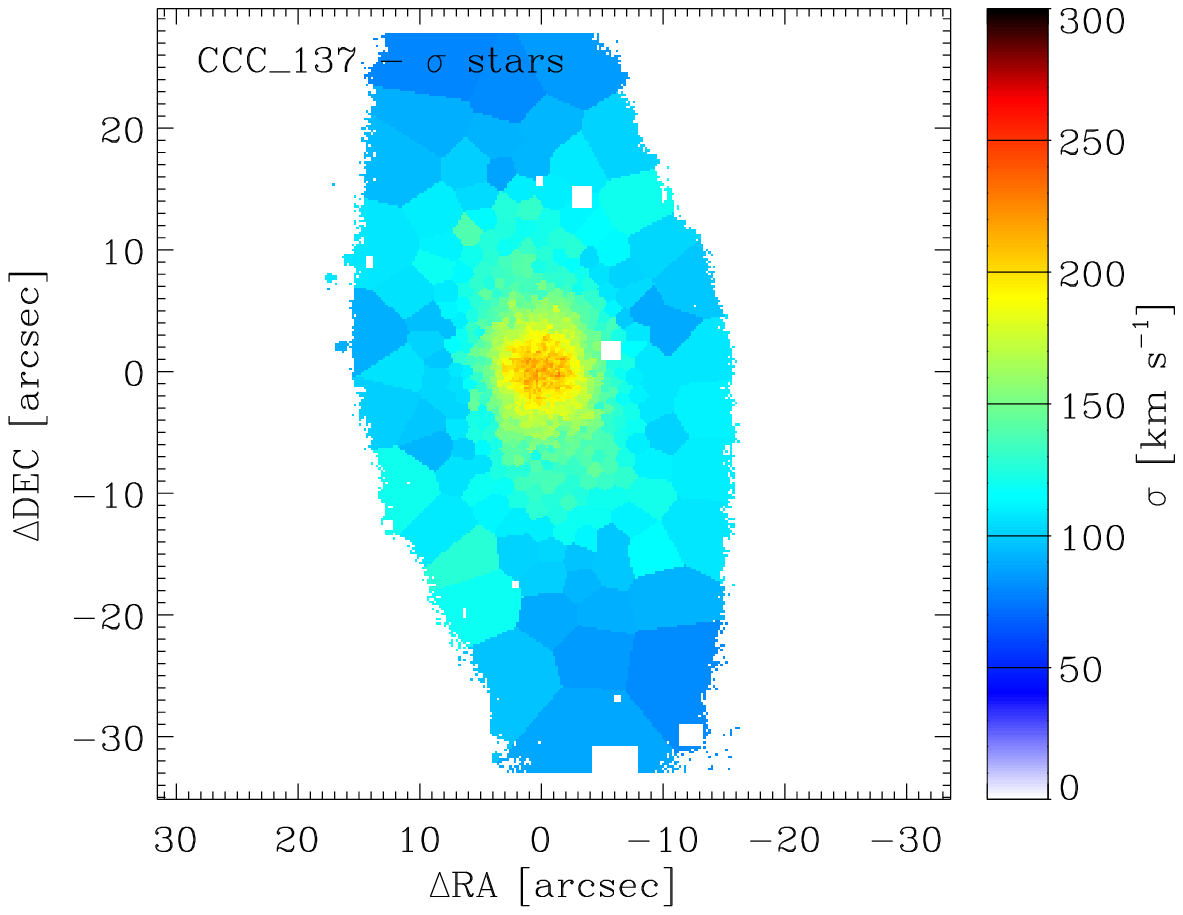}
\includegraphics[width=0.40\textwidth]{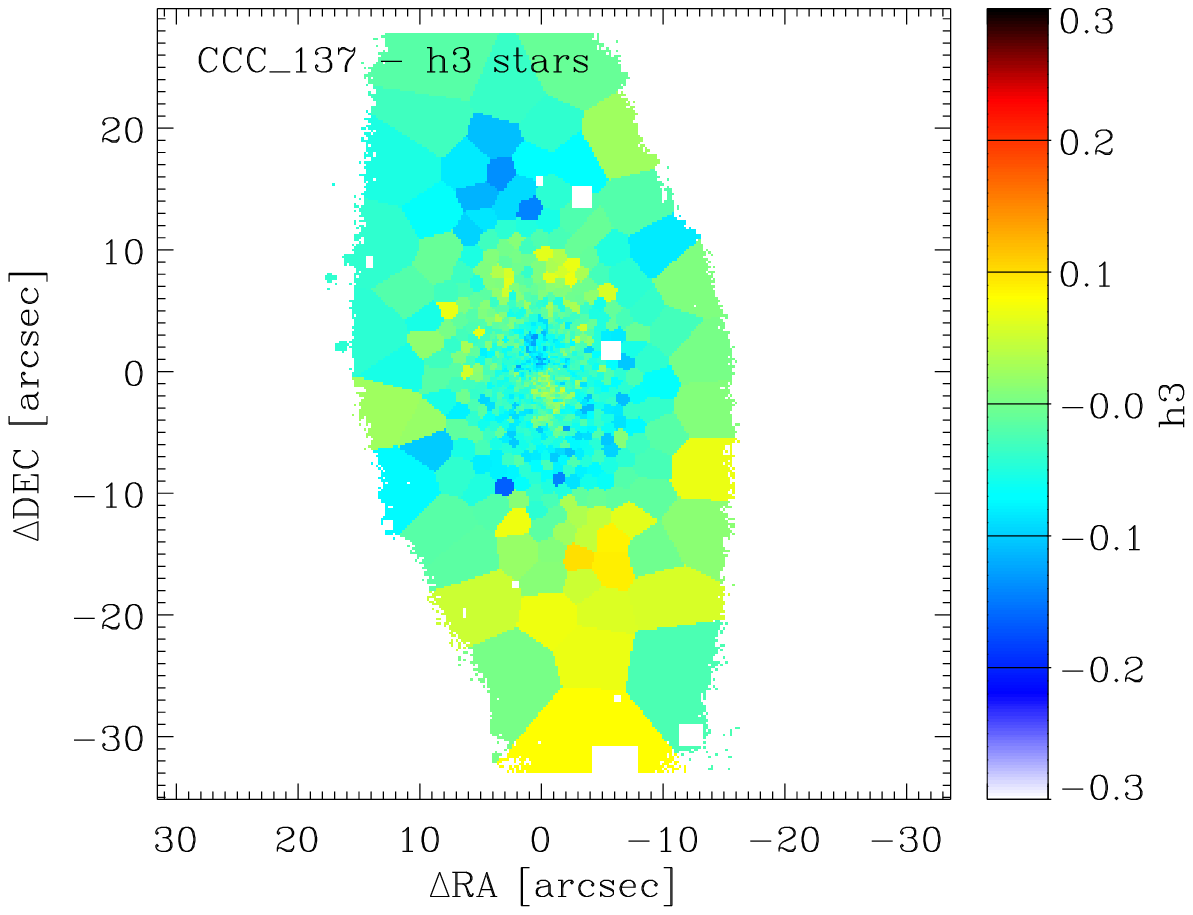}
\includegraphics[width=0.40\textwidth]{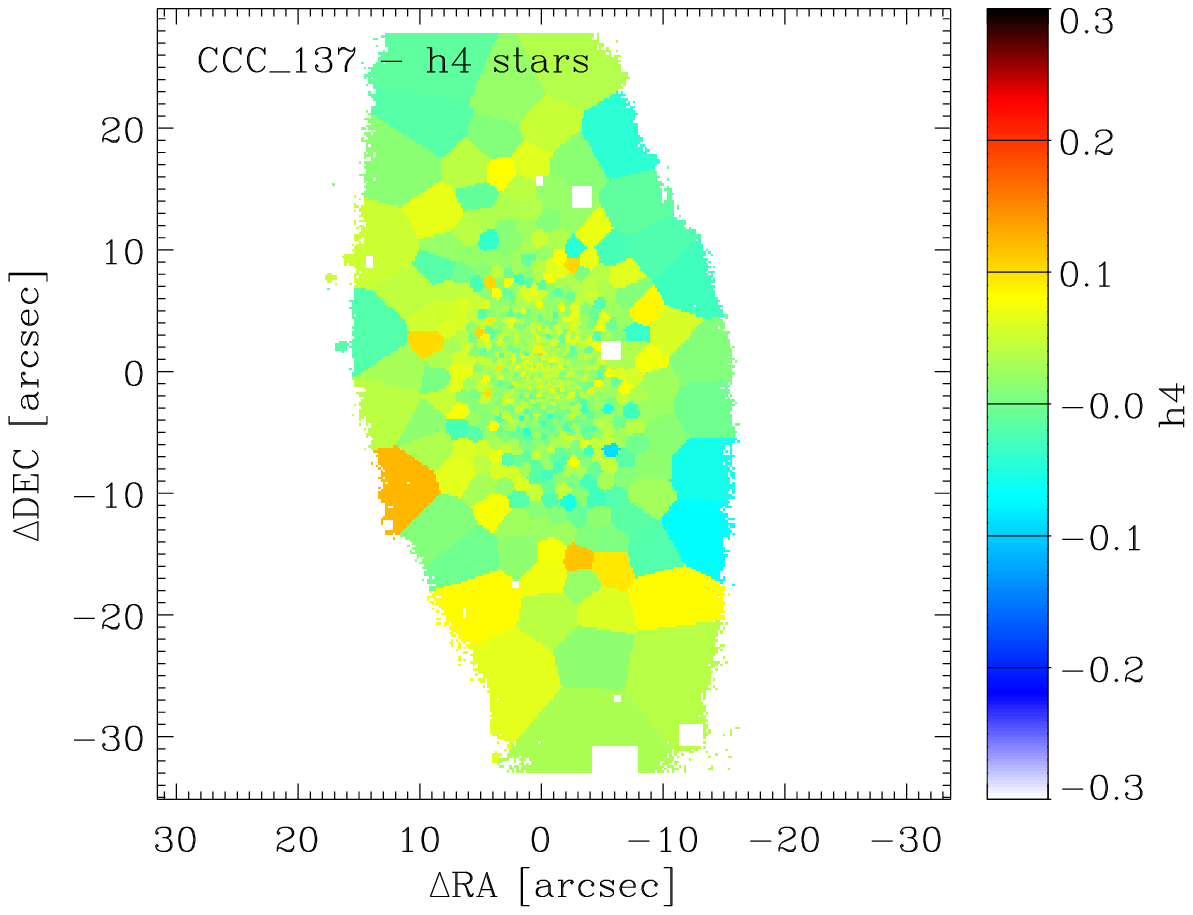}
\includegraphics[width=0.40\textwidth]{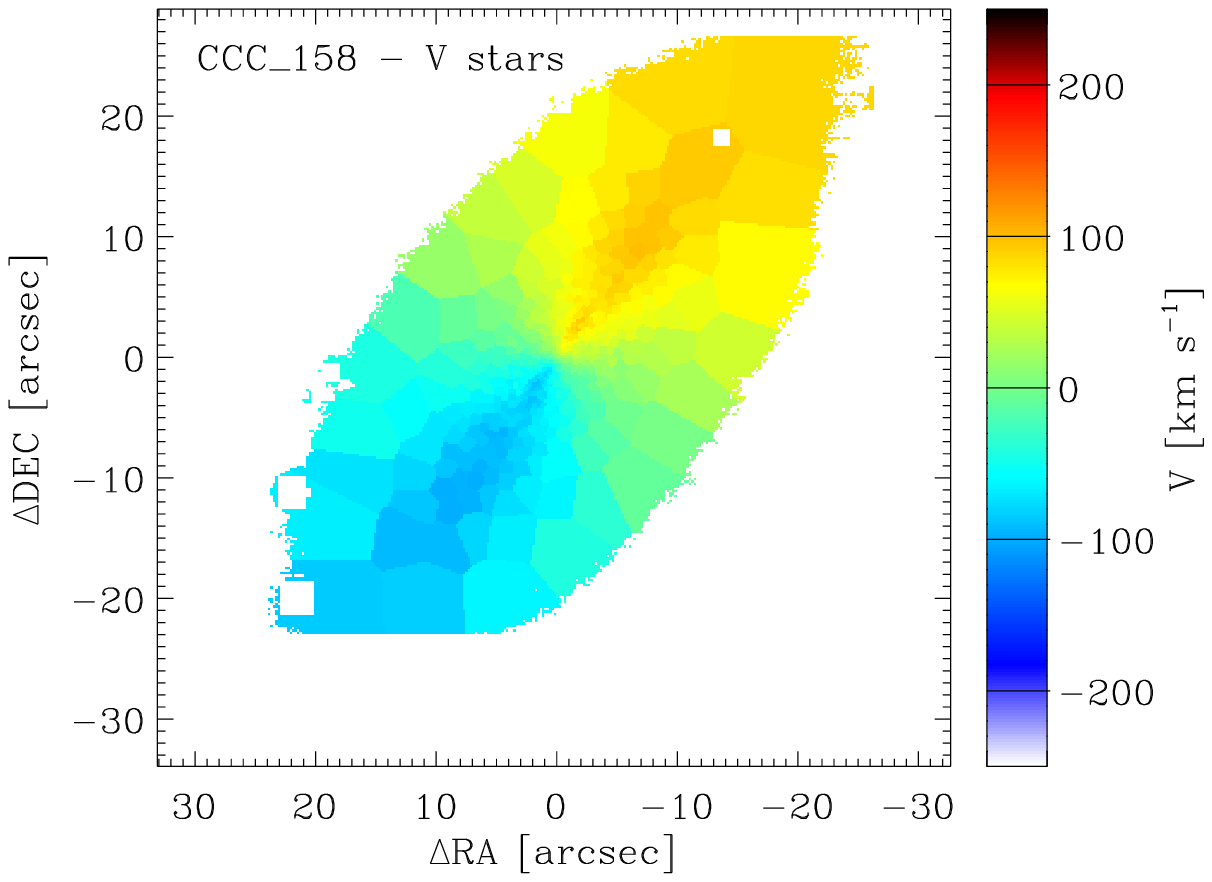}
\includegraphics[width=0.40\textwidth]{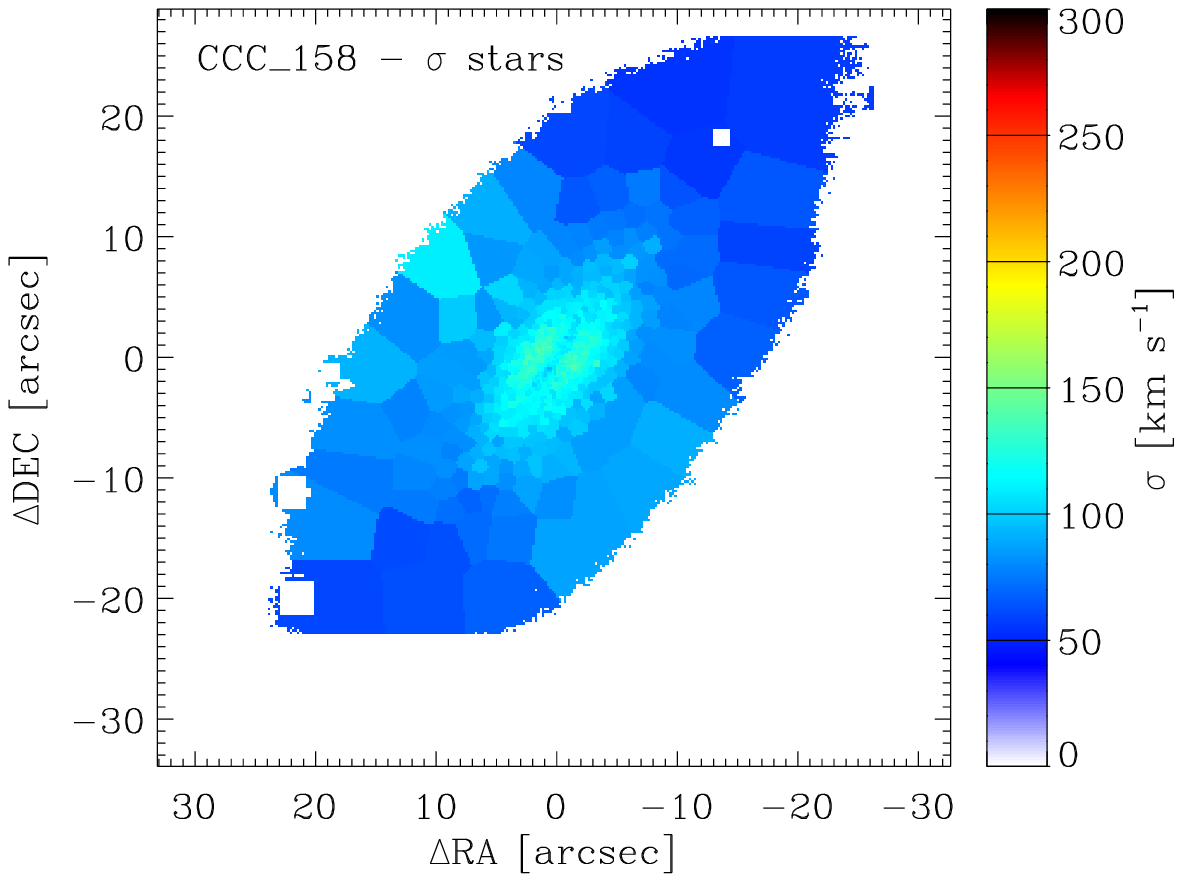}
\includegraphics[width=0.40\textwidth]{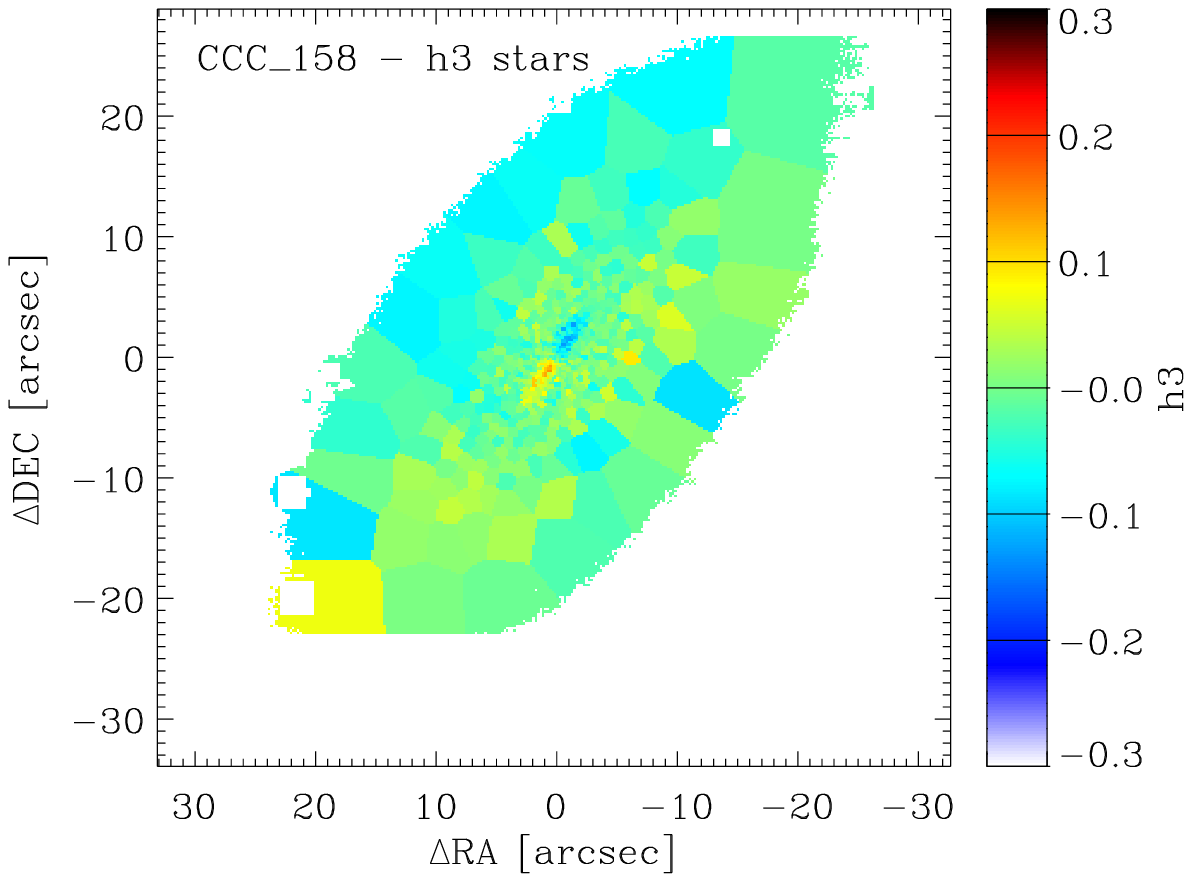}
\includegraphics[width=0.40\textwidth]{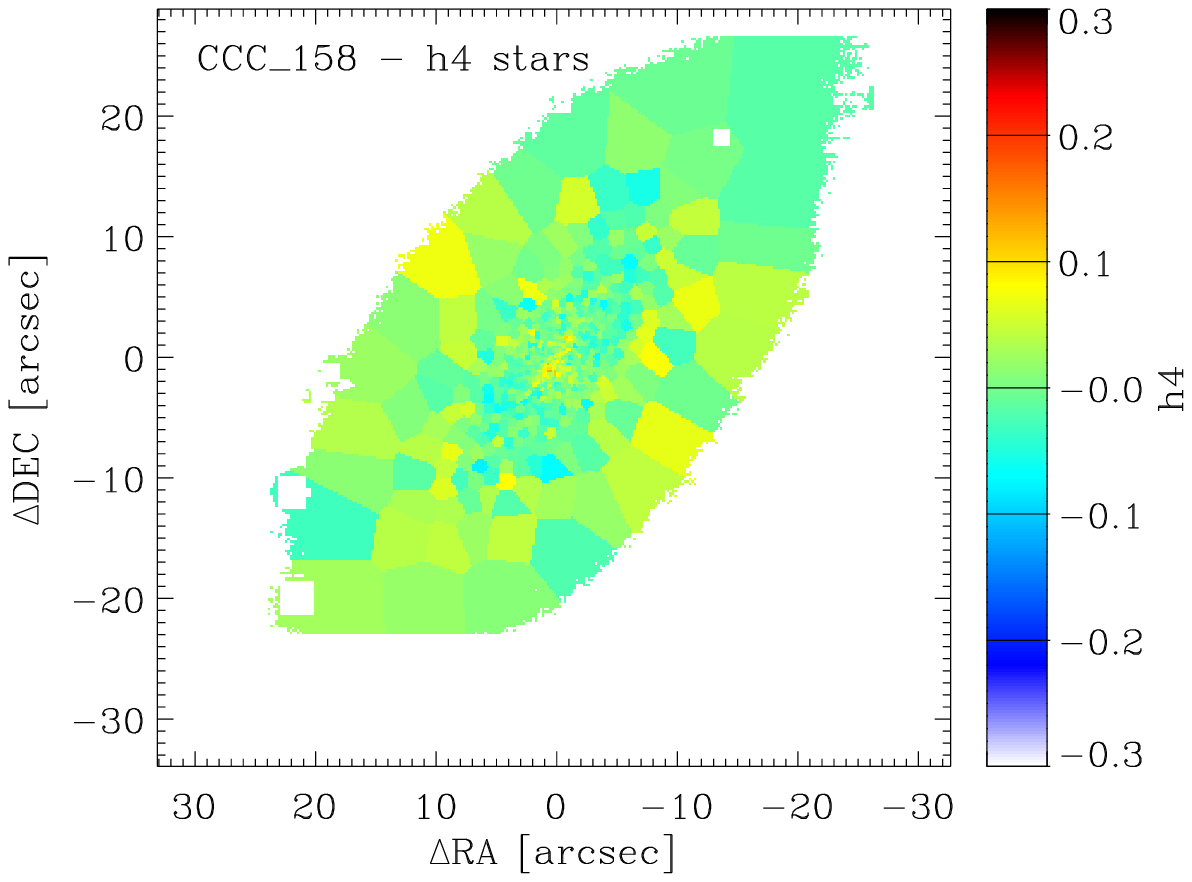}
\caption[]{-- Continued.}
\label{kinemaps_cc}
\end{figure*}

\begin{figure*}
\includegraphics[width=0.40\textwidth]{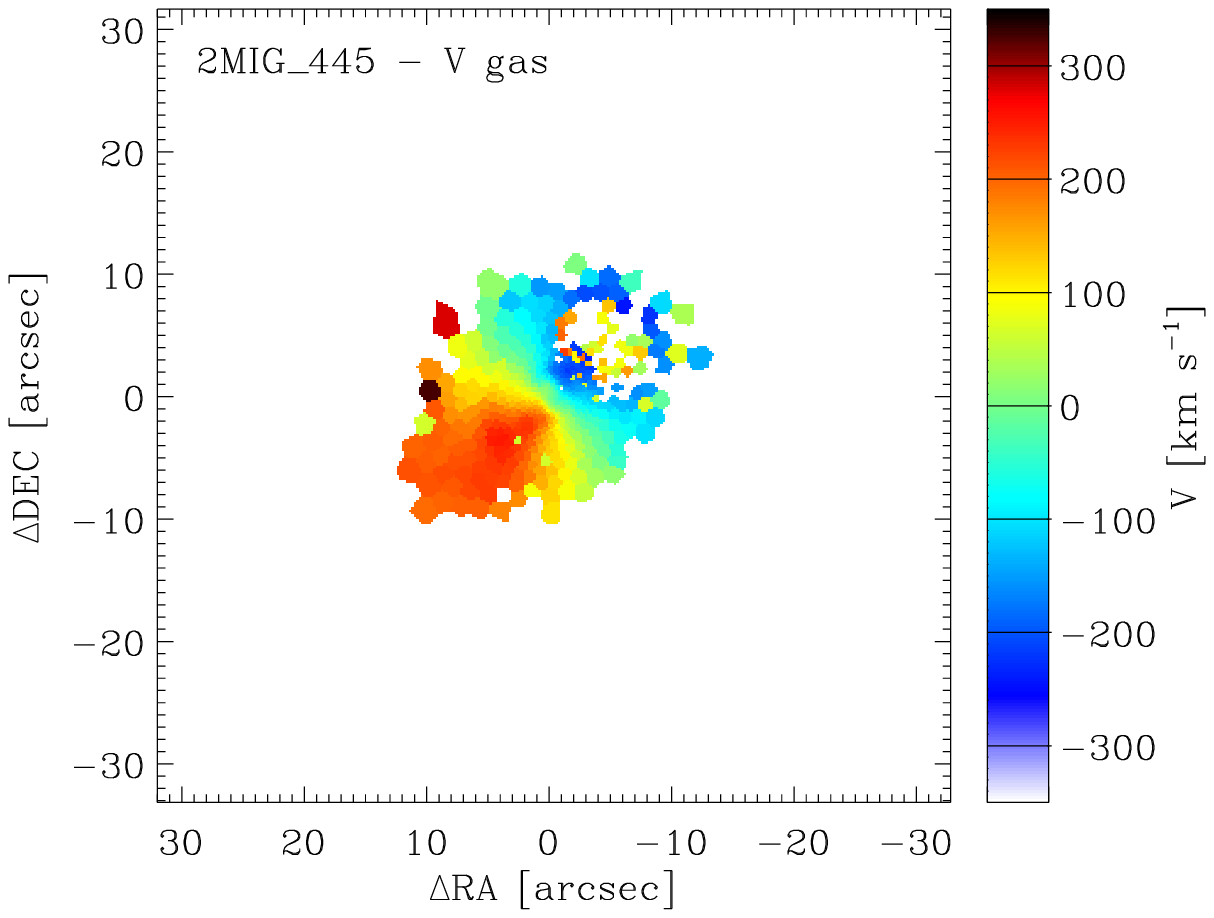}
\includegraphics[width=0.40\textwidth]{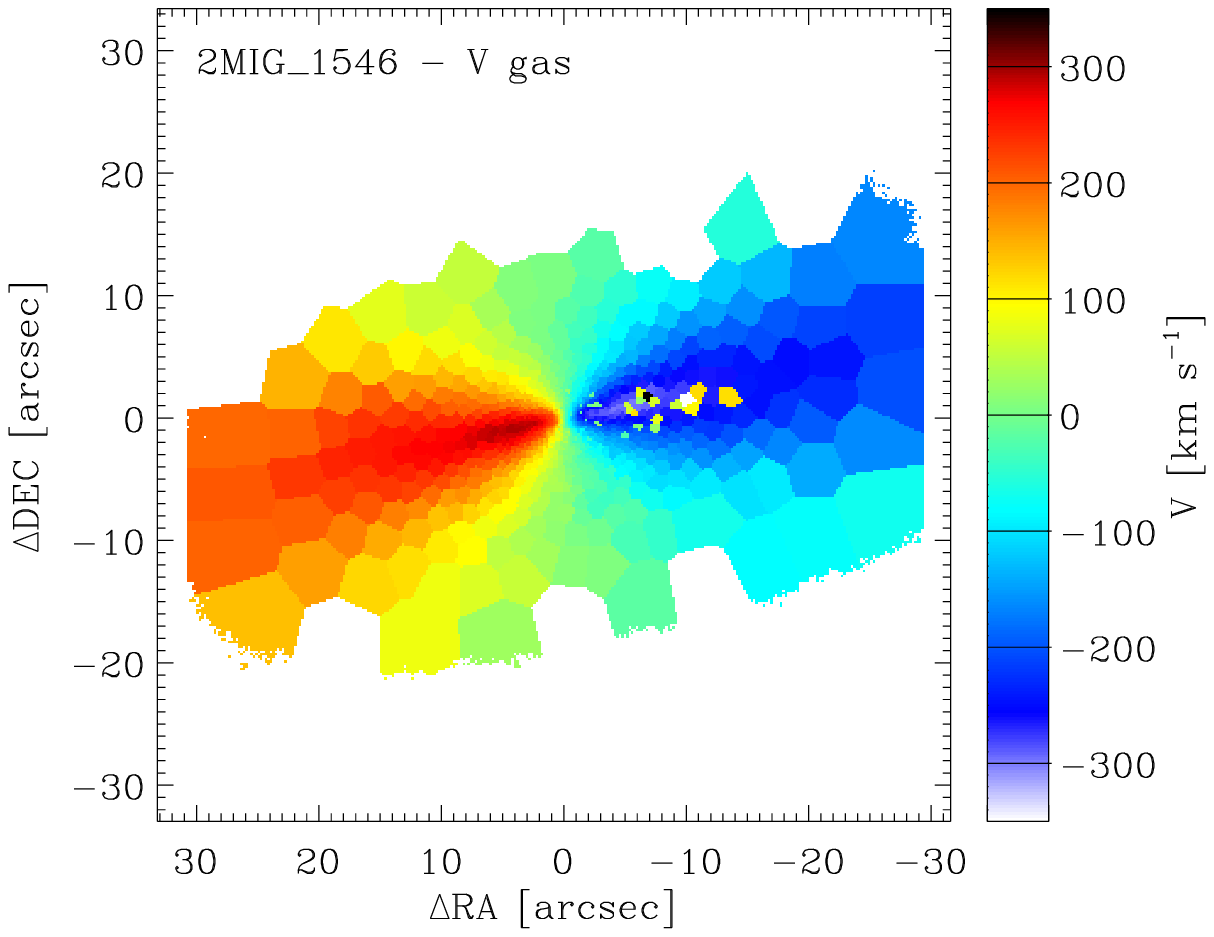}
\includegraphics[width=0.40\textwidth]{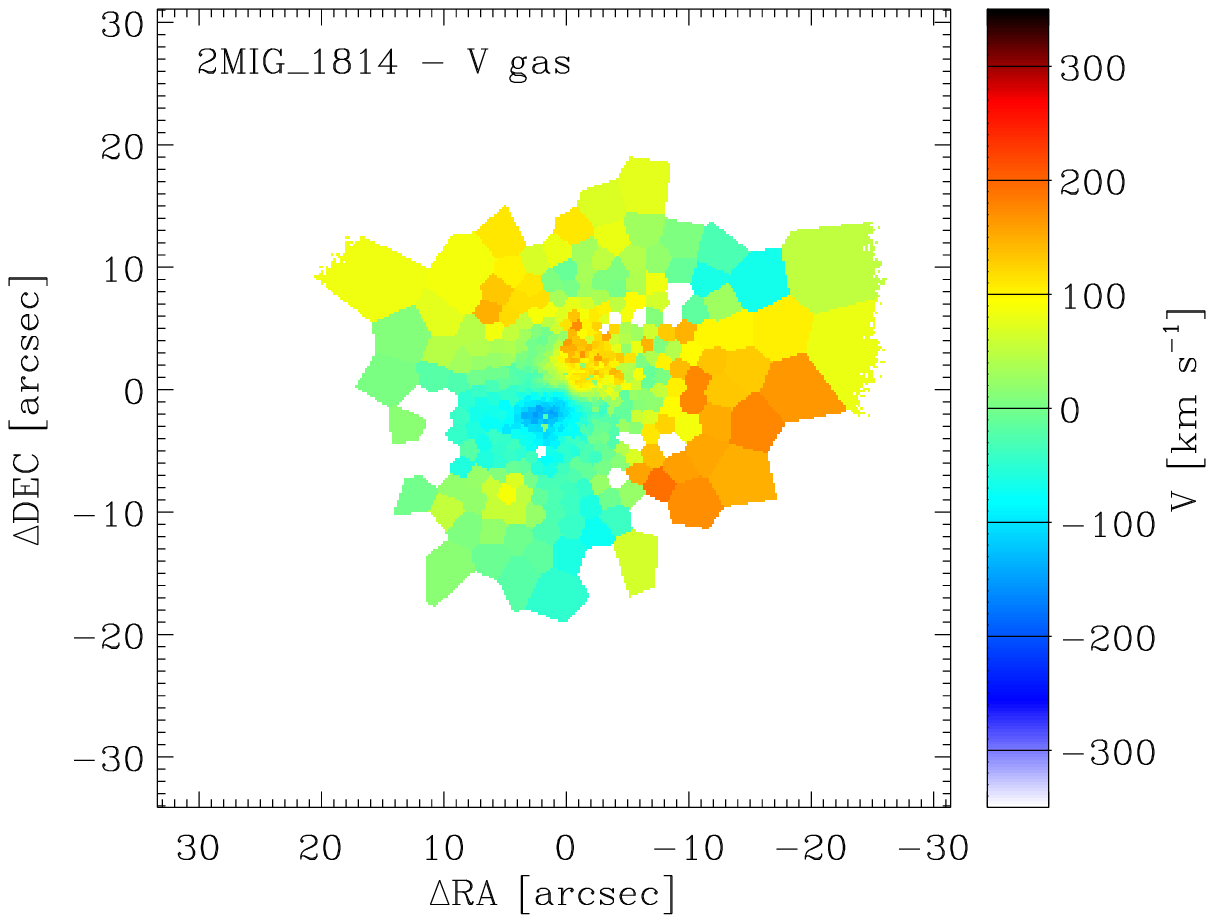}
\includegraphics[width=0.40\textwidth]{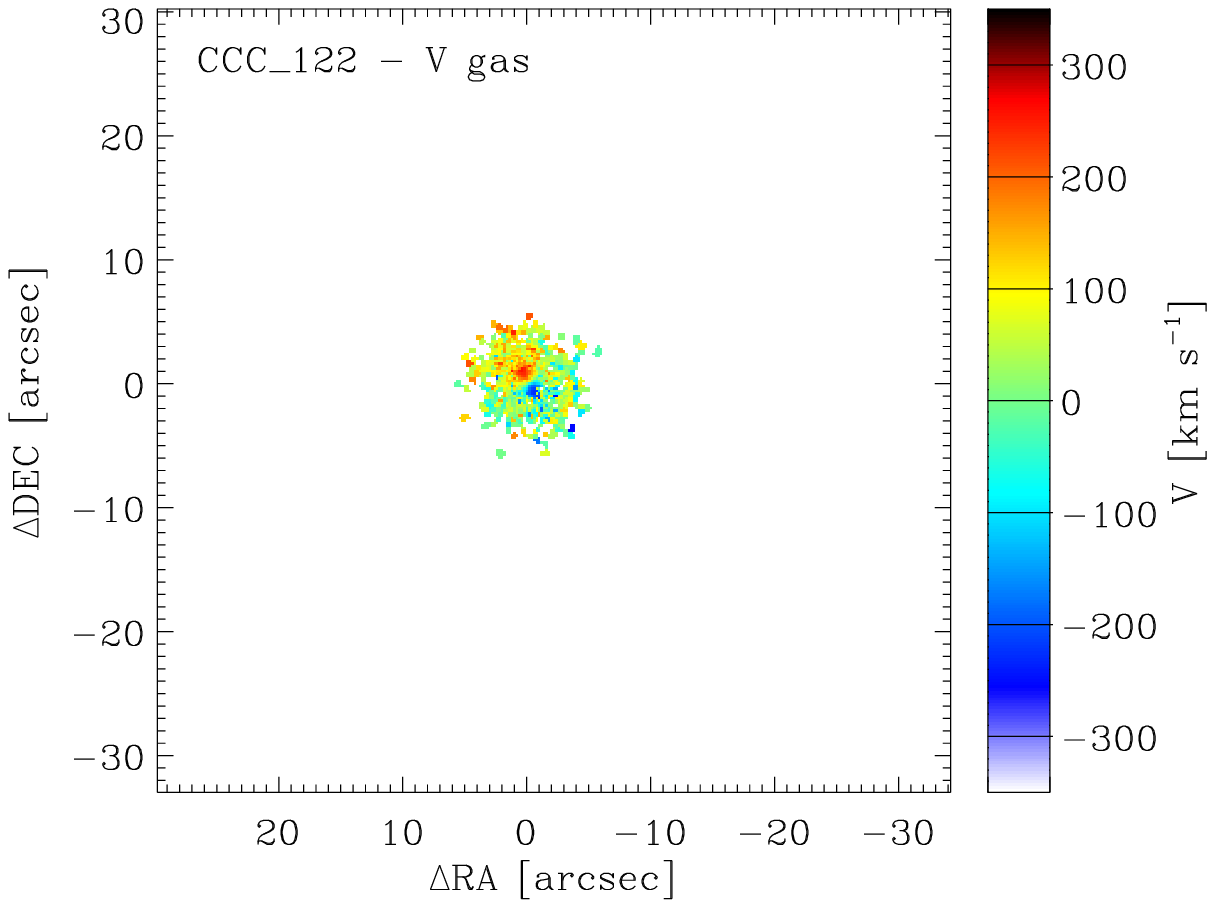}
\includegraphics[width=0.40\textwidth]{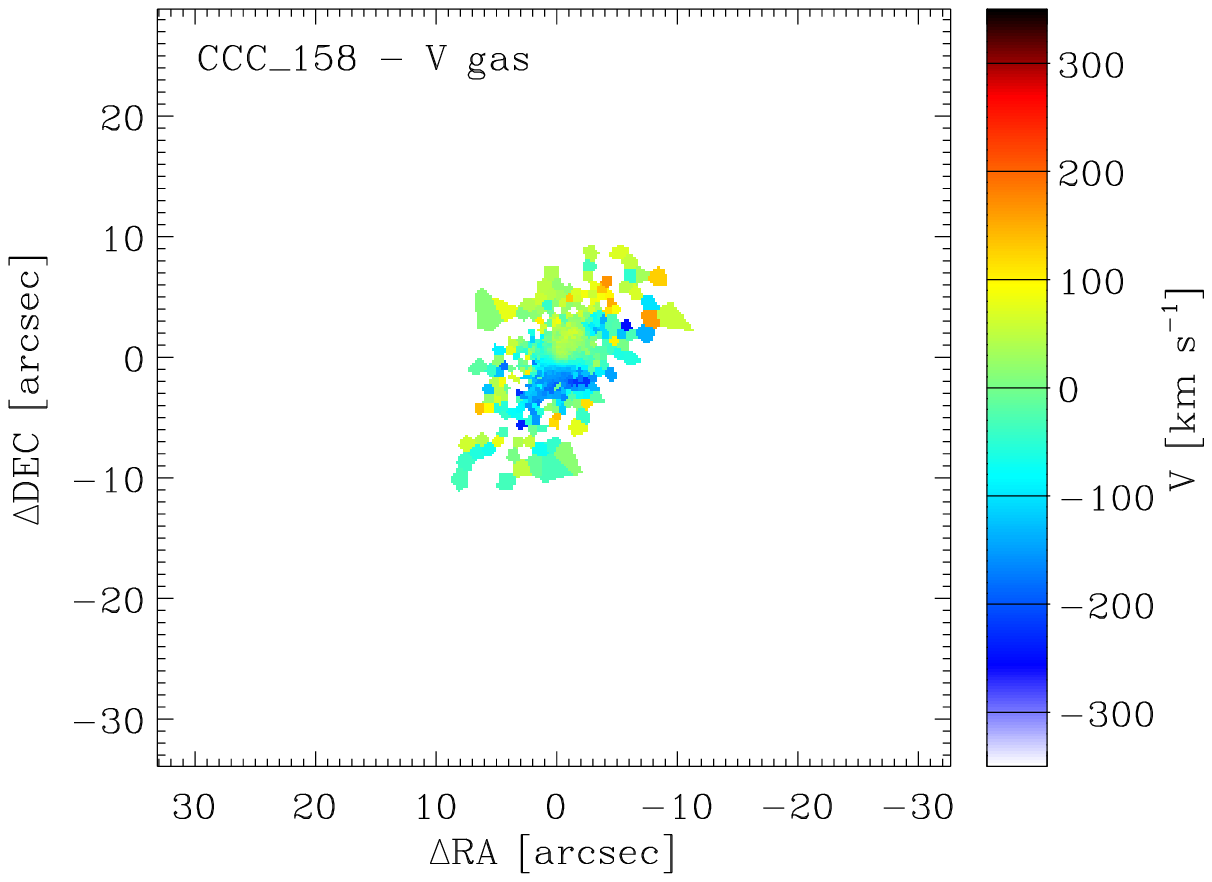}
\caption[]{Ionised-gas velocity maps for the galaxies in our MUSE sample for which there is a significant detection of ionised gas.}
\label{kinemaps_gas}
\end{figure*}

\label{lastpage}
\end{document}